\documentclass[10pt]{elsarticle}
\usepackage{hyperref}
\usepackage[utf8]{inputenc}
\usepackage{wrapfig}
\usepackage{amsmath}
\usepackage{amsthm}
\usepackage{amssymb}
\usepackage[tmargin=1in,bmargin=1in]{geometry}
\usepackage{caption}
\usepackage{subcaption}
\usepackage{graphicx}
\usepackage{latexsym}
\usepackage[boxed]{algorithm}
\usepackage{algorithm}
\usepackage{algpseudocode}
\usepackage{multirow}
\usepackage{rotating}
\usepackage{url}
\usepackage{bigstrut}
\usepackage[utf8]{inputenc}
\usepackage[overload]{empheq}
\usepackage{cases}  
\usepackage{ulem}
\usepackage{mleftright}
\usepackage{algpseudocode}
\usepackage{lineno}

\usepackage[dvipsnames]{xcolor}
\newcommand{\reviewerOne}[1]{\textcolor{black}{#1}}
\newcommand{\reviewerTwo}[1]{\textcolor{black}{#1}}
\newcommand{\reviewerThree}[1]{\textcolor{black}{#1}}

\newcommand{\hide}[1]{}

\algdef{SE}[DOWHILE]{Do}{doWhile}{\algorithmicdo}[1]{\algorithmicwhile\ #1}

\hypersetup{
    colorlinks=true,              
    linkcolor=blue,               
    citecolor=blue,               
    filecolor=black,              
    urlcolor=red,               
    bookmarks=false,
    pdffitwindow=true,
    pdfpagelayout=SinglePage
}

\newcommand{\pd}[2]{\frac{\partial #1}{\partial #2}}

\newcommand{\vect}[1]{\boldsymbol{#1}}

\newcommand{\mat}[1]{\mathcal{#1}}

\newcommand{\hf}{{\frac12}}

\newcommand{\linf}{$L^{\infty}$-norm }

\newcommand{\diff}[1]{\, d#1}
\newcommand{\Oof}[1]{\mathcal{O}\mleft( #1 \mright)}

\newcommand{\daniil}[1]{\textcolor{black}{#1}}

\newcommand{\lap}{\nabla^2}
\newcommand{\ddt}[1]{\partial_t #1}
\newcommand{\ddn}[2][]{\partial_{\vect{n}_{#1}} #2}
\newcommand{\del}[1]{\partial_{#1}}
\newcommand{\jump}[1]{\left[ #1 \right]}
\newcommand{\at}[2]{\left. #1 \right|_{#2}}
\newcommand{\leftbrace}[1]{\left\{\begin{aligned}#1\end{aligned}\right.\end{align*}}
\newcommand{\fd}[2]{\frac{\delta #1}{\delta #2}}

\newcommand{\att}[2]{#1^{#2}}
\newcommand{\set}[3]{\left\{ #1 \right\}_{#2}^{#3}}

\newcommand{\ubcolor}[3]{{\color{#1}\underbrace{\color{black}{#2}}_{#3}}}
\newcommand{\obcolor}[3]{{\color{#1}\overbrace{\color{black}{#2}}^{#3}}}
\newcommand{\range}[2]{\left[ #1, #2 \right]}
\newcommand{\rin}[1][\vect{r}]{\vect{r} \in}

\newcommand{\functlof}[1]{\mleft[ #1 \mright]}

\newcommand{\heatcond}[1]{{\lambda_{#1}}}
\newcommand{\heatdiff}[1]{{\alpha_{#1}}}
\newcommand{\heatcap}[1]{{c_p}_{#1}}
\newcommand{\heatcapv}[1]{s_{#1}}
\newcommand{\den}[1]{\rho_{#1}}
\newcommand{\phase}{\nu}
\newcommand{\partcoeff}{k}

\newcommand{\vn}{v_{\vect{n}}}
\newcommand{\vecr}{\vect{r}}

\newcommand{\J}{J}
\newcommand{\T}[1]{T_{#1}}
\newcommand{\C}[1]{C_{#1}}
\newcommand{\CC}[2]{C_{{#1}{#2}}}
\newcommand{\D}[1]{D_{#1}}
\newcommand{\DD}[2]{D_{{#1}{#2}}}
\newcommand{\lph}{\mathit{l}}
\newcommand{\sph}{\mathit{s}}
\newcommand{\dom}[1]{\Omega_{#1}}
\newcommand{\of}[1]{\mleft( #1 \mright)}

\newcommand{\Tl}{\T{\lph}}
\newcommand{\Ts}{\T{\sph}}
\newcommand{\Tph}{\T{\phase}}
\newcommand{\Cl}[1]{\CC{\lph}{#1}}
\newcommand{\Cs}[1]{\CC{\sph}{#1}}

\newcommand{\ml}[1]{m_{\lph #1}}
\newcommand{\Cg}{{C^\ast_{1}}}

\newcommand{\lam}{\Lambda}
\newcommand{\Tliq}{T_{liq}}
\newcommand{\curv}{\kappa}
\newcommand{\latheat}{L_f}
\newcommand{\PANDAT}{PANDAT\texttrademark }

\newcommand{\kg}{\textrm{kg}}
\newcommand{\cm}{\textrm{cm}}
\newcommand{\kel}{\textrm{K}}
\newcommand{\jou}{\textrm{J}}

\newcommand{\secs}{\textrm{s}}

\def \auxColor {black!50!white}

\newenvironment{myalign}%
{\csname linenomath*\endcsname \align}%
{\endalign \csname endlinenomath*\endcsname }

\newenvironment{myalign*}%
{\csname linenomath*\endcsname \csname align*\endcsname}%
{\csname endalign*\endcsname \csname endlinenomath*\endcsname }

\newenvironment{mymultline}%
{\csname linenomath*\endcsname \multline}%
{\endmultline \csname endlinenomath*\endcsname }

\newenvironment{mymultline*}%
{\csname linenomath*\endcsname \csname multline*\endcsname}%
{\csname endmultline*\endcsname \csname endlinenomath*\endcsname }

\newenvironment{myalignat}[1]%
{\csname linenomath*\endcsname \alignat{#1}}%
{\endalign \csname endlinenomath*\endcsname }

\newenvironment{myalignat*}[1]%
{\csname linenomath*\endcsname \csname alignat*\endcsname{#1}}%
{\csname endalignat*\endcsname \csname endlinenomath*\endcsname }

\begin{document}

\title{A Numerical Method for Sharp-Interface Simulations of Multicomponent Alloy Solidification}

\cortext[cor]{Corresponding author: bochkov.ds@gmail.com}

\address[MECHE]{Department of Mechanical Engineering, University of California, Santa Barbara, CA 93106}
\address[CS]{Department of Computer Science, University of California, Santa Barbara, CA 93106}
\address[MATRL]{Materials Department, University of California, Santa Barbara, CA 93106}

\author[MECHE]{Daniil Bochkov}
\author[MATRL]{Tresa Pollock}
\author[MECHE,CS]{Frederic Gibou}

\begin{abstract}
We present a computational method for the simulation of the solidification of multicomponent alloys \daniil{in the sharp-interface limit}. Contrary to the case of binary alloys where a fixed point iteration is adequate, we hereby propose a Newton-type approach to solve the non-linear system of coupled PDEs arising from the time discretization of the governing equations, \daniil{allowing for the first time sharp-interface simulations of the multialloy solidification}. A combination of spatially adaptive quadtree grids, Level-Set Method, and sharp-interface numerical methods for imposing boundary conditions is used to accurately and efficiently resolve the complex behavior of the solidification front. \daniil{The convergence behavior of the Newton-type iteration is theoretically analyzed in a one-dimensional setting and further investigated numerically in multiple spatial dimensions.} We validate the overall computational method on the case of axisymmetric radial solidification admitting an analytical solution and show that the overall method's accuracy is close to second order. Finally, we perform numerical experiments for the directional solidification of a Co-Al-W ternary alloy with a phase diagram obtained from the \PANDAT database and analyze the solutal segregation dependence on the processing conditions and alloy properties. 
\end{abstract}


\maketitle

\section{Introduction} 

Control of solidification is important for a wide range of manufacturing processes for metallic materials.  A current challenge for solidification modeling is the complex environment encountered in additive manufacturing processes. Additive manufacturing has enormous potential for the design of novel three-dimensional complex geometries and offers the potential for site-specific control of properties, particularly mechanical properties \cite{gu2012laser, pollock2020design, dutta2019science, debroy2018additive, sames2016metallurgy, das2016metallic}.  Achieving this unprecedented level of control requires a fundamental understanding of the complex multi-physics heat, mass and fluid flow phenomena of the printing process, as well as their influence on aspects of the final printed structure that govern properties.   Among the features important to properties are final solute distribution, grain size, morphology, distribution of grain orientation and defects such as pores and cracks. Therefore, understanding and controlling the processing-microstructure relationship in additive manufacturing is key to build materials free of defects and with tailored mechanical properties at specific locations within real world components. Microstructure in additive parts is highly dependent on two crucial quantities: the velocities ($R$) and the thermal gradients ($G$) at the solid-liquid interface, which both can vary by several orders of magnitude during the solidification process within a single melt pool \cite{raghavan2016numerical}. In order to design materials with desired properties, it is essential to predict alloy-dependent solidification maps, which describe regions of planar, cellular, columnar and equiaxed growth in the $(G, R)$ plane, as well as the location of these transitions in structure. While there exist models that attempt to predict the columnar to equiaxed transition \cite{raghavan2016numerical,kurz2001columnar}, 
they depend on parameters that are not trivial to evaluate or measure experimentally and were developed for unidirectional growth, which departs significantly from the typical melt pool environment and thus are unlikely to  be predictive enough in the range of parameters imposed by the beam source and the scan strategy. Therefore, there is a significant gap in knowledge between heat transfer, mass transfer, transport in fluid flow and the structures that develop, especially for multicomponent alloys.

Given the importance of the predictive modeling of solidification phenomena a great number of numerical approaches have been reported in the literature. These computational methods can be categorized in three categories: cellular automata methods \cite{reuther2014perspectives}, phase-field methods \cite{karma2001phase,kim2007phase,steinbach2009phase}\daniil{/diffuse-interface models \cite{tan2007level}}, and sharp-interface methods \cite{yang2005sharp,Theillard;Gibou;Pollock:15:A-sharp-computationa}. Each of these frameworks has its own advantages and disadvantages. Cellular automata methods are computationally efficient; however they are not based on physical equations of solidification but rather on special rules for interactions between automata. In phase field models the solid-liquid interface is described as a smooth transition of a ``phase-field'' variable, which allows efficient numerical implementations that do not require any specialized methods for dealing with moving interfaces. The phase field theory of solidification is mathematically well-justified and guarantees convergence to the sharp-interface equations as the transition width of the solid-liquid interface approaches zero; however, in practice the transition width is far greater than what can be considered the zero limit. The sharp-interface methods are expected to be the most accurate mathematically and consistent with the macroscopic description; however, they are harder to develop \daniil{and typically computationally more expensive}. They require numerical capabilities for explicit handling of evolving interfaces and solving nonlinear systems of PDEs in irregular domains. \daniil{In \cite{tan2007level} a diffuse-interface model was introduced. While it tracks the solidification front explicitly using the level-set approach the underlying heat and species transport equations are solved by artificially smearing the solid-liquid interface and, additionally, enforcing the Gibbs-Thomson condition only approximately.} \daniil{To the best of our knowledge,} so far only cases of binary alloys have been successfully modeled \daniil{in the sharp-interface fashion}  \cite{Theillard;Gibou;Pollock:15:A-sharp-computationa,yang2005sharp}; the current research addresses that gap in the literature.

In this paper we introduce a computational approach that can consider the diffusion in multicomponent alloys coupled with the temperature field and the motion of the solid-liquid interface. The engine also takes into account the crystallographic details, the effects of surface tension and the solute rejection at the solid front in a discretely sharp manner, i.e. the jump in compositions and other quantities that can be only modeled as a discontinuities at the macroscopic level are indeed enforced as discontinuities at the discrete level. Importantly, the computational approach considers the dependence of the liquidus slopes and of the partition coefficients on the time-dependent local composition obtained by the PANDAT$^{\textrm{TM}}$ thermodynamic data base. To the best of our knowledge, this is the only computational engine that can consider ternary or higher order multicomponent systems \daniil{in the sharp-interface limit}. The computational framework is based on state-of-the-art numerical algorithms on adaptive grids that are implemented for massively parallel architectures so that realistic simulations are readily possible. The method is applied to the solidification of a Co-Al-W alloy under \daniil{cooling rates and thermal gradients} relevant to additive manufacturing.

\hide{rev1:1}\reviewerOne{The rest of this manuscript is organized as follows. In section \ref{sec:model}, we summarize the governing equations describing the solidification process of multicomponent alloys. In section \ref{sec:derivation}, we present the derivation and analysis of an approximate Newton method proposed for solving the coupled system of PDEs. Section \ref{sec:numerics} discusses spatial discretization methods used in this work and summarizes the overall solution procedure. Finally, section \ref{sec:results} contains the results of numerical tests and the application of the method to the directional solidification of a Co-Al-W alloy.}

\section{Physical Model}\label{sec:model}

In this section we briefly present a mathematical model of the alloy solidification used in this work. For a detailed discussion on the theory of crystallization processes, we refer the interested reader to the monograph \cite{davis2001theory}.

\begin{figure}[!h]
  \centering
  \begin{subfigure}[b]{.49\textwidth}
    \centering
    \includegraphics[width=0.99\textwidth]{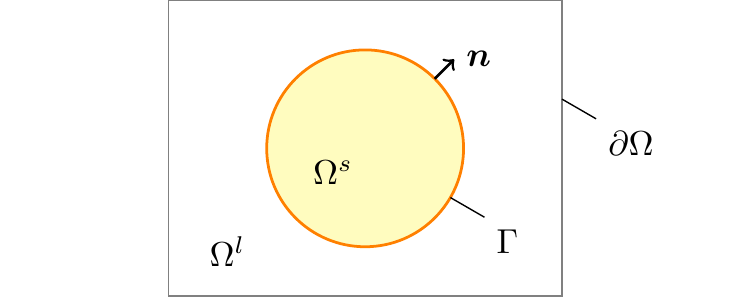}
  \end{subfigure}
  \begin{subfigure}[b]{.49\textwidth}
    \centering
    \includegraphics[width=0.99\textwidth]{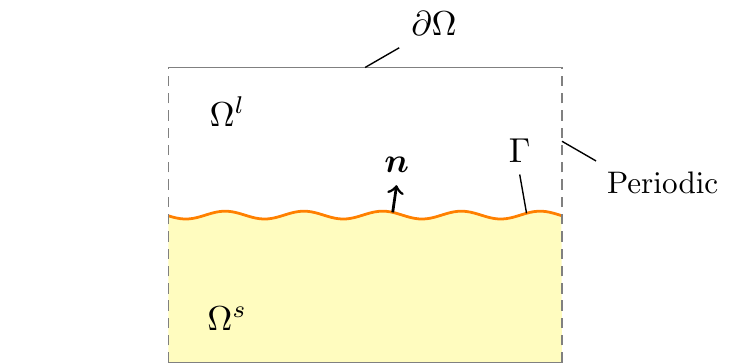}
  \end{subfigure}
  \caption{Notation used in this work demonstrated on examples of crystal growth from a seed (left) and directional solidification (right).}
  \label{fig:model}
\end{figure}

Consider the solidification of an alloy containing $N+1$ different elements: a solvent that constitutes majority of the alloy and $N$ solutes. Specifically, we assume that the process occurs in a rectangular domain $\Omega$ (possibly periodic in some directions) and we consider a mathematical model describing crystallization processes at the macroscopic level without resolving atomistic details. Thus, the transition between solid and liquid phases is assumed to be sharp. We denote this interface boundary as $\Gamma$ and the regions of $\Omega$ occupied by solid and liquid phases as $\dom{\sph}$ and $\dom{\lph}$, respectively (see Figure \ref{fig:model}). The outward normal vectors to the boundaries of $\dom{\lph}$ and $\dom{\sph}$ are denoted as $\vect{n}_l$ and $\vect{n}_s$, respectively. The normal vector to interface $\Gamma$ directed from the solid to the liquid regions is denoted as $\vect{n}$. Note that $\vect{n} = \vect{n}_s = - \vect{n}_l$ on $\Gamma$.

As time $t$ proceeds, the solidification front, $\Gamma = \Gamma\of{t}$, evolves with a normal velocity $\vn = \vn \of{t, \vecr}$, $\rin \Gamma$, according to the crystallization kinetics. In the case of pure substances the process is mainly governed by the thermal transport: the phase transition occurs at the freezing temperature (that may depend on the curvature and normal velocity of the solid-liquid interface) and releases the latent heat which is transported away by the thermal diffusion and, possibly, advection. The case of multicomponent substances, like metal alloys, is complicated by the transport of species in a two-way coupling: on the one hand, freezing temperatures depend on the local alloy composition, and on the other hand, the advancing crystallization front affects the concentration fields via solute-rejection at the interface. 

Thus, at any given moment of time $t$ at every point $\rin \Omega$ the alloy is characterized by the local temperature $T = T\of{t,\vecr}$ and the composition $\C{\J} = \C{\J} \of{t,\vecr}$, $\J \in \range{1}{N}$, where $\C{\J}$ denotes the $J^{\rm th}$ solute's concentration. For convenience, since temperature and concentration fields are not generally smooth and/or continuous across phase boundaries, we use a separate set of fields for each of the two phases, that is:
\begin{myalignat*}{2}
  T\of{t,\vecr} &= 
  \left\{
  \begin{alignedat}{2}
    &\T{\lph}\of{t,\vecr},\quad &&\rin \dom{\lph}\of{t} \\
    &\T{\sph}\of{t,\vecr},\quad &&\rin \dom{\sph}\of{t}
  \end{alignedat}
  \right.
  ,
  \\
  \C{\J}\of{t,\vecr} &= 
  \left\{
  \begin{alignedat}{2}
    &\CC{\lph}{\J}\of{t,\vecr},\quad &&\rin \dom{\lph}\of{t} \\
    &\CC{\sph}{\J}\of{t,\vecr},\quad &&\rin \dom{\sph}\of{t}
  \end{alignedat} 
  \right. 
  ,\quad \J \in \range{1}{N},
\end{myalignat*}
where subscripts $s$ and $l$ denote quantities in solid and liquid phases, respectively.

Suppose, at some initial time $t=t_0$ the state of the system is described by the following initial conditions:
\begin{myalign}\label{eq:model:ic}
\begin{alignedat}{2}
\Gamma \of{t_0} &= \Gamma_0, \\
\T{\phase}\of{t_0,\vecr} &= {T_0}_{\phase}\of{\vecr}, \quad &&\rin \dom{\phase}\of{t_0}, \quad \phase = \sph,\lph, \\
\CC{\phase}{\J}\of{t_0,\vecr} &= {C_0}_{\phase J}\of{\vecr}, \quad &&\rin \dom{\phase}\of{t_0}, \quad \phase = \sph,\lph, \quad \J \in \range{1}{N},
\end{alignedat}
\end{myalign}
where $\Gamma_0$, ${T_0}_{\phase}\of{\vecr}$, and ${C_0}_{\phase J}\of{\vecr}$ describe the initial solid-liquid interface, temperature field and concentration fields.
In the absence of convective effects the transport of heat and species is described by diffusion equations:
\begin{myalignat}{4}\label{eq:model:diffusion_T}
\den{\phase} \heatcap{\phase} \ddt{\T{\phase}} - \heatcond{\phase} \lap \T{\phase} &= 0, \quad \rin \dom{\phase}\of{t}, \quad &\phase& = \sph,\lph,
\\
\label{eq:model:diffusion_C}
\ddt{\CC{\phase}{\J}} - \DD{\phase}{\J} \lap \CC{\phase}{\J} &= 0, \quad \rin  \dom{\phase}\of{t}, \quad &\phase& = \sph,\lph, \quad &\J& \in \range{1}{N},
\end{myalignat}
where $\den{\phase}$, $\heatcap{\phase}$, and $\heatcond{\phase}$, $\phase = \sph,\lph$, are the density, the specific heat, and the heat conductivity of liquid and crystallized alloys; $\DD{\phase}{\J}$, $\phase = \sph,\lph$, are the $J^{\rm th}$ solute's diffusivity coefficient in the liquid and solid phases, respectively. We assume that the alloy parameters $\den{\phase}$, $\heatcap{\phase}$, $\heatcond{\phase}$, $\set{\DD{\phase}{\J}}{\J=1}{N}$, $\phase = \sph,\lph$, are constant.

Since for typical metal alloys the diffusion of solutes in the solid phase is several orders of magnitude slower than in the liquid phase we neglect the species transport in the solid, that is, $\DD{\sph}{\J} = 0$, $\J \in \range{1}{N}$. As a result, the diffusion equations for the concentration fields \eqref{eq:model:diffusion_C} only need to be solved in the liquid.

Temperature and concentration fields must satisfy several conditions on the solidification front $\Gamma$. We assume that the phase transition occurs at the thermodynamic equilibrium, that is, the temperature is continuous across the solidification front:\footnote{Square brackets denote the jump in the value of a quantity across the solidification front, i.e. $\jump{T} = \Ts - \Tl$.}
\begin{myalign}\label{eq:model:jump_t}
	\jump{T} = 0, \quad \rin \Gamma\of{t}, 
\end{myalign}
and satisfies the Gibbs-Thomson relation:
\begin{myalign}\label{eq:model:gibbs}
	\Tl = \Tliq \of{ \Cl{1}, \ldots, \Cl{N} } + \epsilon_v(\vect{n}) \vn +\epsilon_c(\vect{n}) \curv , \quad \rin \Gamma\of{t},
\end{myalign}
where $\Tliq = \Tliq \of{ \Cl{1}, \ldots, \Cl{N} }$ describes the liquidus surface of the alloy (i.e., melting temperature at a given composition), terms $\epsilon_c(\vect{n})$ and $\epsilon_v(\vect{n})$ account for the curvature and kinetic undercoolings, and $\curv$ is the front's mean curvature.
\hide{rev2:8}\reviewerTwo{Sometimes $\Tliq$, see, for example, \cite{Theillard;Gibou;Pollock:15:A-sharp-computationa, davis2001theory}, is assumed to be a linear function of solutal concentrations:
\begin{myalign*}
	\Tliq \of{ \Cl{1}, \ldots, \Cl{N} } = T_m + \ml{1} \Cl{1} + \ldots + \ml{N} \Cl{N},
\end{myalign*}
where $T_m$ is the melting temperature of the pure solvent and $\ml{1}$, $\ldots$, $\ml{N}$ are constants called the \textit{liquidus slopes} corresponding to each of the solutes. The current work is not restricted to such a case and considers $\Tliq \of{ \Cl{1}, \ldots, \Cl{N} }$ to be an arbitrary function, i.e., the liquidus slopes $\ml{\J} = \pd{\Tliq}{\Cl{J}}$, $\J \in \range{1}{N}$, are no longer constants but functions of the local composition as well.} Specifically, for the simulation results presented later in this paper the data from the \PANDAT   thermodynamic database are used.

Note that the undercooling coefficients $\epsilon_c(\vect{n})$ and $\epsilon_v(\vect{n})$ may depend on the normal vector $\vect{n}$ to the solidification front, accounting in such a way for specific crystalline structures of alloys. For example, a two-dimensional four-fold crystalline structure is commonly described as:
\begin{myalign*}
	\epsilon_c (\vect{n}) &= \varepsilon_c ( 1 - 15 \varepsilon \cos(4 \cos^{-1}(\vect{n} \cdot \vect{n}_0))), 
	\\
	\epsilon_v (\vect{n}) &= \varepsilon_v ( 1 - 15 \varepsilon \cos(4 \cos^{-1}(\vect{n} \cdot \vect{n}_0))),
\end{myalign*}
where $\varepsilon_c$ and $\varepsilon_v$ are curvature and kinetic undercooling magnitudes, $\varepsilon$ is the degree of anisotropy and $\vect{n}_0$ is the preferred crystal growth direction.

The thermal balance at the interface leads to the following (Stefan) condition:
\begin{myalign}\label{eq:model:stefan}
	\jump{ \heatcond{} \ddn{T} } &= \vn \latheat , \quad \rin \Gamma\of{t},
\end{myalign}
where $\latheat$ is the latent heat of fusion of the alloy. Note that, as commonly done, the change in surface energy due to stretching/contraction of the curved front's surface in the velocity field $\vn$ is neglected in the above expression. 

At the solidification front the compositions of liquid and solid phases are related to each other through chemical equilibrium. Typically such a relation is described by parameters called \textit{partition coefficients} $\partcoeff_\J$, $\J \in \range{1}{N}$, which represent the ratios of component concentrations in solid and liquid phases, that is:
\begin{myalign*}
	\Cs{\J} = \partcoeff_\J \Cl{\J}, \quad \rin \Gamma\of{t}, \quad \J \in \range{1}{N}.
\end{myalign*}
Since in this work we do not restrict ourselves to linearized liquidus and solidus surfaces the partition coefficients are also assumed to depend on the local composition of the solidifying material, that is:
\begin{myalign*}
  \partcoeff_\J = \partcoeff_\J \of{\Cl{1}, \ldots, \Cl{N}}, \quad \J \in \range{1}{N}.
\end{myalign*}
The conservation of species at the solidification front lead to the following so-called \textit{solute-rejection equations}
\begin{myalign}\label{eq:model:rejection}
	\DD{\lph}{\J} \ddn[l]{\Cl{\J}} - (1-\partcoeff_\J) \vn \Cl{\J} &= 0, \quad  \rin \Gamma\of{t}, \quad \J \in \range{1}{N}.
\end{myalign}

The type of boundary conditions (Dirichlet, Neumann or Robin) on the boundary of the solidification region $\Omega$, denoted as $\partial\Omega$, depends on the particular physical setup. We assume that the total heat flux is specified and the boundary is impermeable to solutes:
\begin{myalign}\label{eq:model:bc}
\begin{alignedat}{3}
	\heatcond{\phase} \ddn[\phase]{\T{\phase}} &= g_{\T{\phase}}, \quad &&\rin \dom{\phase} \cap \partial\Omega, \quad &&\phase = \sph,\lph,
	\\
	\DD{\lph}{\J} \ddn[l]{\Cl{\J}} &= 0, \quad &&\rin \dom{\lph} \cap \partial\Omega, \quad  &&\J \in \range{1}{N},
\end{alignedat}
\end{myalign}
where $g_{\T{\phase}} = g_{\T{\phase}}\of{t,\vecr}$, $\phase = \sph,\lph$, are prescribed heat fluxes for the liquid and solid phases. We note, however, that switching to boundary conditions of another type (Dirichlet or Robin) has minimal consequences on the computational method presented in this work.

To summarize, in this work we present a computational method for solving a multialloy solidification model in which the crystallization process is described by the temporal evolution of temperature fields $\T{\phase} = \T{\phase}\of{t,\vecr}$, $\phase = \sph,\lph$, solutes' concentration fields $\Cl{\J} = \Cl{\J}\of{t,\vecr}$, $\J \in \range{1}{N}$, and an evolving solidification front $\Gamma = \Gamma\of{t}$ that satisfy the partial differential equations \eqref{eq:model:diffusion_T}-\eqref{eq:model:diffusion_C} with the interface conditions \eqref{eq:model:jump_t}-\eqref{eq:model:rejection} on $\Gamma$ and the boundary conditions \eqref{eq:model:bc} on $\partial\Omega$.

\section{\reviewerOne{Approximate Newton Method Derivation and Analysis}}\label{sec:derivation}

\hide{rev1:2}
\reviewerOne{
In this section we focus on the derivation and analysis of a Newton iteration scheme for solving coupled system of equation describing the solidification process, specifically: 
\begin{enumerate}
  \item We begin with discussing the temporal discretization of the system of governing equations and identify specific tasks needed to be performed during each time step (section \ref{sec:derivation:time}).
  \item Second, in section \ref{sec:derivation:nonlinear}, we present a numerical method for solving the nonlinear system of elliptic PDEs resulted from the temporal discretization. The method is based on breaking down the system of nonlinearly coupled equations into a set of separate boundary value problems subject to classical boundary (Dirichlet, Neumann or Robin) and interface conditions.
\end{enumerate}
}

\reviewerOne{
The overall solution procedure as well as a detailed description of specific methods we use for spatial discretization of the computational domain, for evolving the solidification front in time, for solving elliptic partial differential equations with different boundary and for imposing the interface conditions on irregular interfaces are discussed in section \ref{sec:numerics}.
}

\subsection{Discretization in time}\label{sec:derivation:time}

Consider a non-uniform discretization of time $\set{t_j}{j \ge 0}{}$ with time steps $\set{\Delta t_j = t_j - t_{j-1}}{i \ge 1}{}$ and denote the state of the system (i.e. the temperature and concentration fields, and the location of the solidification front) at a time $t_j$ as $\att{\Tph}{j}$, $\att{\Cl{\J}}{j}$ and $\att{\Gamma}{j}$, $j \ge 0$. Given states of the system for $t_j$, $j < n$, the numerical solution at time instant $t_n$ is computed in the following fashion. 


First, the new front's location $\att{\Gamma}{n}$ is obtained from $\att{\Gamma}{n-1}$ in an \textit{explicit} way based on values of the normal velocity at previous time moments $\att{\vn}{j}$, $j < n$, as discussed in section \ref{sec:numerics:level_set}. 

Secondly, equations \eqref{eq:model:ic}-\eqref{eq:model:bc} are solved \textit{implicitly} for $\att{\Ts}{n}$, $\att{\Tl}{n}$, $\set{\att{\Cl{\J}}{n}}{\J=1}{N}$ and $\att{\vn}{n}$ in geometry defined by $\att{\Gamma}{n}$. To this end, we use a second-order accurate implicit (BDF2) discretization in time. Let us write the approximation of the temporal derivative of a quantity $A$ at a time instant $t=t_n$ as:
\begin{myalign}\label{eq:time_approx:general}
  \att{\ddt{A}}{n} = \frac{1}{\Delta t_n} \sum_{j \geq 0} a_j \att{A}{n-j} + \Oof{\Delta t_\text{max}^q},
\end{myalign}
where $\Delta t_\text{max} = \max\limits_{j \ge 0} \left( \Delta t_{n-j} \right)$ and the coefficients $\{ a_j \}_{j \geq 0}$ are given by:
\begin{myalign*}
  a_0 = \frac{1+2r}{1+r}, \quad a_1 = -(1+r), \quad a_2 = \frac{r^2}{1+r}, \quad a_j = 0,\, j \geq 3, \quad \text{where } r = \frac{\Delta t_n}{\Delta t_{n-1}} \text{ and } q = 2.
\end{myalign*}
Using approximation \eqref{eq:time_approx:general} in the diffusion equations \eqref{eq:model:diffusion_T} and \eqref{eq:model:diffusion_C} we get:
\begin{myalignat}{3}
\label{eq:discretized:diffusion_T}
\left(\den{\phase} \heatcap{\phase} \frac{1}{\Delta t_n} a_0 - \heatcond{\phase} \lap \right) \att{\T{\phase}}{n} &= - \den{\phase} \heatcap{\phase} \frac{1}{\Delta t_n}\sum_{j \geq 1} a_\J \att{\T{\phase}}{n-j}, \quad &&\rin \att{\dom{\phase}}{n}, \quad &\phase &= \sph,\lph,
\\
\label{eq:discretized:diffusion_C}
\left( \frac{1}{\Delta t_n} a_0 - \DD{\lph}{\J} \lap \right) \att{\Cl{\J}}{n} &= - \frac{1}{\Delta t_n} \sum_{j \geq 1} a_\J \att{\Cl{\J}}{n-j},  \quad &&\rin \att{\dom{\lph}}{n}, \quad &\J &\in \range{1}{N},
\end{myalignat}
where known quantities are collected in the right-hand side. The above two expressions are simple linear Poisson-type equations, however they must be solved subject to the non-linear interface and boundary conditions \eqref{eq:model:jump_t}-\eqref{eq:model:rejection} on $\att{\Gamma}{n}$. The source of non-linearity is in the Robin-type boundary conditions \eqref{eq:model:rejection} that contain the product of two unknowns -- the concentration $\att{\Cl{\J}}{n}$ and the velocity $\att{\vn}{n}$.\footnote{Note that even when the so-called Frozen Temperature Approximation is applied (i.e., the temperature field is not solved for but prescribed by an analytical expression) the system of equations is still non-linearly coupled for $N > 1$.} Once a method for solving \eqref{eq:discretized:diffusion_T}-\eqref{eq:discretized:diffusion_C} subject to \eqref{eq:model:jump_t}-\eqref{eq:model:rejection} is available, then it is relatively easy to construct a time-stepping procedure for solving the entire dynamic problem.
Thus, the solution of \eqref{eq:discretized:diffusion_T}-\eqref{eq:discretized:diffusion_C} subject to \eqref{eq:model:jump_t}-\eqref{eq:model:rejection} is the cornerstone problem in simulating multialloy solidification processes.


\subsection{Solving the non-linearly coupled system of Poisson-type equations}\label{sec:derivation:nonlinear}

For clarity of presentation, we write the system of the coupled Poisson-type equations \eqref{eq:discretized:diffusion_T}-\eqref{eq:discretized:diffusion_C} subject to \eqref{eq:model:jump_t}-\eqref{eq:model:bc} in a generic fashion \hide{rev3:1} as:
\begin{myalignat}{4}
\label{eq:nonlinear:diffusion_T}
\text{\textit{Heat transport:}} && \quad
\left( \heatcapv{\phase} - \heatcond{\phase} \lap \right) \T{\phase} &= f_{\T{\phase}}, \quad &&\rin \dom{\phase}, \quad &&\phase = \sph,\lph,
\\
\label{eq:nonlinear:diffusion_C}
\text{\textit{Species transport:}} && \quad
\left( a - \D{\J} \lap \right) \C{\J} &= f_{\C{\J}}, \quad &&\rin \dom{\lph}, \quad &&\J \in \range{1}{N},
\\
\text{\textbf{Conditions on} $\Gamma$:}\notag
\\
\label{eq:nonlinear:jump_t}
\text{\textit{Temperature continuity:}} && \quad
\jump{ T } &= h_T,  \span \span \span
\\
\label{eq:nonlinear:stefan}
\text{\textit{Stefan condition:}} && \quad
\jump{ \heatcond{} \ddn{T} } &= h_S + \vn \latheat, \span \span \span \span
\\
\label{eq:nonlinear:gibbs}
\text{\textit{Gibbs-Thompson:}} && \quad
\Tl &= h_G + \Tliq\of{\C{1}, \ldots, \C{N}} + \epsilon_v \vn, \span \span \span \span
\\
\label{eq:nonlinear:rejection}
\text{\textit{Solute-rejection:}} && \quad
\D{\J} \ddn[l]{\C{\J}} - (1-\partcoeff_\J) \vn \C{\J} &= h_{\C{\J}}, \quad &&\J \in \range{1}{N}, \span
\\
\text{\textbf{Conditions on} $\partial\Omega$:}\notag
\\
\label{eq:nonlinear:bc_T}
\text{\textit{Heat supply/withdrawal:}} && \quad
\heatcond{\phase} \ddn[\phase]{\T{\phase}} &= g_{\T{\phase}}, \quad &&\phase = \sph,\lph, \span
\\
\label{eq:nonlinear:bc_C}
\text{\textit{Impermeable boundaries:}} && \quad
\D{\J} \ddn[l]{\C{\J}} &= g_{\C{\J}}, \quad  &&\J \in \range{1}{N}, \span
\end{myalignat}
The original system of equations related to the solidification process is recovered by the following substitutions:
\begin{myalign*}
	\heatcapv{\phase} &\rightarrow \den{\phase} \heatcap{\phase} \frac{a_0}{\Delta t_n},
  &
 	\heatcond{\phase} &\rightarrow \heatcond{\phase},
 	&
  \T{\phase} &\rightarrow \att{\T{\phase}}{n},
	&
	f_{\T{\phase}} &\rightarrow - \den{\phase} \heatcap{\phase} \frac{1}{\Delta t_n} \sum_{j \geq 1} a_j \att{\T{\phase}}{n-j},
	\\
  a &\rightarrow \frac{a_0}{\Delta t_n},
  &
  \D{\J} &\rightarrow \DD{\lph}{\J},
  &
	\C{\J} &\rightarrow \att{\Cl{\J}}{n},
	&
	f_{\C{\J}} &\rightarrow - \frac{1}{\Delta t_n} \sum_{j \geq 1} a_j \att{\Cl{\J}}{n-j},
  \\
 	h_T &\rightarrow 0,
 	&
  h_S &\rightarrow 0,
 	&
 	h_{\C{\J}} &\rightarrow 0,
 	&
 	h_G &\rightarrow \epsilon_c (\vect{n}),
  \\
  (1-\partcoeff_\J) &\rightarrow (1-\partcoeff_\J),
  &
  g_{\C{\J}} &\rightarrow g_{\Cl{\J}},
  &
  g_{\T{\phase}} &\rightarrow g_{\T{\phase}},
  &
  \latheat &\rightarrow \latheat.
\end{myalign*}

\subsubsection{Fixed-point iteration}\label{sec:derivation:nonlinear:fixed_point}
In \cite{theillard2015sharp} a simple fixed-point iteration algorithm was used for simulating the solidification of bi-alloys, that is, in the case $N=1$. It is based on breaking down the system \eqref{eq:nonlinear:diffusion_T}-\eqref{eq:nonlinear:bc_C} into separate boundary value problems (BVPs) for the temperature and the concentrations fields with simple boundary conditions (Dirichlet, Neumann or Robin) and interface conditions and iteration procedures. A direct extension of this algorithm to the case of arbitrary $N$ has the following form:
\begin{enumerate}
	\item Let us denote the value of concentration $\C{1}$ at the solidification front during the $q$th iteration as $\Cg^{(q)} \of{\vecr}$, $\rin \Gamma$. Set $q=0$ and some initial guess $\Cg^{(0)}$ (e.g., using its value at the previous time step).
	\item Solve for $\C{1}^{(q)}$ imposing Dirichlet BC on $\Gamma$:
		\begin{myalign}
    \label{eq:spliting:c1}
		\left\{
		\begin{aligned}
			\left( a - D_1 \lap \right) \C{1}^{(q)} &= f_{\C{1}}, \quad &&\rin \dom{\lph},
			\\
			\C{1}^{(q)} &= \Cg^{(q)}, \quad &&\rin \Gamma,
			\\
			D_1\ddn[l]{\C{1}^{(q)}} &= g_{\C{1}}, \quad &&\rin \partial\Omega \cap \overline{\dom{\lph}}.
		\end{aligned}
		\right.
		\end{myalign}
	\item Compute the front's velocity $\vn^{(q)}$ using the solute-rejection equation \eqref{eq:nonlinear:rejection}:
		\begin{myalign}
    \label{eq:spliting:v}
			\vn^{(q)} = \frac{1}{(1-\partcoeff_1) \C{1}} \left( D_1 \ddn[l]{\C{1}^{(q)}} - h_{\C{1}}\right), \quad \rin\Gamma.
		\end{myalign}
	\item Solve for $\C{\J}^{(q)}$, $\J \in \range{2}{N}$, imposing Robin BC on $\Gamma$:
		\begin{myalign}
    \label{eq:spliting:ci}
		\left\{
		\begin{aligned}
			\left( a - \D{\J} \lap \right) \C{\J}^{(q)} &= f_{\C{\J}},\quad &&\rin \dom{\lph},
			\\
			\D{\J} \ddn[l]{\C{\J}^{(q)}} - (1-\partcoeff_\J) \vn^{(q)} \C{\J}^{(q)} &= h_{\C{\J}},\quad  &&\rin \Gamma,
			\\
			\D{\J} \ddn[l]{\C{\J}^{(q)}} &= g_{\C{\J}},\quad &&\rin \partial\Omega \cap \overline{\dom{\lph}}.
		\end{aligned}
  	\right.
		\end{myalign}
	\item Solve for $\Tph^{(q)}$, $\phase =\sph,\lph$, imposing jump conditions on $\Gamma$:
		\begin{myalign}
    \label{eq:spliting:t}
		\left\{
		\begin{aligned}
			\left( \heatcapv{\phase} - \heatcond{\phase} \lap \right) \Tph^{(q)} &= f_T, \quad &&\rin \dom{\phase}, \quad \phase = \sph,\lph,
			\\
			\jump{T^{(q)}} &= h_T, \quad &&\rin \Gamma,
			\\
			\jump{\heatcond{} \ddn{T^{(q)}}} &= h_S + \vn^{(q)} \latheat, \quad &&\rin \Gamma,
			\\
			\heatcond{\phase} \ddn[\phase]{\Tph^{(q)}} &= g_{\T{\phase}}\of{\vecr}, \quad &&\rin \partial\Omega \cap \overline{\dom{\phase}}, \quad \phase = \sph,\lph.
		\end{aligned}
		\right.
		\end{myalign}
  \item Compute error $\mathcal{E}^{(q)}\of{\vecr}$ in satisfying the Gibbs-Thomson relation \eqref{eq:nonlinear:gibbs} on $\Gamma$:
    \begin{myalign}
      \label{eq:spliting:error}
      \mathcal{E}^{(q)}\of{\vecr} = \Tl^{(q)} - h_G - \Tliq\of{\C{1}^{(q)}, \ldots, C_N^{(q)}} - \epsilon_v \vn^{(q)}, \quad \rin \Gamma.
    \end{myalign}
	\item If the maximum error exceeds a user-defined tolerance $\epsilon_\text{tol}$, i.e if
		\begin{myalign*}
			\max_{\rin \Gamma} \left| \mathcal{E}^{(q)}\of{\vecr} \right| > \epsilon_\text{tol},
    \end{myalign*}
    then adjust $\Cg^{(q)}$ by inverting the Gibbs-Thomson relation \eqref{eq:nonlinear:gibbs}:
		\begin{myalign}
      \label{eq:nonlinear:fixed}
      \Cg^{(q+1)} = \Cg^{(q)} + \frac{ \Tl^{(q)} - h_G - \Tliq\of{\C{1}^{(q)}, \ldots, C_N^{(q)}} - \epsilon_v \vn^{(q)}}{\ml{1}\of{C^{(q)}_1, \ldots, C^{(q)}_N}} 
		\end{myalign}
		set $q=q+1$ and go to step 2.
\end{enumerate}

In this procedure $\C{1}$ should denote the component that diffuses the slowest, that is, $D_1 < \D{\J}$, $\J \geq 2$, since it is the slowest diffusing component that limits the velocity of front propagation. The case of arbitrary $N$ differs from the case $N=1$ by the presence of step 4, which is absent for $N=1$. 

Clearly, the above splitting scheme of coupled system \eqref{eq:nonlinear:diffusion_T}-\eqref{eq:nonlinear:bc_C} into separate simpler BVPs is not unique. In fact, perhaps a more ``symmetric'' way is to solve for temperature fields both in solid and liquid phases using Dirichlet boundary conditions on $\Gamma$, compute the front velocity using the Stefan condition \eqref{eq:nonlinear:stefan}, use the computed velocity to solve for concentrations $\{ \C{\J} \}_{\J=1}^{N}$ imposing the solute-rejection equations \eqref{eq:nonlinear:rejection} as Robin boundary conditions and, finally, correct the guessed value for the temperature on $\Gamma$ using the Gibbs-Thompson relation \eqref{eq:nonlinear:gibbs}. However, such an iterative procedure \reviewerTwo{was shown to be}\hide{rev2:4} to be very unstable in numerical experiments for typical parameters of metal alloy, for which the thermal diffusivity is much less than the diffusivity of solutes, i.e. $\displaystyle \frac{\heatcond{\phase}}{\heatcapv{\phase} \heatcap{\phase}} \ll \D{\J}$. It appears to be crucial to compute the front's velocity based on values of the slowest components in the system in order to obtain a stable iterative scheme (as done in the scheme above).

The success, i.e., fast convergence, of the simple iterative scheme presented above in case of binary alloys also seems to owe to the fact that the diffusivity of the quantity that is used for velocity calculations (concentration $\C{1}$) is much less than the diffusivity of the quantity which is computed using the found velocity (temperatures $\Tl$ and $\Ts$).
In the case of multialloys ($N \geq 2$) this is no longer true because the velocity $\vn$ is also used in the Robin boundary conditions when solving for concentrations $\C{\J}$, $\J \geq 2$, which may diffuse at a rate very close to the rate of $\C{1}$, that is, $D_1 \approx \D{\J}$, $i = 2, \ldots, N$. We have found from numerical experiments that in such a case the above iterative scheme exhibits a slow convergence and even often an unstable behavior.

\hide{rev3:12}\reviewerThree{\textbf{Remark.} Indeed, not addressing the inherit stiffness of the problem properly is believed to cause the instability of the fixed-point scheme. This is confirmed, by a linear stability analysis performed in section \ref{sec:derivation:nonlinear:stability}, which shows that the amplification factor in the fixed-point iteration contains terms of orders $\Oof{\D{1}/\heatcond{\phase}}$ and $\Oof{\D{1}/\D{\J}}$. This is similar to the stability issues encountered in fluid-structure interaction problem, see, for example, \cite{causin2005added}.}

\textbf{Remark.} Note that the specification of the function $\Cg\of{\vecr}$, $\rin \Gamma$, uniquely defines the functions $\vn$, $\{\C{\J}\}_{\J=1}^N$ and $\{\T{\phase} \}_{\phase = \sph,\lph}$ through equations \eqref{eq:spliting:c1}-\eqref{eq:spliting:t}. Thus, these functions can be considered as functionals of $\Cg$, that is, $\vn=\vn\functlof{\Cg}$, $\set{\C{\J} = \C{\J}\functlof{\Cg}}{\J=1}{N}$ and $\set{\T{\phase}=\T{\phase}\functlof{\Cg}}{\phase = \sph,\lph}{}$ (here the square brackets denote the functional argument), and the above fixed-point iteration scheme can be expressed as a non-linear functional equation:
\begin{myalign*}
	\Cg = \Phi\functlof{\Cg},
\end{myalign*}
where functional $\Phi\functlof{\varphi}$ is defined as:
\begin{myalign*}
	\Phi\functlof{\varphi} = \varphi + \frac{\Tl\functlof{\varphi} - h_G - \Tliq\of{\C{1}\functlof{\varphi}, \ldots, C_N\functlof{\varphi}}- \epsilon_v \vn\functlof{\varphi}}
  {\ml{1}\of{\C{1}\functlof{\varphi}, \ldots, C_N\functlof{\varphi}}}.
\end{myalign*}

\subsubsection{Approximate Newton iteration}\label{sec:derivation:nonlinear:newton}
In the current work, we use the above scheme as a basis and apply variational calculus to estimate how the error in satisfying the Gibbs-Thomson relation $\mathcal{E}\of{\vecr}$ at any given point changes when the boundary concentration $\Cg\of{\vecr}$ is changed by some $\Delta\Cg\of{\vecr}$. This information is then used to obtain an alternative updating formula for $\Cg$ (instead of \eqref{eq:nonlinear:fixed}) such that $\mathcal{E}\of{\vecr}$ converges to zero efficiently.

In general, the change in error $\mathcal{E}\of{\vecr}$ due to a change in $\Cg$ up to linear order can be expressed as:
\begin{myalign*}
	\Delta \mathcal{E}\of{\vecr} = \int_\Gamma \fd{\mathcal{E}\of{\vecr}}{\Cg (\vect{r}^\prime)} \Delta\Cg(\vect{r}^\prime) \diff{\Gamma}, \quad \vect{r} \in \Gamma,
\end{myalign*}
where the functional derivative $\displaystyle \fd{\mathcal{E}\of{\vecr}}{\Cg (\vect{r}^\prime)}$ represents the sensitivity of $\mathcal{E}$ at point $\vect{r}$ to the change in $\Cg$ at point $\vect{r}^\prime$. We refer the interested reader to \ref{app:derivation} where it is shown how $\displaystyle \fd{\mathcal{E}\of{\vecr}}{\Cg (\vect{r}^\prime)}$ can be expressed as the solution to an adjoint system of PDEs corresponding to \eqref{eq:spliting:c1}-\eqref{eq:spliting:t}. In principle, one could use the above expression to find the optimal ${\Delta \Cg}^{\!,\textrm{best}} \of{\vecr}$ that is expected to reduce $\mathcal{E}$ to zero everywhere on $\Gamma$, that is, $\Delta \mathcal{E}\of{\vecr} = - \mathcal{E}\of{\vecr}$ or:
\begin{myalign}\label{eq:newton:integral}
\mathcal{E}\of{\vecr} + 
\int_\Gamma 
  \fd{\mathcal{E}\of{\vecr}}{\Cg(\vect{r}^\prime)} 
  {\Delta \Cg}^{\!,\textrm{best}} (\vect{r}^\prime)
  \diff{\Gamma} = 0, \quad \vect{r} \in \Gamma.
\end{myalign}
However solution of the above boundary integral equation is not a simple task that requires, first, an efficient calculation of functional derivative $\displaystyle \fd{\mathcal{E}\of{\vecr}}{\Cg (\vect{r}^\prime)}$ for all interface points $\vecr\in\Gamma$ and, second, inversion of the convolution term. We defer the further investigation of this avenue to future works. Instead, we propose to use a greatly simplified approach in which we approximate the boundary integral in \eqref{eq:newton:integral} as:
\begin{myalign*}
\int_\Gamma 
  \fd{\mathcal{E}\of{\vecr}}{\Cg(\vect{r}^\prime)} 
  {\Delta \Cg}^{\!,\textrm{best}} (\vect{r}^\prime)
  \diff{\Gamma} 
&\approx
{\Delta \Cg}^{\!,\textrm{best}} \of{\vecr}
\int_\Gamma 
  \fd{\mathcal{E}\of{\vecr}}{\Cg(\vect{r}^\prime)} 
  \diff{\Gamma}.
\end{myalign*}
Such an approach is reasonable provided that the sensitivity $\displaystyle \fd{\mathcal{E}\of{\vecr}}{\Cg(\vect{r}^\prime)}$ decays fast as the distance between $\vect{r}$ and $\vect{r}^\prime$ increases. Using such an approximation equation \eqref{eq:newton:integral} is trivially solved to obtain:
\begin{myalign}
{\Delta \Cg}^{\!,\textrm{best}} \of{\vecr}
\approx
-
\frac{
\mathcal{E}\of{\vecr}
}
{
\displaystyle
\int_\Gamma 
  \fd{\mathcal{E}\of{\vecr}}{\Cg(\vect{r}^\prime)} 
  \diff{\Gamma}
}.
\end{myalign}
Thus, instead of using \eqref{eq:nonlinear:fixed}, we calculate $\Cg^{(q+1)}$ as:
\begin{myalign}\label{eq:nonlinear:newton}
  \Cg^{(q+1)} = \Cg^{(q)} - \frac{\mathcal{E}^{(q)}\of{\vecr}}{G^{(q)}\of{\vecr}},
\end{myalign}
where $\displaystyle G\of{\vecr} = \int_\Gamma \fd{\mathcal{E}\of{\vecr}}{\Cg(\vect{r}^\prime)} \diff{\Gamma}$.
Note that the quantity $\displaystyle G\of{\vecr}$ has the meaning of the directional derivative of $\mathcal{E}\of{\vecr}$ in the ``direction'' $\delta \Cg \of{\vecr} \equiv 1$, $\rin \Gamma$. Using results of \ref{app:derivation2}, $\displaystyle G\of{\vecr}$ can be efficiently computed as:
\begin{myalign*}
\displaystyle G^{(q)}\of{\vecr} = \lam_{\Tl} - \sum_{\J=1}^{N} \ml{\J} \of{\C{1}, \ldots, C_N} \lam_{\C{\J}} - \varepsilon_v \lam_{v}
\end{myalign*}
where $\lam_{\Tl}$, $\set{\lam_{\C{\J}}}{\J=1}{N}$ and $\lam_{v}$ are the solutions to the following adjoint system of PDEs:
\begin{myalign}
  &
  \label{eq:adjoint:c1}
  \left\{
    \begin{aligned}
    \left( a - D_1 \lap \right) \lam_{\C{1}} &= 0 &&\text{in } \dom{\lph}
    \\
    \lam_{\C{1}} &= 1 &&\text{on } \Gamma
    \\
    D_1 \ddn[l]{\lam_{\C{1}}} &= 0 &&\text{on } \partial\Omega \cap \overline{\dom{\lph}}
    \end{aligned}
  \right.
  \\&
  \label{eq:adjoint:vn}
  \lam_{v} = \frac{1}{(1-\partcoeff_1) \C{1}} \left( D_1 \ddn[l]{\lam_{\C{1}}} - v_n (1-\partcoeff_1) \lam_{\C{1}}\right) \quad \text{on }\Gamma
  \\&
  \label{eq:adjoint:ci}
  \left\{
  \begin{aligned}
    \left( a - \D{\J} \lap \right) \lam_{\C{\J}} &= 0 &&\text{in } \dom{\lph}
    \\
    \D{\J} \ddn[l]{\lam_{\C{\J}}} - (1-\partcoeff_\J) \vn \lam_{\C{\J}} &= (1-\partcoeff_\J) \lam_{v} \C{\J}  &&\text{on } \Gamma
    \\
    \D{\J} \ddn[l]{\lam_{\C{\J}}} &= 0 &&\text{on } \partial\Omega \cap \overline{\dom{\lph}}
  \end{aligned}
  \right.
  \\&
  \label{eq:adjoint:t}
  \left\{
  \begin{aligned}
    \left( \heatcapv{\phase} - \heatcond{\phase} \lap \right) \lam_{\T{\phase}} &= 0 &&\text{in } \dom{\phase}, \quad \phase = \sph,\lph
    \\
    \left[ \lam_T \right] &= 0 &&\text{on } \Gamma
    \\
    \left[ \heatcond{} \ddn{\lam_T} \right] &= \latheat \lam_v &&\text{on } \Gamma
    \\
    \heatcond{\phase} \ddn[\phase]{\lam_{\T{\phase}}} &= 0 &&\text{on } \partial\Omega \cap \overline{\dom{\phase}}, \quad \phase = \sph,\lph
  \end{aligned}
  \right.
\end{myalign}

For clarity, Algorithm \ref{algo:newton} summarizes the overall Newton-like iterative procedure developed in this work for solving the system of nonlinearly coupled PDEs based on \eqref{eq:nonlinear:newton}. 

\begin{algorithm}[!h]
  \caption{An approximate Newton iteration for solving nonlinear system of equations \eqref{eq:nonlinear:diffusion_T}-\eqref{eq:nonlinear:bc_C}}
  \label{algo:newton}
  \begin{algorithmic}[1]
    \State Provide an initial guess $\Cg^{(0)}$, tolerance $\epsilon_\text{tol}$, and maximum iterations allowed $q_\textrm{max}$ \State Set $q=0$
    \State Solve \eqref{eq:spliting:c1} for $\C{1}^{(q)}$ \label{algo:aN:start} 
    \State Compute velocity $\vn$ using \eqref{eq:spliting:v} 
    \State Solve \eqref{eq:spliting:ci} for $\C{\J}^{(q)}$, $\J=2,N$
    \State Solve \eqref{eq:spliting:t} for $\T{\phase}^{(q)}$, $\phase = \lph, \sph$ 
    \State Compute error $\mathcal{E}\of{\vecr}^{(q)}$ on $\Gamma$ \eqref{eq:spliting:error}
    \If {$\max_{\vect{r}\in\Gamma} \left| \mathcal{E}^{(q)}\of{\vecr} \right| > \epsilon_\text{tol}$ and $q < q_\textrm{max}$}
      \State Solve \eqref{eq:adjoint:c1} for $\lam_{\C{1}}^{(q)}$
      \State Compute $\lam_v^{(q)}$ using \eqref{eq:adjoint:vn}
      \State Solve \eqref{eq:adjoint:ci} for $\lam_{\C{\J}}^{(q)}$, $\J=2,N$
      \State Solve \eqref{eq:adjoint:t} for $\lam_{\T{\phase}}^{(q)}$, $\phase = \lph, \sph$
      \State Compute $\Cg^{(q+1)}$ using \eqref{eq:nonlinear:newton}
      \State Set $q \leftarrow q + 1$
      \State Go to step \ref{algo:aN:start}
    \EndIf
  \end{algorithmic}
\end{algorithm}

\textbf{Remark.} The proposed Newton-type approach requires solving twice as many BVPs compared to the simple fixed-point iteration. However, the above adjoint system of equations has the same structure as the system of equations describing the physical quantities \eqref{eq:spliting:c1}-\eqref{eq:spliting:t}. Consequently, the discretization matrices obtained for \eqref{eq:spliting:c1}-\eqref{eq:spliting:t} can be reused while solving \eqref{eq:adjoint:c1}-\eqref{eq:adjoint:t}, resulting in computational time savings. In addition, the quantity $\lam_{\C{1}}$ does not change form iteration to iteration, thus equation \eqref{eq:adjoint:c1} needs to be solved only once.



\subsubsection{Convergence of iterative schemes}\label{sec:derivation:nonlinear:stability}

In order to gain some insight into the convergence properties of the iterative schemes above, we analyze them from point of view of the linear stability analysis in a simple setting of quasi one dimensional planar geometry. Specifically, we consider an infinite domain with the interface located at $y=0$ such that the $y>0$ and $y<0$ half-spaces are occupied by liquid and solid phases, respectively. For simplicity we assume that the constitutional undercooling is linear, that is, $ \Tliq = T_m + \sum_{i = 1}^N \ml{\J} \C{\J}$, that the partition coefficients are constant, and the absence of kinetic and curvature undercoolings ($\epsilon_v = 0$ and $\epsilon_c = 0$).

\begin{figure}[!h]
  \centering
  \begin{subfigure}[T]{0.4\textwidth}
    \centering
    \includegraphics[width=0.99\textwidth]{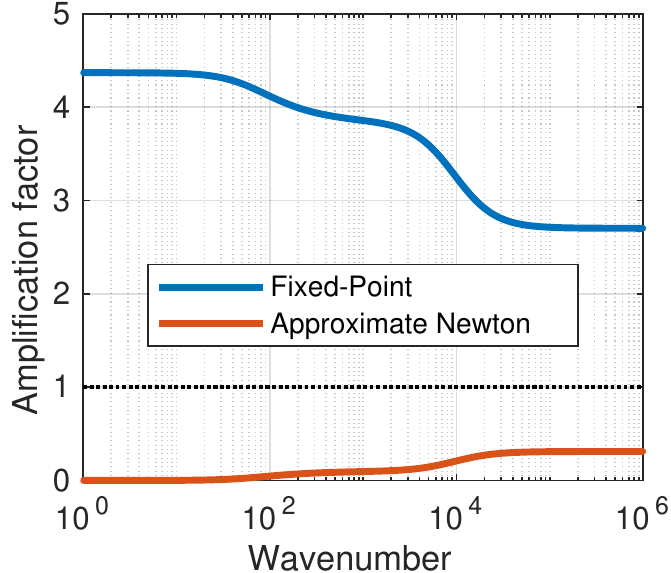}
    \caption{} \label{fig:stability:wavenumber}
  \end{subfigure}
  \begin{subfigure}[T]{0.4\textwidth}
    \centering
    \includegraphics[width=0.99\textwidth]{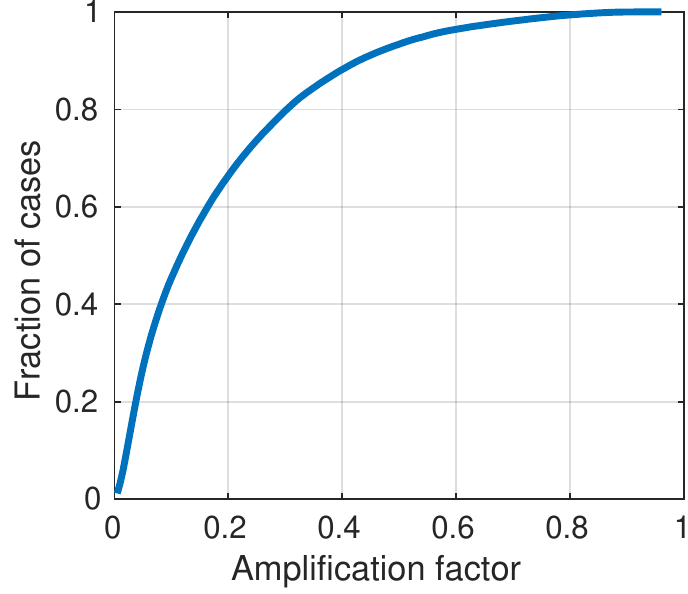}
    \caption{} \label{fig:stability:sweep}
  \end{subfigure}
  \caption{Results of the linear stability analysis of the fixed-point and the approximate Newton iterations: (a) Dependence of the amplification factor on the perturbation's wave number; (b) Cumulative fraction of cases observed in the parameter sweep study with amplification factors equal or less than a given one for the proposed Newton-type method (amplification factor was found to be greater than one for all parameter combinations in case of the fixed-point method).}
\end{figure}

Denote as $\tilde{\C{J}}=\tilde{\C{\J}}\of{y}$, $\J\in\range{1}{N}$, $\tilde{\Ts} = \tilde{\Ts}\of{y}$ and $\tilde{\Tl} = \tilde{\Tl}\of{y}$ the solution to nonlinear system of equation of \eqref{eq:nonlinear:diffusion_T}-\eqref{eq:nonlinear:bc_C}. Let us consider an infinitesimally perturbed boundary concentration with magnitude $\delta_C$ and spatial frequency $\omega_x$, that is:
\begin{myalign*}
  \Cg^{(0)}  = \tilde{\C{\J}}\of{0} + \delta_C e^{-i \omega_x x}
\end{myalign*}
We seek solutions satisfying iterative equations \eqref{eq:nonlinear:fixed} and \eqref{eq:nonlinear:newton} up to linear order in $\delta_C$ of the form:
\begin{myalign*}
  \Cg^{(q)} &= \tilde{\C{\J}}\of{0} + r_{\textrm{f.p.}}^q \delta_C \exp \of{-i \omega_x x} + \Oof{\delta_C^2} \quad \textrm{and} \\
  \Cg^{(q)} &= \tilde{\C{\J}}\of{0} + r_{\textrm{a.N.}}^q \delta_C \exp \of{-i \omega_x x} + \Oof{\delta_C^2},
\end{myalign*}
respectively, where $r_{\textrm{f.p.}} = r_{\textrm{f.p.}}\of{\omega_x}$ and $r_{\textrm{a.N.}} = r_{\textrm{a.N.}}\of{\omega_x}$ denote amplification factors for disturbances of frequency $\omega_x$ in cases of fixed-point and approximate Newton iterations, respectively. An iterative scheme is expected to be unstable if its amplification factor is greater than one and stable otherwise where a smaller amplification factor indicates faster convergence. Substitution of the above expressions into \eqref{eq:nonlinear:fixed} and \eqref{eq:nonlinear:newton} gives (see \ref{app:stability}):
\begin{mymultline}\label{eq:amplification:fp}
  r_{\textrm{f.p.}}\of{\omega_x} = 
  \left(
  \frac
  {\latheat}
  {\ml{1} (1-\partcoeff_1) \tilde{\C{1}}}
  \right)
  \left( 
  \frac{\D{1} \Omega_1\of{\omega_x} - (1-\partcoeff_1) \tilde{\vn}}{\heatcond{\sph} \dom{\sph} \of{\omega_x} + \heatcond{\lph} \dom{\lph} \of{\omega_x}}
  \right)
  \\
  - 
  \sum_{i = 2}^{N}
  \left(
  \frac
  {\ml{\J}}
  {\ml{1}} 
  \right)
  \left(
  \frac
  {1-\partcoeff_\J}
  {1-\partcoeff_1} 
  \right)
  \left(
  \frac
  {\tilde{\C{\J}}}
  {\tilde{\C{1}}} 
  \right)
  \left(
  \frac
  {\D{1} \Omega_1 \of{\omega_x} - (1-\partcoeff_1) \tilde{\vn}}
  {\DD{\lph}{\J} \Omega_\J \of{\omega_x} - (1-\partcoeff_\J) \tilde{\vn}}
  \right),
\end{mymultline}
and
\begin{myalign}\label{eq:amplification:an}
  r_{\textrm{a.N.}}\of{\omega_x} = 1 - \frac{1 - r_{\textrm{f.p.}}\of{\omega_x}}{1 - r_{\textrm{f.p.}}(0)}.
\end{myalign}
where
\begin{myalign}\label{eq:frequencies}
\begin{alignedat}{2}
\Omega_\J\of{\omega_x} &= \sqrt{\omega_x^2 + \frac{a}{\Delta t \DD{\lph}{\J}}}, \quad &\J &\in \range{1}{N}, \\
\dom{\phase}\of{\omega_x} &= \sqrt{\omega_x^2 + \frac{\heatcapv{\phase}}{\Delta t \heatcond{\phase}}}, \quad &\phase &= \sph,\lph.
\end{alignedat}
\end{myalign}
First, note that the second term in the expression for the amplification factor in the case of the fixed-point iteration is $\Oof{1}$, depending on problem parameters it can be greater or less than one. Thus, it follows immediately that the fixed-point iteration is not a robust approach for solving the nonlinear system of PDEs at hand. The expression for the amplification factor in the case of the proposed approximate Newton method is quite involved, however it appears that for typical physical parameters the value of $r_{\textrm{f.p.}}$ is always negative and reaches its maximum value at $\omega_x =0$. It is straightforward to show that under this conditions $r_{\textrm{a.N.}} < 1$. Figure \ref{fig:stability:wavenumber} shows the dependence of amplification factors in cases of the fixed-point iterative scheme and the approximate Newton one for a ternary alloy with parameters 
$\C{1} = 10.7$ at\%, $\C{2} = 9.4$ at\%, 
$T_m = 1900$ K, $\ml{1} = -5.43$ K/at\%, $\ml{2} = -10.4$ K/at\%,
$\partcoeff_1 = 0.94$, $\partcoeff_2 = 0.83$, 
$\D{1} = 10^{-5}$ cm$^2$/s, $\D{2} = 2\cdot10^{-5}$ cm$^2$/s,
$\vn = 0.01$ cm/s, $\latheat = 2600$, $\den{\lph} = \den{\sph} =  9.24 \cdot 10^{-3}$,
$\heatcond{\lph} = \heatcond{\sph} = 1.3$, $\heatcap{\lph} = \heatcap{\sph} = 356$ (motivated by the Co-W-Al alloy simulated later in this work).

In order to further investigate the robustness of the proposed approximate Newton approach we perform a parameter sweep in ranges $\C{1} \in \range{1}{20}$ at\%, $\C{2} \in \range{1}{20}$ at\%, 
$\ml{1} \in \range{-20}{-1}$ K/at\%, $\ml{2} \in \range{-20}{-1}$ K/at\%,
$\partcoeff_1 \in \range{0.1}{0.9}$, $\partcoeff_2 \in \range{0.1}{0.9}$, 
$\D{1} \in \range{10^{-6}}{10^{-4}}$ cm$^2$/s, $\D{2} \in \range{10^{-6}}{10^{-4}}$ cm$^2$/s and compute the maximum amplification factor in each case. The worst case is found to have an amplification factor of $0.9536$, which still indicates convergence although a very slow one. However, the total number of cases with relatively high amplification factors is found to be small. As demonstrated in figure \ref{fig:stability:sweep} the number of cases having the amplification factor $0.5$ or less is more than $90$\%; thus we expect the proposed method to perform well for a wide range of alloys. A further investigation into developing a more accurate Newton-type method as outlined in section \ref{sec:derivation:nonlinear:newton} in future works would likely result in a method performing well for alloys with any parameters.
 
\section{\reviewerOne{Numerical Methods and Solution Procedure}}\label{sec:numerics}
\hide{rev1:3}
Besides dealing with a system of \reviewerThree{nonlinear coupled}\hide{rev3:3} partial differential equations, the simulation of the multialloy solidification also poses several additional challenges. First of all, the solidification front evolves in time and undergoes large transformations (e.g., from a planar front into a forest of dendritic structures). Second, the solution of PDEs for the temperature and concentration fields requires enforcing Dirichlet, Robin, and ``jump'' interface conditions on this evolving irregular interface between the liquid and solid phases. Third, during the solidification of multicomponent alloys, the solute-rejection phenomena leads to the development of a steep solutal boundary layer ahead of the crystallization front, which must be accurately captured numerically since they directly influence the dynamics of the process. 
Similarly to  
\cite{theillard2015sharp}, 
we address these challenges by a combination of adaptive Cartesian quad-tree grids (to resolve steep gradients), Level-Set Method (to describe the front and its evolution) and sharp numerical methods for imposing boundary and interface conditions (to accurately solve BVPs). Specifically, the present work is based on a second-order accurate Level-Set framework on adaptive quadtree grids presented 
\cite{min2007second} 
and parallelized in 
\cite{mirzadeh2016parallel}
using the scalable \verb|p4est| grid management library \cite{burstedde2011p4est}. 

\textbf{Remark.} Note that the proposed above Newton-type approach for solving the nonlinear system of governing equations can also be implemented in other numerical frameworks (e.g., interface-fitted finite element method).

\reviewerOne{
For the sale of clarity, we begin this section with summarizing the overall computational procedure: \hide{rev1:4}
\begin{enumerate}
\item Set $n=0$ and $t_0 = 0$.
\item Provide initial conditions for the front's shape $\att{\Gamma}{0}$ and velocity $\att{\vn}{0}$, temperature field $\set{\att{\T{\phase}}{0}}{\phase = \lph, \sph}{}$, and concentration fields $\set{\att{\C{\J}}{0}}{\J=1}{N}$.
\item Set $n \leftarrow n+1$.
\item Compute time step $\Delta t_{n}$ according to \eqref{eq:numerics:timestep} and set $t_n = t_{n-1} + \Delta t_n$.
\item Advance solidification front $\Gamma^{n-1} \rightarrow \Gamma^{n}$ by solving \eqref{eq:numerics:advection}.
\item Refine/coarsen computational grid according to \eqref{eq:numerics:refinement}.
\item Reinitialize the level-set function by solving \eqref{eq:numerics:reinit} to restore the signed-distance property.
\item Compute properties of the just solidified material as described in section \ref{sec:numerics:solid}.
\item Solve nonlinear system of equations \eqref{eq:nonlinear:diffusion_T}-\eqref{eq:nonlinear:bc_C} for $\set{\att{\T{\phase}}{n}}{\phase = \lph, \sph}{}$, $\set{\att{\C{\J}}{n}}{\J=1}{N}$ and $\att{\vn}{n}$ using Algorithm~\ref{algo:newton} and initial guess $\Cg^{(0)} = \att{\C{1}}{n-1}$.
\item If $t_n < t_\text{final}$, go to step 3.
\end{enumerate}
where specific steps are discussed in the subsections to follow.}

\subsection{Space discretization: Adaptive Quadtree grids}\label{sec:numerics:grid}
\begin{figure}[!h]
  \centering
  \begin{subfigure}[b]{0.6\textwidth}
    \centering
    \includegraphics[width=0.99\textwidth]{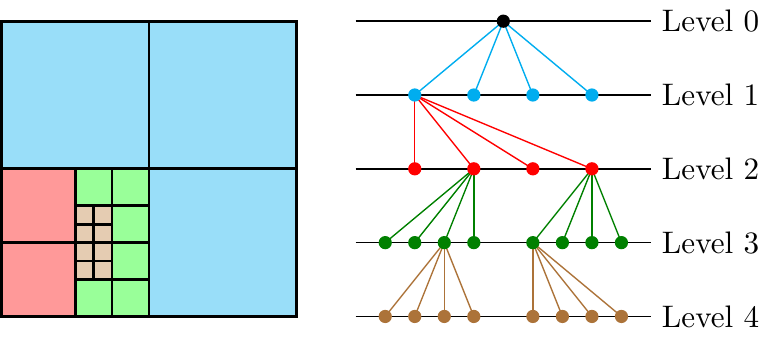}
    \caption{}
  \end{subfigure}
  \begin{subfigure}[b]{0.3\textwidth}
    \includegraphics[width=0.99\textwidth]{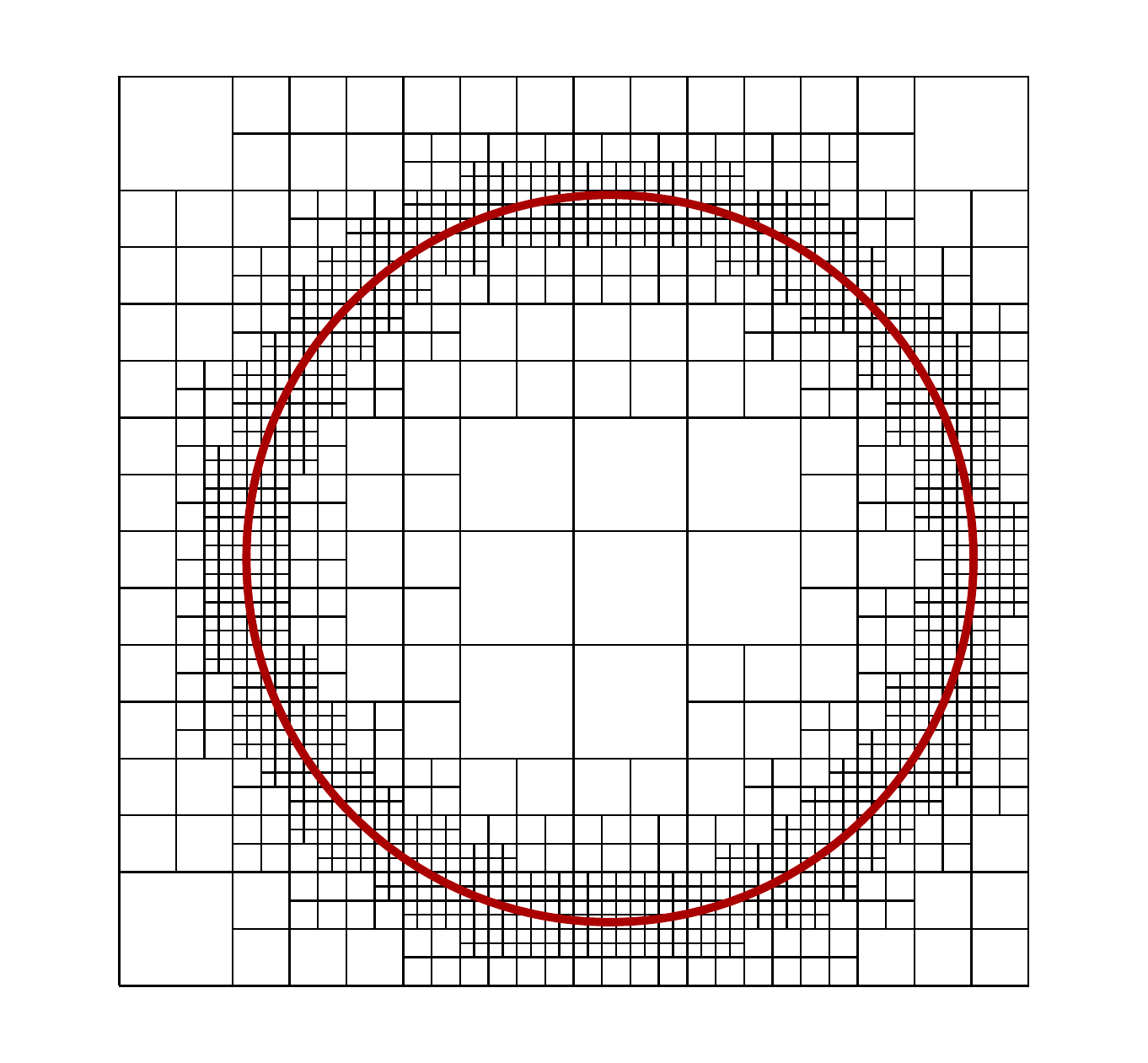}
    \caption{}
  \end{subfigure}
  \caption{(a) illustration of the hierarchical structure of a quad-tree grid. (b) example of adaptive mesh refinement using Cartesian quad-tree grids and the refinement criterion used in this work in case of a circular interface for $l_\text{min} = 3$, $l_\text{max} = 6$, and $K = 1.4$.}
  \label{fig:quadtree}
\end{figure}

In order to efficiently address the multiscale nature of the solidification process without compromising the accuracy of the numerical approximations, we employ adaptive Cartesian quadtree grids to discretize the computational domain. Such grids are constructed using a selective recursive refinement of rectangular cells into four smaller equal cells starting from a root cell that represents the entire computational domain. The size of cells in such a grid are equal to $\left( L_x \times L_y \right)/2^l$, where $\left( L_x \times L_y \right)$ is the root cell's size and $l$ is an integer called \textit{the level of refinement} that is equal to the number of refinements. Usually the resolution of quad-tree grids is specified by the minimum and the maximum levels of refinement, $l_\text{min}$ and $l_\text{max}$ in the computational domain. Fig. \ref{fig:quadtree} (a) illustrates the construction process and the concept of the refinement level for cells.

The spatially adaptive structure of a quad-tree grid allows one to create regions of high densities of computational nodes in areas where the finest resolution is needed for accuracy and for capturing fine spatial details, while keeping the grid relatively coarse outside of such regions. Specifically, in this work we construct the computational grid such that in the band of width $B$ around the solidification front, every grid cell is refined to the highest level $l_{\max}$, while any cell outside of this band is refined if the $K^{\rm th}$ fraction of its diagonal is greater than the distance from that cell to the band. Mathematically this criterion can be expressed as follows: \reviewerThree{a grid cell}\hide{rev3:4} $\mathcal{C}$ is refined if  
\begin{myalign} \label{eq:numerics:refinement}
  \min_{\rin \mathcal{C}} \left| \textrm{dist} \of{\vecr} \right| < B + K \textrm{diag}(\mathcal{C}).
\end{myalign}
This simple refinement criterion creates a computational grid with a band of \reviewerThree{the finest}\hide{rev3:5} grid cells with $l = l_\text{max}$ around the solidification front and gradually coarsen cells to the largest $l=l_\text{min}$ away front the front (see Fig. \ref{fig:quadtree} (b)). Such a refinement strategy is adequate for simulating solidification processes considered in this work since the steepest gradients are expected in the vicinity of interface $\Gamma$. In the simulations, we take $B=2$ and $K=1$.

We choose to represent spatial fields by their values at corners of grid cells. Such a choice allows for an easy calculation of first- and second- Cartesian derivatives as well as interpolation of spatial fields as described in \cite{min2006supra,min2007second}.

\subsection{Description of solidification front: Level-Set Method}\label{sec:numerics:level_set}
\begin{figure}[!h]
  \centering
  \begin{subfigure}[b]{.49\textwidth}
    \centering
    \includegraphics[width=\textwidth]{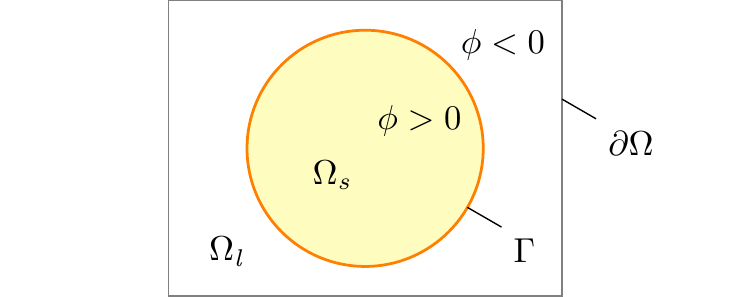}
  \end{subfigure}
  \begin{subfigure}[b]{.49\textwidth}
    \centering
    \includegraphics[width=.55\textwidth]{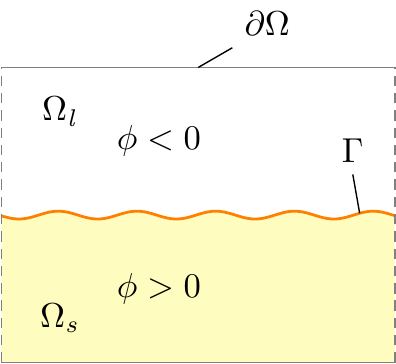}
  \end{subfigure}
  \caption{Illustration of representing irregular domains by the Level-Set Method on examples of crystal growth from a seed (left) and directional solidification (right).}
  \label{fig:level_set}
\end{figure}

In the Level-Set Method \cite{sethian1999level, osher2003level, gibou2018review}, the boundary of an irregular domain is implicitly defined by the zero-isocontour of a Lipschitz-continuous function $\phi\of{\vecr}$, called the level-set function, such that it has one sign inside the domain and the opposite one outside. We set the level-set function $\phi\of{\vecr}$ to be negative in the liquid phase and positive in the solid phase (see Fig. \ref{fig:level_set}), i.e:
\begin{myalign*}
  \phi\of{\vecr} &< 0 \quad \forall\ \rin \dom{\lph}, \\
  \phi\of{\vecr} &= 0 \quad \forall\ \rin \Gamma, \\
  \phi\of{\vecr} &> 0 \quad \forall\ \rin \dom{\sph}.
\end{myalign*}
Following the common practice we choose the level-set function to be the signed distance to the interface $\Gamma$.

Among advantages of the Level-Set Method is that it provides an easy way to compute the normal vector to the boundary and its mean curvature:
\begin{myalign*}
  \vect{n} &= \vect{n}_s = - \vect{n}_l = - \frac{\nabla \phi}{| \nabla \phi |},
  \\
  H &= - \nabla \cdot \vect{n} = \nabla \cdot \frac{\nabla \phi}{| \nabla \phi |}.
\end{myalign*}

The evolution of an interface represented by a level-set function $\phi\of{\vecr}$ under a velocity field $\vect{v}$ is described by a simple advection equation:
\begin{myalign}
  \label{eq:numerics:advection}
  \ddt{\phi} + \vect{v} \cdot \nabla \phi = 0.
\end{myalign}
We use a second-order accurate semi-Lagrangian method to solve the above equation. That is, the value of the level-set functions $\phi^{n} (\vect{r}_\J)$ at a location $\vect{r}_\J$ and time instant $t_n$ is computed as the value of the level-set function at time instant $t_{n-1}$ at the departing point $\vect{r}_d$ of the characteristic that passes through the point $(\vect{r}_\J, t_n)$:
\begin{myalign*}
  \phi^{n}(\vect{r}_\J) = \phi^{n-1}(\vect{r}_d),
\end{myalign*}
where the departure point $\vect{r}_d$ is computed by tracing the characteristic backward in time using the midpoint rule as described, e.g., in \cite{min2007second}.

To restore the signed distance property of the level-set function, which usually deteriorates during the advection steps, we solve the reinitialization equation for a few fictitious time steps $\tau$:
\begin{myalign}\label{eq:numerics:reinit}
  \partial_\tau \phi^{n}(\tau, \vect{r}) + \mathtt{sgn}(\phi^n(0, \vect{r})) (| \nabla \phi^n(\tau, \vect{r}) | - 1) = 0,
\end{myalign}
where $\verb|sgn|$ is the signum function. Specifically, we employ a method based on using the Godunov Hamiltonian for the discretization of $| \nabla \phi |$, a TVD RK-2 time-stepping scheme and the sub-cell fix of 
\cite{russo2000remark}
as described in 
\cite{min2007second}.

During the course of the simulation, the solidification front may evolve in such a way that it leads to under-resolved geometries. A typical situation is the slow solidification of narrow inter-dendritic gaps. The poor resolution by computational grid of such regions may cause an unstable behavior of the computational scheme. To avoid such issues, underresolved region (if any) are regularized after each motion of the front as described in \ref{app:removing}.

\subsection{Solving BVPs}\label{sec:numerics:solvers}
The advantages of using adaptive grids and the level-set method come at the price of (1) discretizing PDEs on complex nonuniform grid structures and (2) imposing boundary conditions on implicitly defined interfaces that cut through grid lines \reviewerTwo{arbitrarily}\hide{rev2:3}. We note, however, that the refinement criterion used in this work ensures that the computational grid is locally uniform near the solidification front, thus, nonuniform node arrangements (such as T-junctions and missing neighbors) are present only away from it and one needs to deal with the two tasks separately. Specifically, for the discretization of Poisson-type equations \eqref{eq:nonlinear:diffusion_T}-\eqref{eq:nonlinear:bc_C} on adaptive quadtree structures we use the second-order accurate approach of 
\cite{min2006supra}. 
Dirichlet boundary conditions (for $\C{1}$) are imposed using the Shortley-Weller method 
\cite{shortley1938numerical,ng2009guidelines}. 
Robin boundary conditions (for $\{ \C{\J} \}_{\J=2}^{N}$) are imposed using the finite volume method described in 
\cite{bochkov2019solving}. 
Finally, to impose jump conditions (for $\{ \T{\phase} \}_{\phase = \sph,\lph}$) we use a finite-volume approach of \cite{bochkov2020solving}. All the aforementioned numerical methods reduce to the standard five-point discretization on uniform grids. This allows their seamless combination on adaptive Cartesian grids provided grids are locally uniform in the vicinity of boundaries and interfaces.

After solving each BVP, its numerical solution is smoothly extended across the interface $\Gamma$. This simplifies the calculation of the front's velocity \eqref{eq:spliting:v} and also defines valid values at grid points near the interface that will become part of solution domain during the next time step. Specifically, we use the PDE-based quadratic extension presented in \cite{bochkov2020pde}. This approach consists in computing the first- and second- order Cartesian derivatives $\vect{q} = \nabla u$ and $\mat{Q} = \nabla\nabla u$ of a given spatial field $u$ inside the domain where $u$ is well-defined, followed by the hierarchical extension of these quantities as well as the numerical values of $u$ themselves to grid nodes outside of the domain by sequentially solving the following three advection-type equations for 50 time steps until steady state is reached in vicinity of the interface:
\begin{myalign*}
&\partial_\tau \mat{Q}  + \chi_{\nabla\nabla} \left( \nabla \mat{Q}             \right) \cdot \vect{n} = 0, \\
&\partial_\tau \vect{q} + \chi_{\nabla}       \left( \nabla \vect{q} - \mat{Q}  \right) \cdot \vect{n} = 0, \\
&\partial_\tau u        + \chi                \left( \nabla u        - \vect{q} \right) \cdot \vect{n} = 0,
\end{myalign*}
where $\tau$ is a fictitious time and $\chi$, $\chi_{\nabla}$ and $\chi_{\nabla\nabla}$ denote characteristic functions representing grid nodes that did not contain valid values of $u$, $\nabla u$ and $\nabla\nabla u$, correspondingly. The above advection-type equations are solved using an explicit upwind scheme (see \cite{bochkov2020pde} for details).

\subsection{Determining the simulation time-step}\label{sec:numerics:timestep}

From the point of view of accuracy it is reasonable to select the time step $\Delta t$  such that the solidification front advances no more than a predefined fraction $f_{\textrm{CFL}}$ of the smallest cell size. From the point of view of stability one has to choose the time step small enough to satisfy the stability constraint $\Delta t < \Delta t_{\textrm{crit}}$ due to an explicit discretization of the curvature dependent evolution of the solidification front. In case of the solidification of pure materials such a criterion has the form (see \cite{hou1994removing}) $\Delta t_{\textrm{crit}} = B \left( \Delta x \right)^{\frac{3}{2}}$, where constant $B$ depends on the parameters of the problem but not on the mesh size $\Delta x$. For the problem at hand, we expect a similar dependence; however we do not attempt to analytically establish such a relation and simply find the critical time step $\Delta t_\text{crit}$ by a trial and error approach for each run. The overall value of the time step is thus given by:
\begin{myalign} \label{eq:numerics:timestep}
  \Delta t^n = \min \left( f_{\textrm{CFL}} \frac{\Delta x}{\max_{\vecr\in\Gamma} \of{\vn^n}}, \Delta t_\textrm{crit} \right)
\end{myalign}
In numerical experiments presented in this work we use $f_\text{CFL} = 0.4$ unless stated otherwise.

\hide{rev2:7}\reviewerTwo{In typical cases of directional dendritic solidification, the time step is usually restricted by $\Delta t_\textrm{crit}$ in early simulation stages when the front velocity is low, but once the dendrites are fully developed, the dendrite's tip velocity reaches high enough values so that the CFL-like restriction becomes dominant.}

\subsection{Composition of the solid phase}\label{sec:numerics:solid}
The mathematical model of the solidification process presented in this paper assumes an infinitely slower transport of solutes in the solid phase than in the liquid phase, which is a good approximation for typical metal alloys. As result the solid phase's composition has no influence on the evolution of the crystallization front. While it is not necessary to solve for and keep track of concentration fields inside the solid phase $\set{ C^s_\J\of{t,\vecr} }{\J=1}{N}$ for simulating the solidification process, such information is of greatest importance for the analysis of the resulting crystals.

Assuming negligible diffusion in the solid, the governing equations for solutes' concentrations can be formally written as:
\begin{myalignat*}{2}
  \left\{
  \begin{alignedat}{2}
  \ddt{C^s_\J} &= 0, \quad &&\text{in } \dom{\sph}, \\
  C^s_\J &= \partcoeff_\J \Cl{\J}, \quad &&\text{on } \Gamma.
  \end{alignedat}
  \right.
  , \quad \J\in\range{1}{N}
\end{myalignat*}
In other words, once the alloy at a point in space $\vect{r}_* \in \Omega$ has crystallized at some time $t=t_*$, that is, $\vect{r}_* \in \Gamma(t_*)$, its compositions at this point, given by $\set{C^s_\J(t, \vect{r}_*) = \partcoeff_\J \Cl{\J}(t_*, \vect{r}_*)}{\J=1}{N}$, stays unchanged for $t>t_*$.

Our numerical method for determining the solid composition mimics this behavior. Specifically, after each motion of the solidification front, for each grid node $\vect{r}_p$ belonging to the solid phase, i.e, $\vect{r}_p \in \dom{\sph}$, it is checked whether the phase transition happened during the last front's movement, that is, whether $\phi^{n} (\vect{r}_p) > 0$ and $\phi^{n-1} (\vect{r}_p) < 0$. If so, the time moment of the phase transition at node $\vect{r}_p$ is approximated as:
\begin{myalign*}
  t_*^{(p)} = t_{n-1} + \frac{|\phi_{n-1}|}{|\phi_{n-1}|+|\phi_{n}|} \Delta t_n,
\end{myalign*}
which is used to estimate the interface composition when it swept across the grid node:
\begin{myalign*}
  \Cs{\J} = \partcoeff_\J \Cl{\J}(t_*^{(p)}) = \partcoeff_\J \left( \frac{t_n - t_*^{(p)}}{\Delta t_n} \att{\Cl{\J}}{n-1} + \frac{t_*^{(p)} - t_{n-1}}{\Delta t_n} \att{\Cl{\J}}{n}  + \Oof{\Delta t^2} \right),
\end{myalign*}
where we used a linear interpolation between time instants $t_{n-1}$ and $t_{n}$ to compute $\Cl{\J}(t_*^{(p)})$. Once the solid's composition is obtained at a ``just solidified'' grid node it is stored unchanged for the remaining of a simulation run. In addition to the composition values, we also compute and store the front's normal velocity, temperature, curvature, and orientation at the moment of crystallization in to obtain a comprehensive data set of the solidification process. \hide{rev2:5}\reviewerTwo{Note that the accuracy with which the solid phase properties are recorded does not influence the overall accuracy of the simulation. This is because in this mathematical solidification model the solid phase is decoupled from the dynamics of the solidification process due to the assumption of an infinitely slow diffusion in the solid phase. Thus, using a linear interpolation is sufficient to obtain solid phase characteristics with the second order of accuracy and to not degrade the overall simulation accuracy.}

Since there is no species transport in the solid, any frozen steep concentration gradients persist in time even when the solidification front has moved significantly far (in some sense the concentration profiles in the solid phase remember the history of all front's locations). Thus, the refinement strategy used for simulating the dynamics of the solidification process, namely, a very fine grid only near the interface $\Gamma$, is not adequate for representing the composition fields in the solid. We address this issue by having a second grid that is refined to $l_\text{max}$ everywhere in $\dom{\sph}$ for storing the values of $\set{C^s_\J}{\J=1}{N}$. Because no other mathematical operations are performed on this grid, such an approach has negligible effect on the overall computational time.

\newpage
\section{Results}\label{sec:results}

In this section we present a number of examples related to the solidification of a Co-W-Al alloy with initial composition $C_{\textrm{W}}^\infty = 10.7$ at\% and $C_{\textrm{Al}}^\infty = 9.4$ at\% studied experimentally in \cite{tsunekane2011single}. First, we present tests for the validation of the computational approach and then analyze the solutal segregation during directional solidification.  We then extend the modeling to conditions relevant to additive manufacturing. The parameter values used to describe the alloy are listed in Table \ref{table:CoWAl:params}.
\begin{table}[!h]
\centering
\begin{tabular}{l|c|c}
Parameter & Value & Units \\
\hline
Density of liquid alloy, $\den{\lph}$ &  $9.24 \cdot 10^{-3}$ & $\kg\cdot\cm^{-3}$ \\
Density of solidified alloy, $\den{\sph}$ & $9.24 \cdot 10^{-3}$ & $\kg\cdot\cm^{-3}$ \\
Heat capacity of liquid alloy, $\heatcap{\sph}$ & 356 & $\jou\cdot\kg^{-1}\cdot\kel^{-1}$ \\
Heat capacity of solidified alloy, $\heatcap{\sph}$ & 356 & $\jou\cdot\kg^{-1}\cdot\kel^{-1}$ \\
Thermal conductivity of liquid alloy, $\heatcond{\lph}$ &  $1.3$ & $W\cdot\cm^{-1}\cdot\kel^{-1}$ \\
Thermal conductivity of solidified alloy, $\heatcond{\sph}$ & $1.3$ & $W\cdot\cm^{-1}\cdot\kel^{-1}$ \\
Latent heat of fusion, $\latheat$ & 2590 & $\jou\cdot\cm^{-3}$ \\
Degree of anisotropy, $\varepsilon$ & $0.05$ & -- \\
Curvature undercooling, $\epsilon_c$ & $10^{-5}$ & $\cm$ \\
Kinetic undercooling, $\epsilon_v$ & $0$ & $\secs\cdot\cm^{-1}$ \\
Diffusivity of W in liquid, $\D{\text{W}}$ & $10^{-5}$ & $\cm^2\cdot\secs^{-1}$ \\
Diffusivity of Al in liquid, $\D{\text{Al}}$ & $2 \cdot 10^{-5}$ & $\cm^2\cdot\secs^{-1}$. 
\end{tabular}
\caption{Parameters used for description of the Co-W-Al alloy.} \label{table:CoWAl:params}
\end{table}
Note that the density, the heat capacity, the thermal conductivity and the latent heat of fusion are estimated as weighted averages of those of pure elements. The undercooling parameters and diffusion coefficients are chosen in the range of typical magnitudes. 
\begin{figure}[!h]
\begin{center}
\includegraphics[width=.99\textwidth]{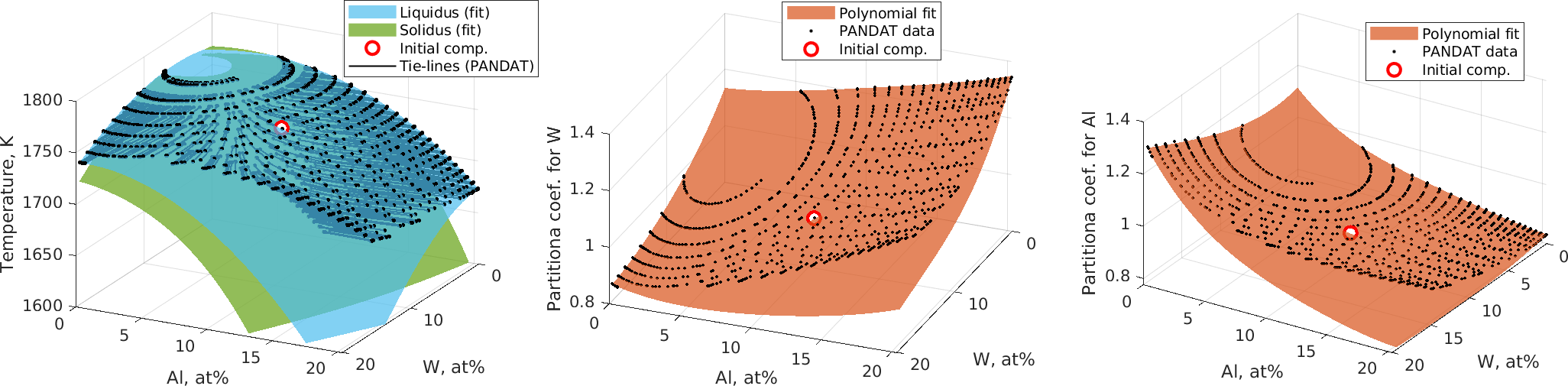}
\end{center}
\vspace{-.3cm}\vspace{-.3cm}\caption{\it Phase-diagram of Co-W-Al as predicted by the \PANDAT database (dots) and polynomial approximations (solid surface) used in this work.} \label{fig:CoWAlphase}
\end{figure}
The dependence of the liquidus surface and partition coefficients on the alloy composition are approximated by the fourth-order polynomials 
\begin{myalign*}
  \Delta T_C (C_{\textrm{W}}, C_{\textrm{Al}}) &= \sum_{p=0}^{4} \sum_{q=0}^{p} a^{\Delta T}_{p-q, q} C_{\textrm{W}}^{p-q} C_{\textrm{Al}}^{q}, \\
  \partcoeff_{\textrm{W}}  (C_{\textrm{W}}, C_{\textrm{Al}}) &= \sum_{p=0}^{4} \sum_{q=0}^{p} a^{\textrm{W}}_{p-q, q} C_{\textrm{W}}^{p-q} C_{\textrm{Al}}^{q}, \\
  \partcoeff_{\textrm{Al}} (C_{\textrm{W}}, C_{\textrm{Al}}) &= \sum_{p=0}^{4} \sum_{q=0}^{p} a^{\textrm{Al}}_{p-q, q} C_{\textrm{W}}^{p-q} C_{\textrm{Al}}^{q},
\end{myalign*}
fitted to data obtained from the \PANDAT database (see Figure \ref{fig:CoWAlphase}). The coefficients of these approximations are listed in Table \ref{table:CoWAl}.

\begin{table}[!h]
\centering
\begin{tabular}{c|c|c|c}
& $\Delta T_C (C_{\textrm{W}}, C_{\textrm{Al}})$ & $\partcoeff_{\textrm{W}}  (C_{\textrm{W}}, C_{\textrm{Al}})$ & $\partcoeff_{\textrm{Al}} (C_{\textrm{W}} C_{\textrm{Al}})$ \\
\hline
$a^{[\cdot]}_{0,0}$ & $ 0                $ & $ 1.135              $ & $ 1.114              $ \\          
$a^{[\cdot]}_{1,0}$ & $ 2.369              $ & $-3.118 \cdot 10^{-2}$ & $-9.187 \cdot 10^{-3}$ \\         
$a^{[\cdot]}_{0,1}$ & $ 1.771              $ & $ 2.239 \cdot 10^{-3}$ & $-4.804 \cdot 10^{-2}$ \\          
$a^{[\cdot]}_{2,0}$ & $-2.238 \cdot 10^{-1}$ & $ 1.463 \cdot 10^{-3}$ & $ 1.733 \cdot 10^{-3}$ \\          
$a^{[\cdot]}_{1,1}$ & $ 1.041 \cdot 10^{-1}$ & $-1.917 \cdot 10^{-3}$ & $-1.406 \cdot 10^{-4}$ \\         
$a^{[\cdot]}_{0,2}$ & $-3.046 \cdot 10^{-1}$ & $ 1.135 \cdot 10^{-3}$ & $ 4.313 \cdot 10^{-3}$ \\          
$a^{[\cdot]}_{3,0}$ & $ 1.358 \cdot 10^{-3}$ & $-4.768 \cdot 10^{-5}$ & $-5.248 \cdot 10^{-5}$ \\
$a^{[\cdot]}_{2,1}$ & $-1.568 \cdot 10^{-2}$ & $ 9.953 \cdot 10^{-5}$ & $-8.236 \cdot 10^{-5}$ \\
$a^{[\cdot]}_{1,2}$ & $-1.457 \cdot 10^{-2}$ & $ 4.394 \cdot 10^{-5}$ & $ 2.716 \cdot 10^{-5}$ \\
$a^{[\cdot]}_{0,3}$ & $ 3.317 \cdot 10^{-3}$ & $-6.706 \cdot 10^{-5}$ & $-2.159 \cdot 10^{-4}$ \\          
$a^{[\cdot]}_{4,0}$ & $-3.283 \cdot 10^{-6}$ & $ 8.710 \cdot 10^{-7}$ & $ 5.841 \cdot 10^{-7}$ \\
$a^{[\cdot]}_{3,1}$ & $ 3.559 \cdot 10^{-4}$ & $-2.235 \cdot 10^{-6}$ & $ 9.961 \cdot 10^{-7}$ \\
$a^{[\cdot]}_{2,2}$ & $ 1.228 \cdot 10^{-4}$ & $-5.952 \cdot 10^{-7}$ & $ 1.741 \cdot 10^{-6}$ \\
$a^{[\cdot]}_{1,3}$ & $ 1.531 \cdot 10^{-4}$ & $-3.904 \cdot 10^{-7}$ & $-9.305 \cdot 10^{-7}$ \\
$a^{[\cdot]}_{0,4}$ & $-1.866 \cdot 10^{-4}$ & $ 1.545 \cdot 10^{-6}$ & $ 4.122 \cdot 10^{-6}$   
\end{tabular}
\caption{Coefficients in polynomial approximations of $\Delta T_C (C_{\textrm{W}}, C_{\textrm{Al}})$, $\partcoeff_{\textrm{W}}  (C_{\textrm{W}}, C_{\textrm{Al}})$ and $\partcoeff_{\textrm{Al}} (C_{\textrm{W}} C_{\textrm{Al}})$ for Co-W-Al alloy.} \label{table:CoWAl}
\end{table}

\clearpage
\subsection{Validation of the numerical approach: Axisymmetric stable solidification } \label{sec:results:validation}
To validate the proposed computational approach we consider the problem of an axisymmetric solidification of a ternary alloy in the infinite domain due to a line sink, which has an analytical similarity solution (see \ref{app:analytical}) in the absence of kinetic and curvature undercoolings, that is, $\epsilon_c = 0$ and $\epsilon_v = 0$. In order to avoid the necessity of simulating a singular heat source and an infinite domain, we confine the solidification process into an annular region with the internal radius $r_{\textrm{in}} = 0.1L$ and external radius $r_{\textrm{out}} = 0.45L$, impose on boundaries of this region Dirichlet boundary conditions for the heat and concentration fields based on the analytical solution, and set the starting position of the solidification front at $r_{\textrm{start}} = 0.2L$, where $L = 0.02$ cm denotes the simulation scale. We select the analytical solution that satisfies the following conditions: at the infinity concentration of solutes approaches the nominal alloy composition, that is, $C_{\textrm{W}}^\infty = 10.7$ at\% and $C_{\textrm{Al}}^\infty = 9.4$ at\%; the normal front's velocity at the initial moment is equal to $\vn^0 = 0.01$ cm/s; and the ratio of the compositional to thermal gradients at the solidification front is $M_0 = 0.75$ (see \ref{app:analytical} for details).

\begin{figure}[!h]
\begin{center}
\begin{tabular}{ c c c c c c }
\includegraphics[width=.15\textwidth]{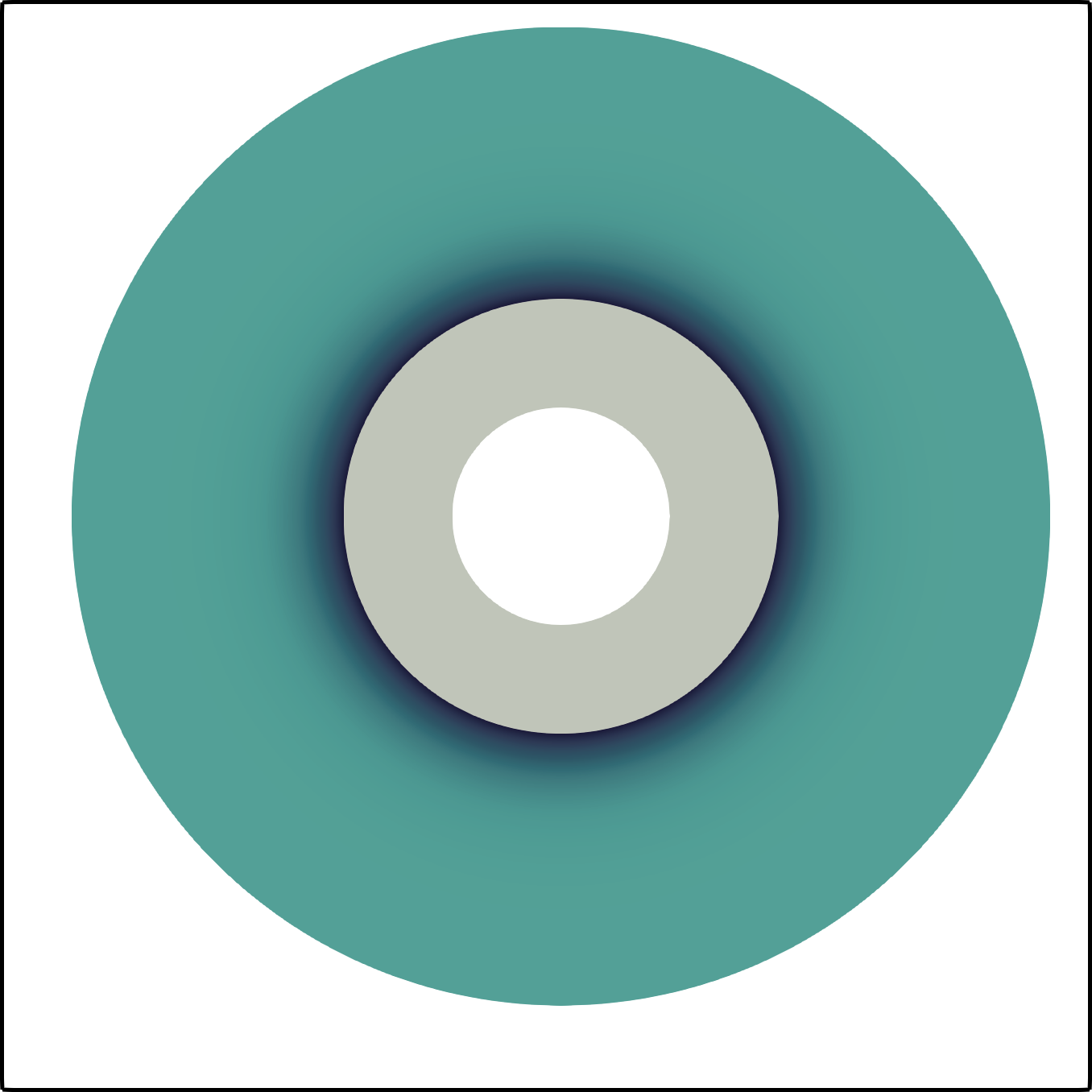}&
\includegraphics[width=.15\textwidth]{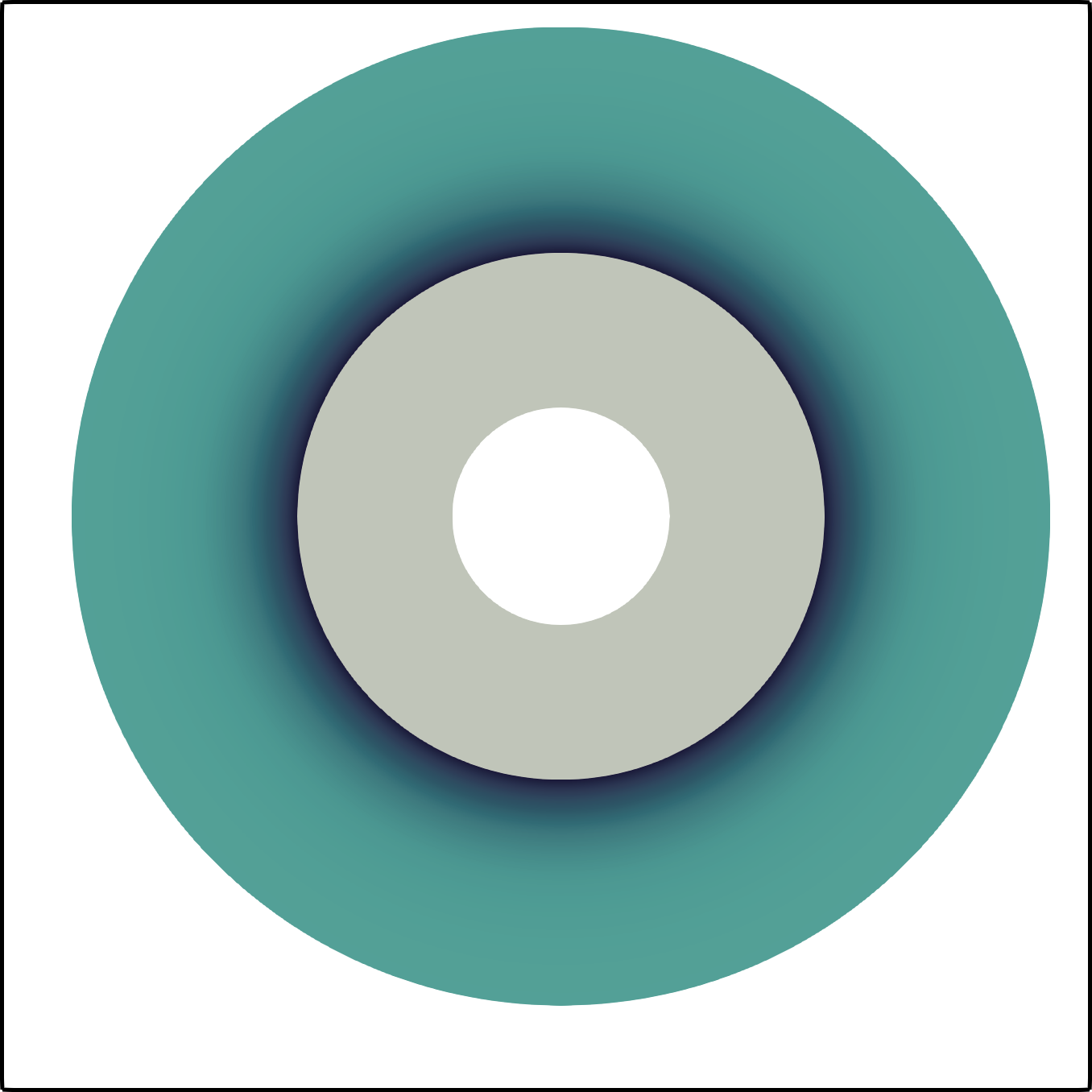}&
\includegraphics[width=.15\textwidth]{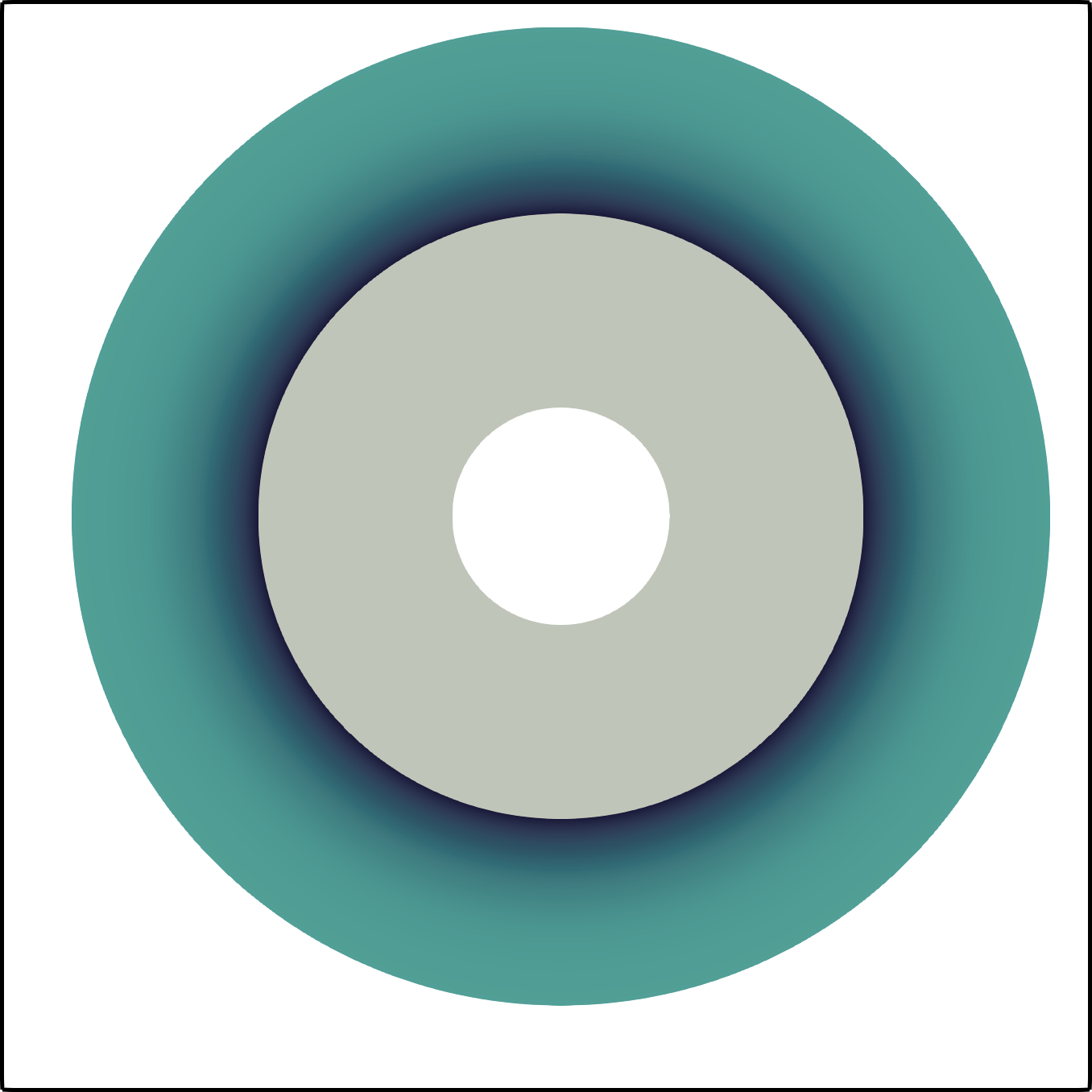}&
\includegraphics[width=.15\textwidth]{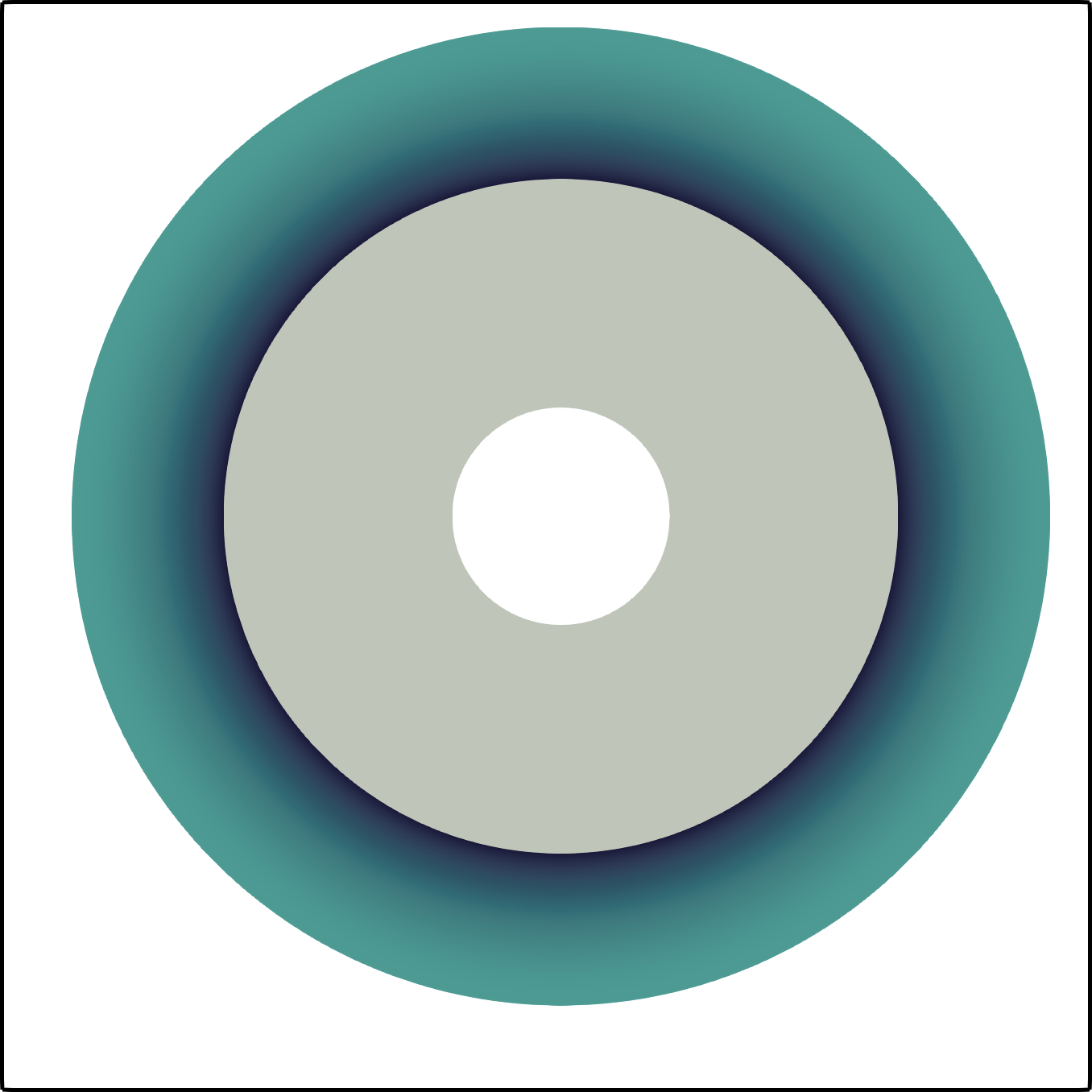}&
\includegraphics[width=.15\textwidth]{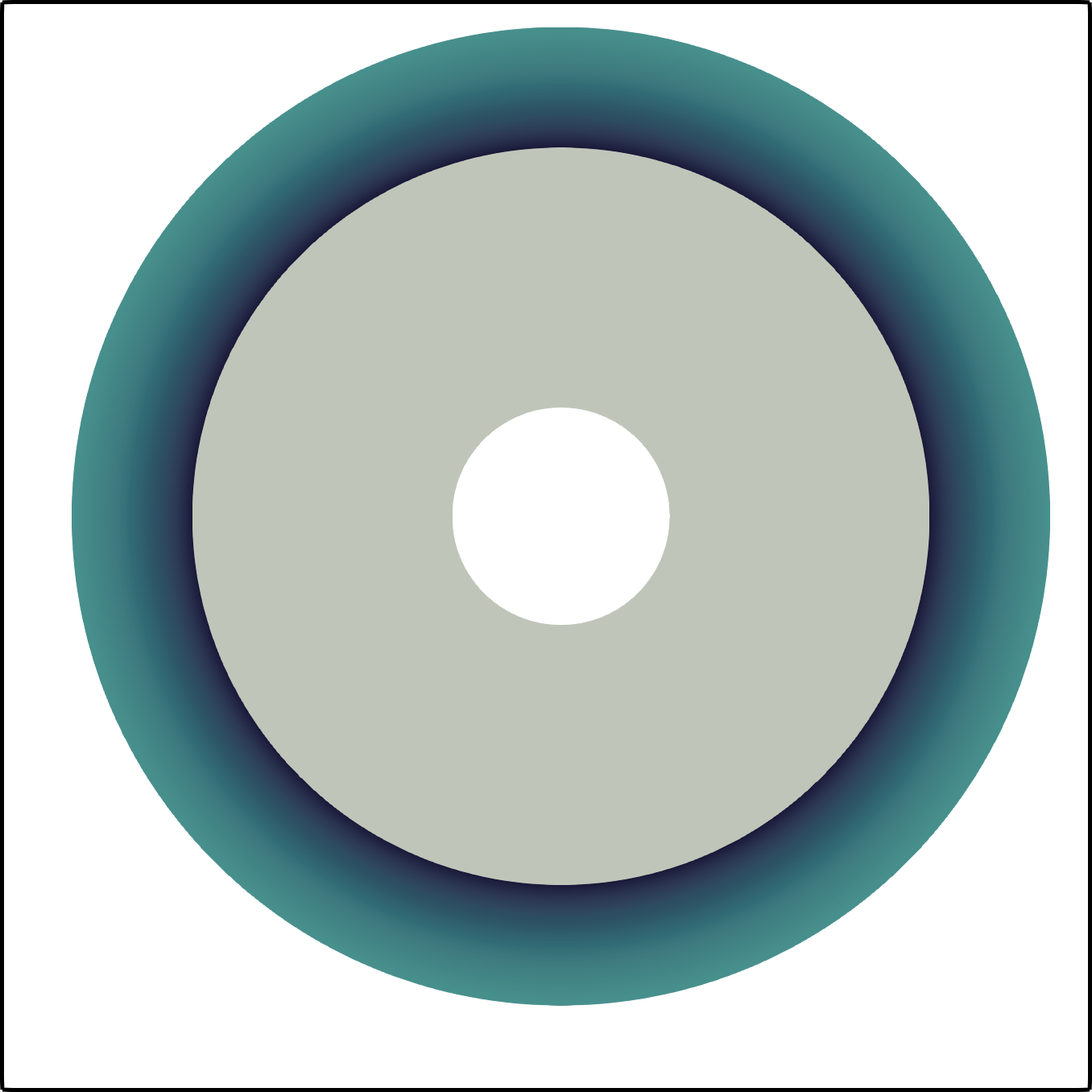}&
\includegraphics[width=.055\textwidth]{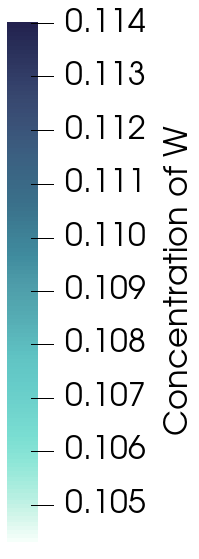}
\\
$t = 0$ ms &
$t = 93.75$ ms &
$t = 187.50$ ms &
$t = 281.25$ ms &
$t = 375.00$ ms
\end{tabular}
\end{center}
\vspace{-.3cm}\vspace{-.3cm}\caption{\it Visualization of the W concentration field obtained in the case of axisymmetric solidification at several moments of time.} \label{fig:validation:visual}
\end{figure}

Figure \ref{fig:validation:visual} illustrates simulation results for the distribution of W in the solid and liquid phases at several moments of time.

\begin{figure}[!h]
\begin{center}
\includegraphics[height=2.2in]{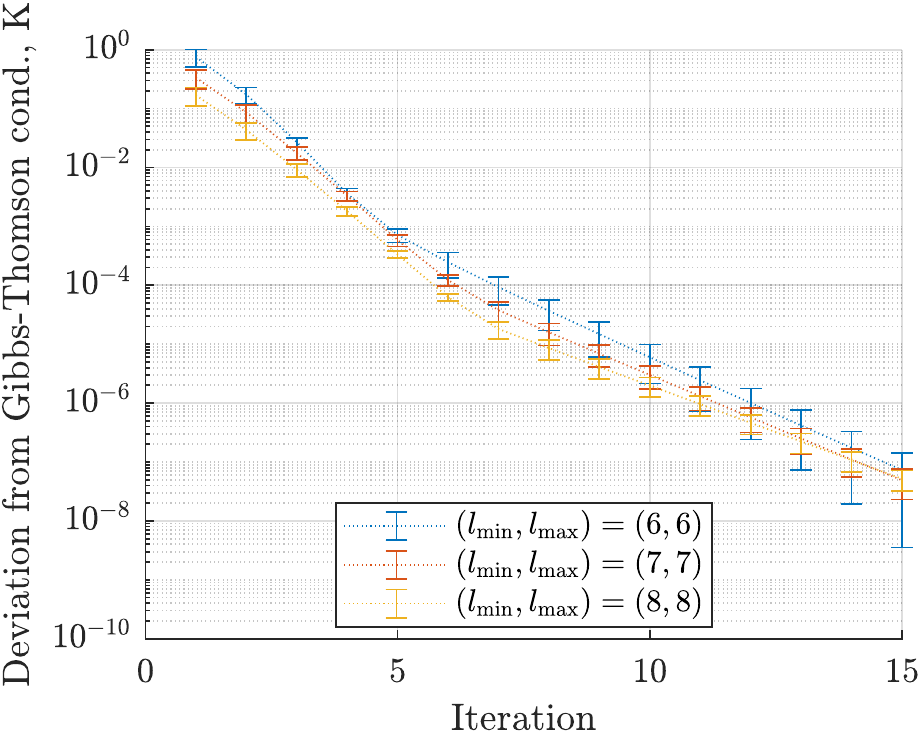}
\includegraphics[height=2.2in]{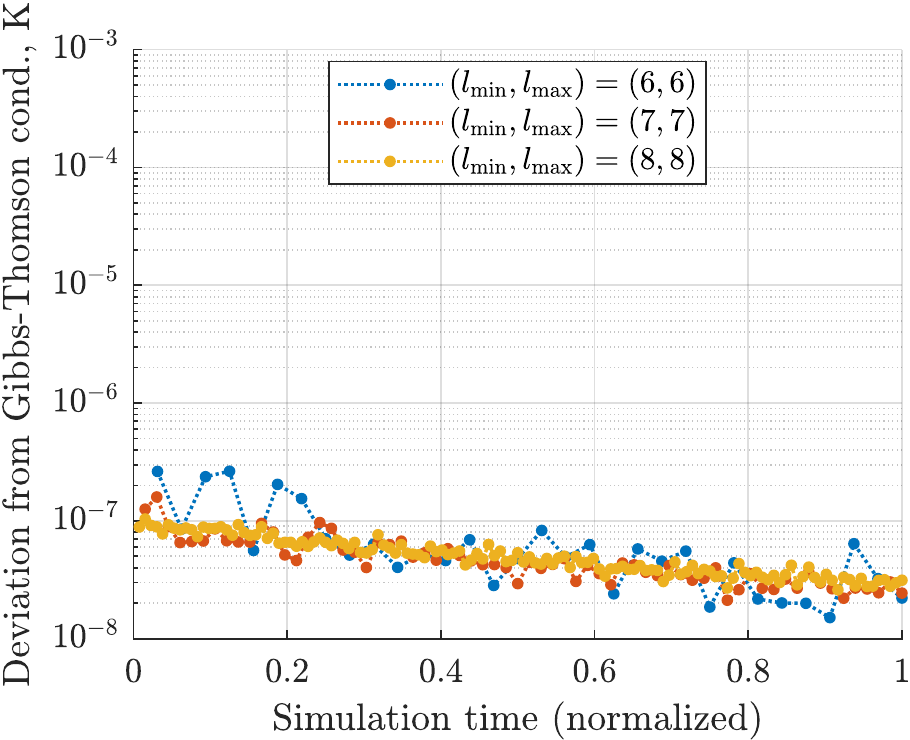}
\end{center}
\vspace{-.3cm}\vspace{-.3cm}\caption{\it Performance of the proposed approximate Newton method in the case of stable axisymmetric solidification for several grid resolutions. Left: convergence to the Gibbs-Thomson condition during one time step averaged among all time steps (error bars represent the standard deviation). Right: maximum deviation from the Gibbs-Thomson condition at all time steps.} \label{fig:results:convergence}
\end{figure}

Figure \ref{fig:results:convergence} shows the convergence of the Newton-type approach to the Gibbs-Thomson condition during a single time step and the deviation from the Gibbs-Thomson condition for all time steps for several grid resolutions (specifically, $64\times64$, $128\times 128$, and $256\times256$). As one can see, for well-resolved simulations the proposed approach shows little dependence on the grid resolution, converges rather quickly (deviation from the Gibbs-Thomson conditions is reduced by about 5 orders of magnitude in 10 iterations), and is able to maintain the error in satisfying interface conditions under $10^{-6}$ K throughout the entire course of simulation.

\begin{figure}[!h]
\begin{center}
\includegraphics[width=.75\textwidth]{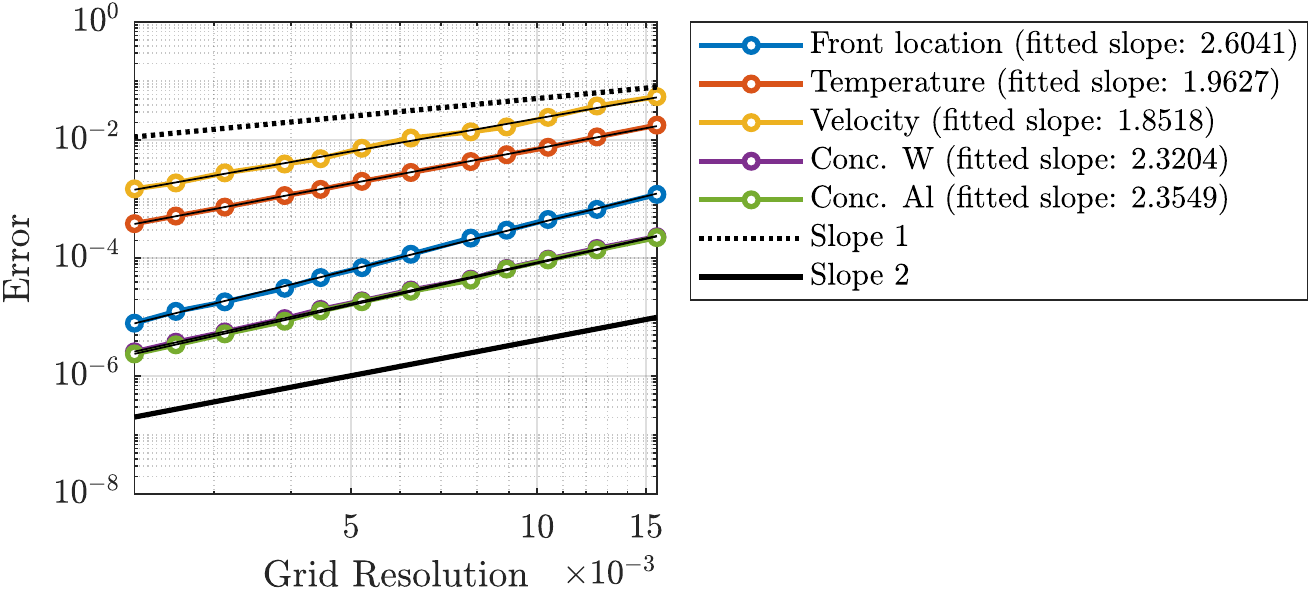}
\end{center}
\vspace{-.3cm}\vspace{-.3cm}\caption{\it Overall accuracy of the computational method in the case of stable axisymmetric solidification} \label{fig:results:accuracy}
\end{figure}

The accuracy analysis for the temperature field, the concentration fields, the front's velocity and the front's location is presented in Figure \ref{fig:results:accuracy}. The error of each quantity is defined as the maximum error in the \linf that occurred throughout the entire course of the simulation. The observed convergence rates are close to 2, which is consistent with the fact that second-order accurate numerical approximations are used in all components of the overall computational approach.

\begin{figure}[!h]
\begin{center}
\includegraphics[height=2.2in]{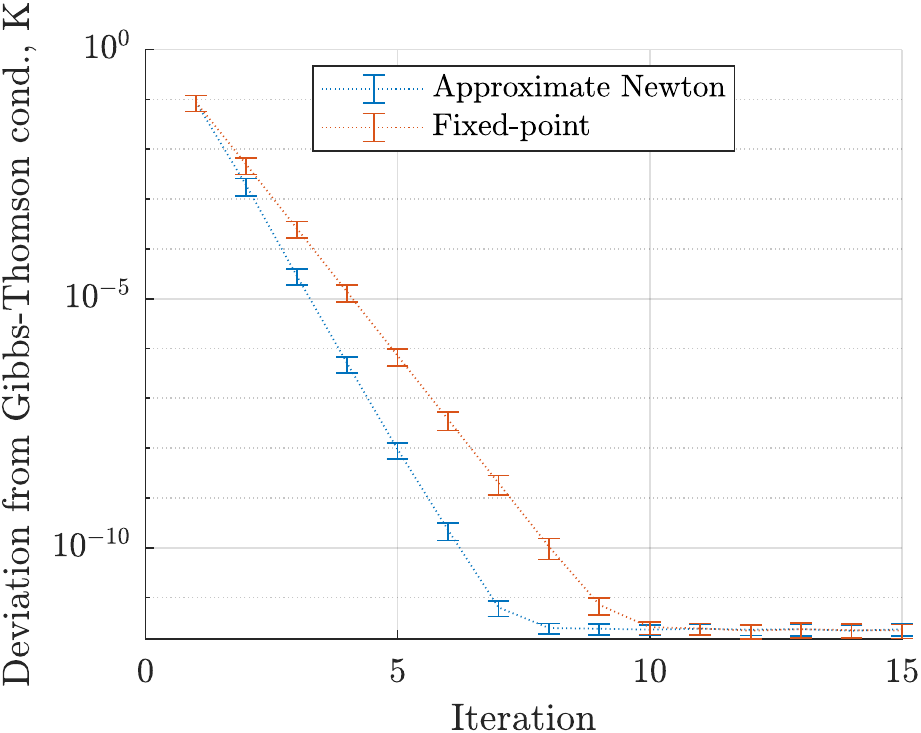}
\includegraphics[height=2.2in]{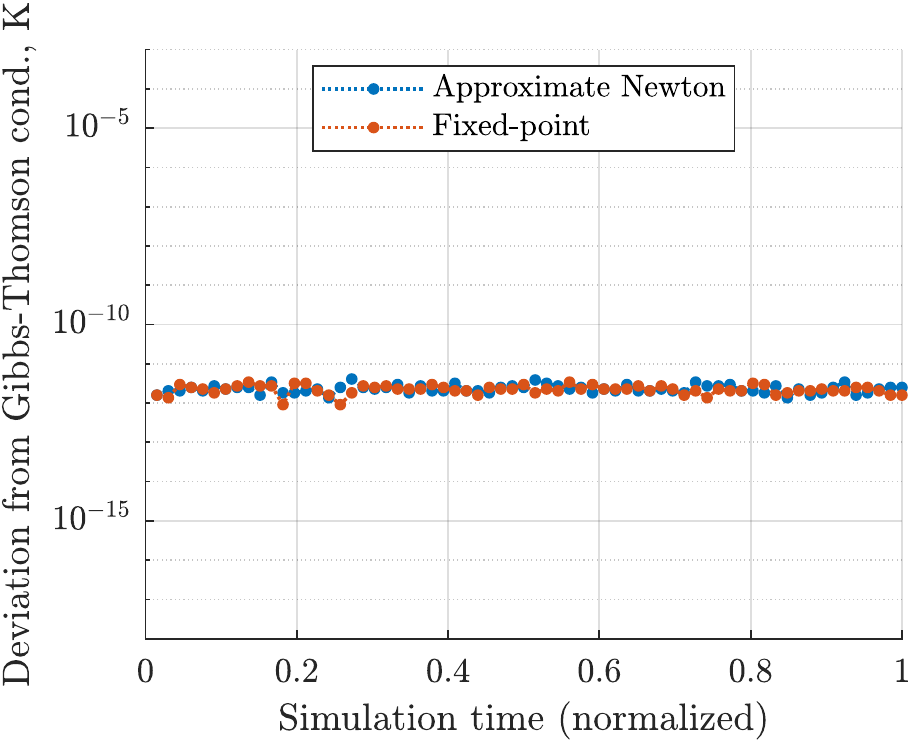}
\end{center}
\vspace{-.3cm}\vspace{-.3cm}\caption{\it Comparison between the proposed approximate Newton method and the fixed-point method of \cite{Theillard;Gibou;Pollock:15:A-sharp-computationa}. Left: convergence to the Gibbs-Thomson condition during one time step averaged among all time steps (error bars represent the standard deviation). Right: maximum deviation from the Gibbs-Thomson condition at all time steps.} \label{fig:results:convergence:compare}
\end{figure}

\hide{rev1:5}\reviewerOne{For the sake of comparison, we consider the solidification of a Co-W bi-alloy (just by removing Al component from the Co-W-Al system) in the same setting and apply both the proposed approximate Newton method and the fixed point method of \cite{Theillard;Gibou;Pollock:15:A-sharp-computationa}. The convergence for a single time step and the resulting deviation for all time steps for both methods are compared in figure \ref{fig:results:convergence:compare}. As one can see, both methods reduce the error in satisfying Gibbs-Thomson condition very quickly within 10 iterations and manage to maintain its level around $10^{-12}$ K throughout the entire simulation run. Note that the lower overall level of errors compared to the ternary alloy simulation above can be attributed to the more amenable nature of bi-alloy solidification to numerical simulations. Comparing performance of the two methods, one can notice that the proposed approximate Newton method slightly outperforms the fixed-point iteration method. However, one must remember that the approximate Newton method requires solution of an adjoint system of PDEs, which almost doubles the cost of a single iteration and makes implementation more involved. Thus, the main advantage of the approximate Newton method over the fixed-point iteration is the stabilization for multicomponent systems. If only the bi-alloy solidification is of interest, then the fixed-point method is preferred both from implementation and performance stand points.
}

\clearpage
\subsection{Directional solidification of a Co-W-Al ternary alloy} \label{sec:results:CoWAl}
In order to simulate the directional solidification of the Co-W-Al alloy, we consider a rectangular computational domain periodic in the $x$-direction and having dimensions $L\times 4 L$. The solidification front travels along the positive $y$-direction. The processing conditions are modeled by imposing thermal fluxes at the top and bottom boundaries of the computational domain, such that 
\begin{myalign*}
\heatcond{\lph} \ddn{\Tl} &= \heatcond{\lph} G_T, \quad y = pL, \\
\heatcond{\sph} \ddn{\Ts} &= - \heatcond{\lph} G_T + V \left(\latheat + \den{\lph} \heatcap{\lph} G_T pL \right), \quad y = 0,
\end{myalign*}
where $G_T$ is the desired temperature gradient and $V$ has the meaning of the approximate front velocity if the solidification were to occur in the planar regime. As initial conditions, we take a stationary planar solidification front in a prescribed temperature gradient $G_T$. Specifically, the initial location of the front is $y_0=0.1L$, the concentration fields are uniform in the solid and liquid:
\begin{myalign*}
\C{\textrm{W}} = 
\begin{cases}
\partcoeff_{\textrm{W}} \of{\C{\textrm{W}}^\infty, \C{\textrm{Al}}^\infty} \,\C{\textrm{W}}^\infty, &y < y_0, \\
\C{\textrm{W}}^\infty, &y > y_0, 
\end{cases}
\quad
\C{\textrm{Al}} = 
\begin{cases}
\partcoeff_{\textrm{Al}} \of{\C{\textrm{W}}^\infty, \C{\textrm{Al}}^\infty} \, \C{\textrm{Al}}^\infty, &y < y_0, \\
\C{\textrm{Al}}^\infty, &y > y_0,
\end{cases}
\end{myalign*}
and the temperature field is defined as:
\begin{myalign*}
T = 
\begin{cases}
\Tliq\of{\C{\textrm{W}}^\infty, \C{\textrm{Al}}^\infty} + (y-y_0) \, G_T \,\dfrac{\heatcond{\lph}}{\heatcond{\sph}}, &y > y_0, \\
\Tliq\of{\C{\textrm{W}}^\infty, \C{\textrm{Al}}^\infty} + (y-y_0)\, G_T, &y > y_0.
\end{cases}
\end{myalign*}

\begin{figure}[!h]
\begin{center}
\begin{subfigure}[b]{.99\textwidth}
\centering
\includegraphics[width=.99\textwidth]{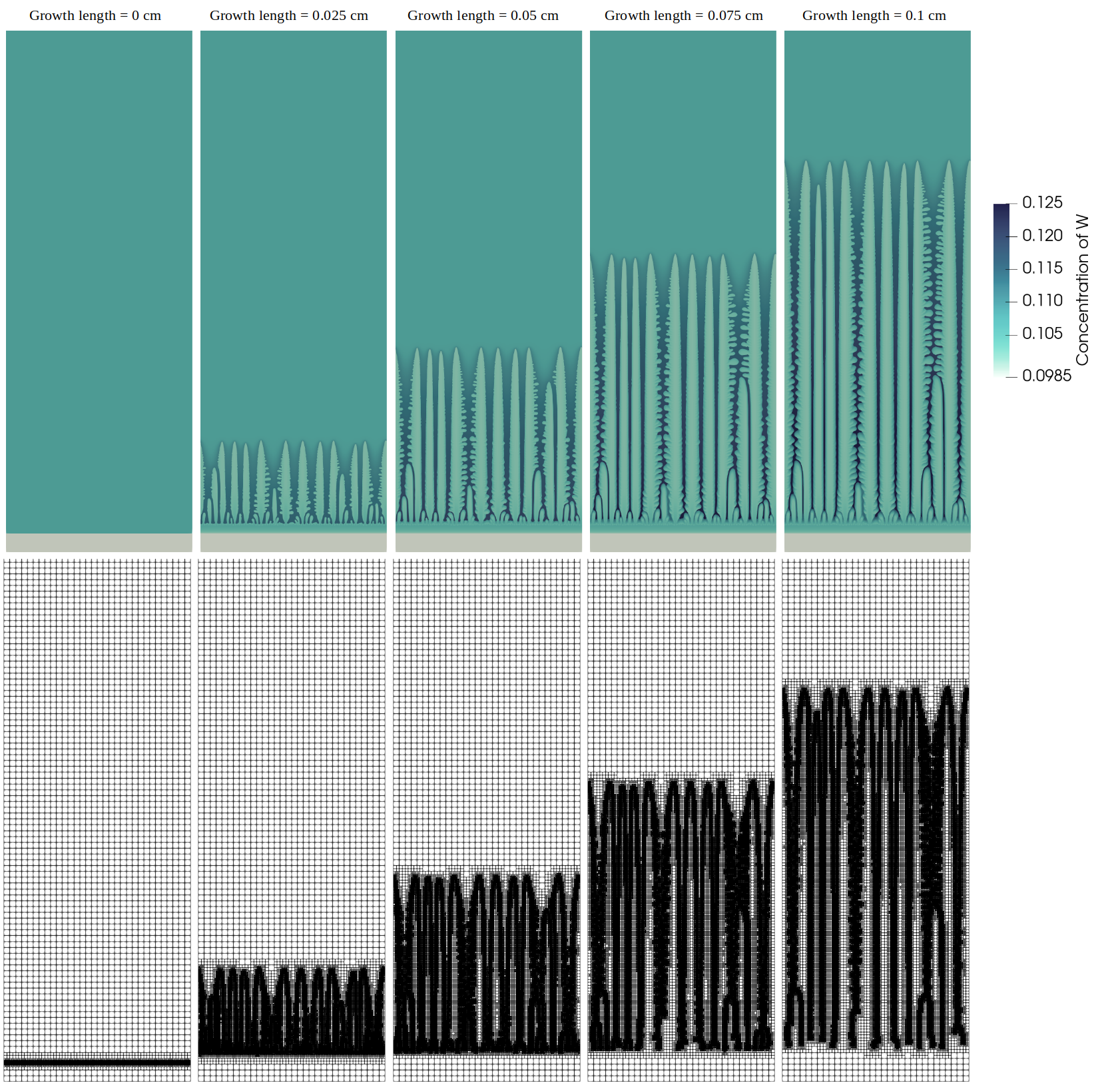}
\end{subfigure}
\end{center}
\vspace{-.3cm}\vspace{-.3cm}\caption{\it Progression in time of a representative simulation run (specifically, $L = 0.05$~cm, $V = 0.03$~cm/s, $G_T=$500~K/cm). Top row: concentration field of W. Bottom row: computational grid.} \label{fig:directional:representative}
\end{figure}

\begin{figure}[!h]
\begin{center}
\centering
\includegraphics[width=.75\textwidth]{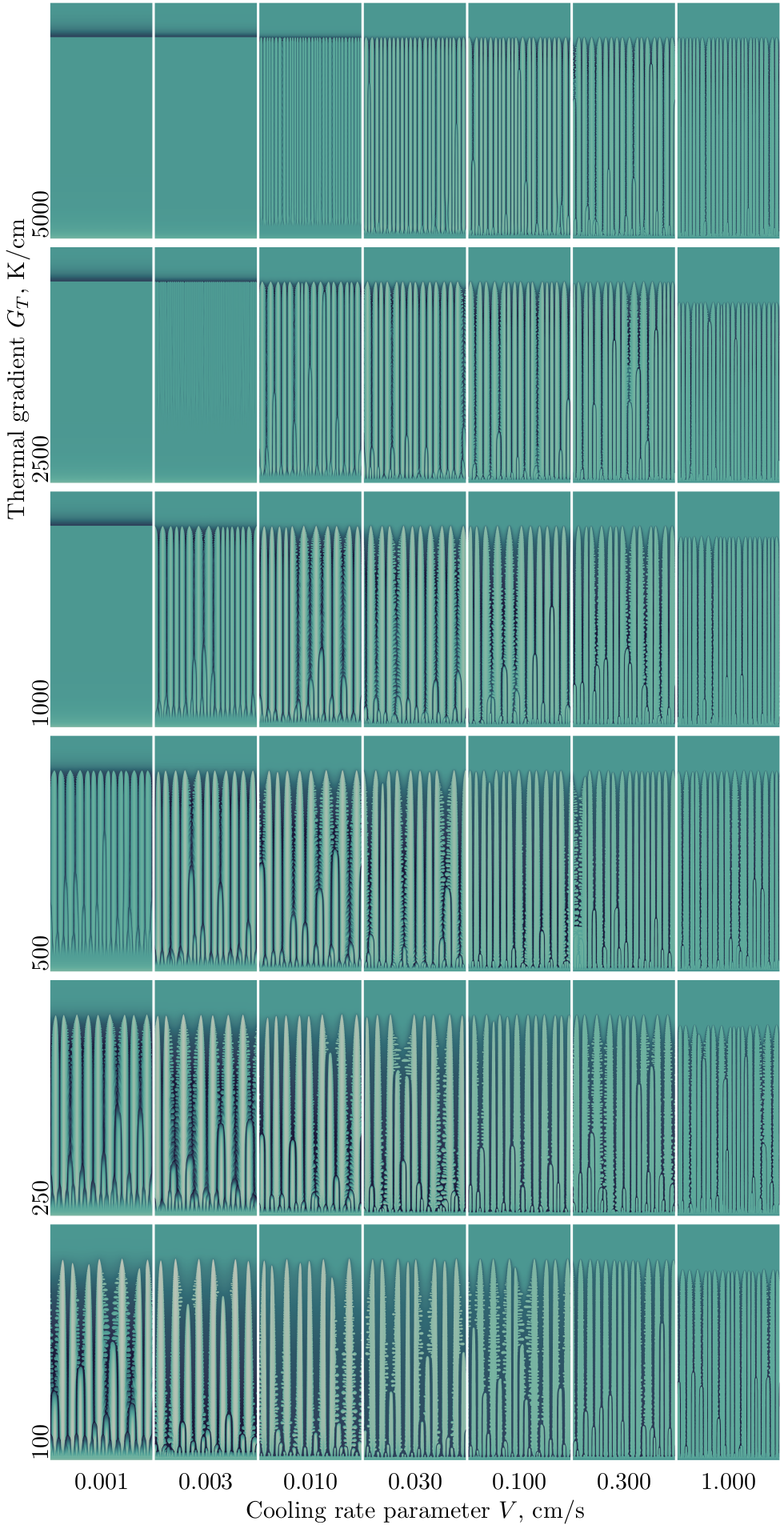}
\end{center}
\vspace{-.3cm}\vspace{-.3cm}\caption{\it Solidification microstructures obtained for the range of processing conditions $G_T$ from 100 K/cm to 5000 K/cm and $V$ from 0.001 cm/s to 1 cm/s. Relative size of simulation domains is not portrayed to scale (see Figure~\ref{fig:directional:size}). Physical domain size in each column is $L = 0.16$ cm, 0.092 cm, 0.05 cm, 0.03 cm, 0.016 cm, 0.0092 cm, 0.005 cm from left to right. Displayed in color is the concentration field of W.} \label{fig:directional:visual:full}
\end{figure}

\begin{figure}[!h]
\begin{center}
\begin{subfigure}[b]{.99\textwidth}
\centering 
\includegraphics[width=.6\textwidth]{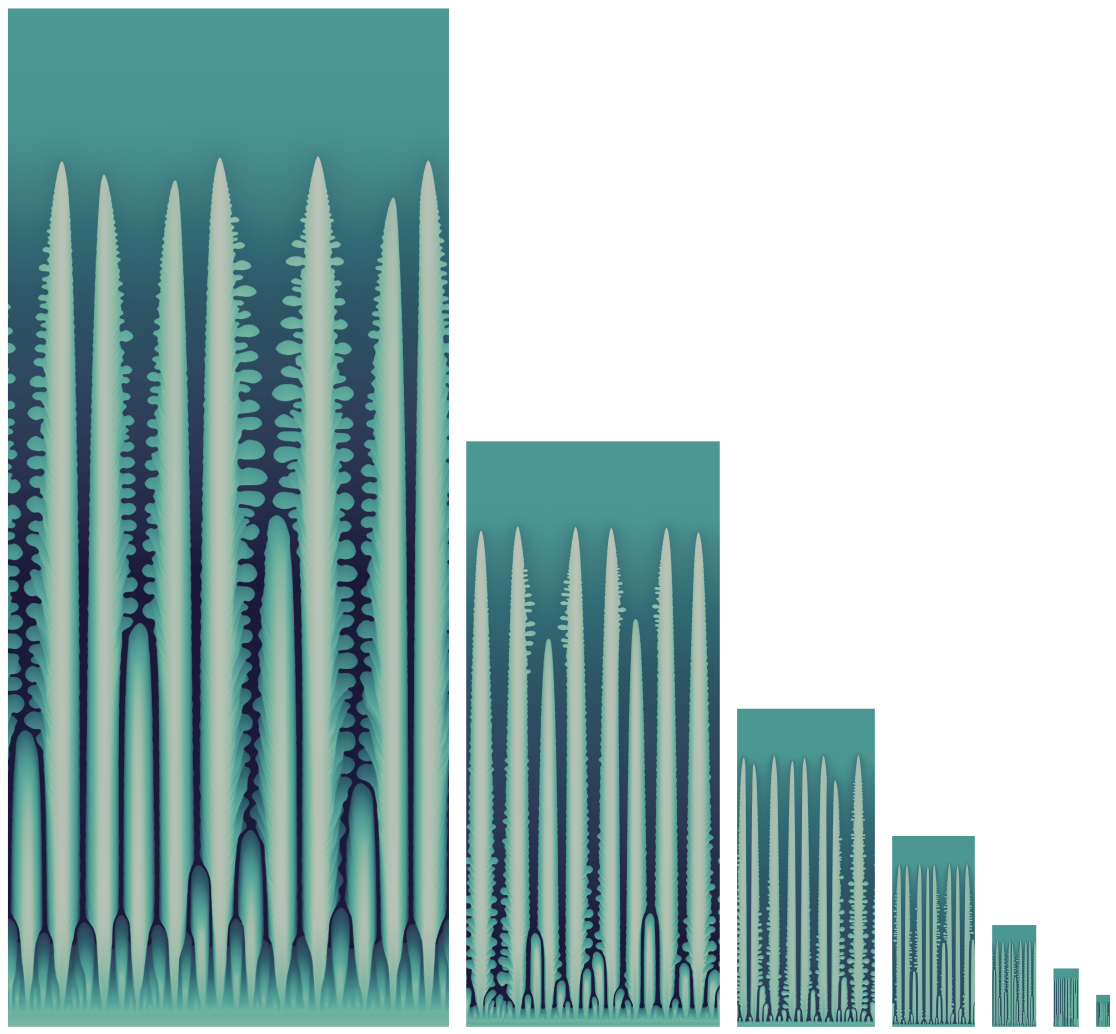}
\end{subfigure}
\end{center}
\vspace{-.3cm}\vspace{-.3cm}\caption{\it Demonstration of simulation domain dimensions used for different values of cooling rate parameter $V$ (left to right: 0.001 cm/s, 0.003 cm/s, 0.01 cm/s, 0.03 cm/s, 0.1 cm/s, 0.3 cm/s, 1 cm/s) relatively to each other. } \label{fig:directional:size}
\end{figure}

We start with performing simulation runs for a range of processing conditions of $G_T$ from 100 K/cm to 5000 K/cm and $V$ from 0.001 cm/s to 1 cm/s. Figure~\ref{fig:directional:representative} demonstrates the progression in time of a representative simulation run while Figure~\ref{fig:directional:visual:full} shows the resulting solidification microstructures for moments of time when the solid phase reaches $y=2L$ for all considered processing conditions. Note that for convenience the simulation domains are pictured to be of the same size, however their actual dimensions vary with the parameter $V$. Specifically, we choose the simulation domain size depending on the cooling rate parameter $V$ such that $L = 0.16$ cm, 0.092 cm, 0.05 cm, 0.03 cm, 0.016 cm, 0.0092 cm, 0.005 cm for $V=0.001$ cm/s, 0.003 cm/s, 0.01 cm/s, 0.03 cm/s, 0.1 cm/s, 0.3 cm/s, 1 cm/s, respectively. This choice is made in order to have the wavelength of the faster growing mode naturally emerging in the beginning of a simulation run to span approximately the same number of grid points in all runs. Figure~\ref{fig:directional:size} demonstrates the domain sizes for different values of $V$ relatively to each other.  Lower thermal gradients and solidification velocities (lower left of Figure~\ref{fig:directional:visual:full}) represent the conditions encountered in Bridgman single crystal growth. \hide{rev2:6}The simulations here predict dendritic growth with average dendrite arm spacings between 153 - 230 microns at the lower two velocities, in the same range as dendrite arm spacings measured experimentally \cite{tsunekane2011single}. 

\hide{rev2:9}\reviewerTwo{The total simulation times in case of $V=0.001$, $0.01$, and $0.3$ cm/s at $G_T = 1000$ K/cm, which are representative planar, cellular, and dendritic cases, on 40 cores of an Intel KNL type CPU (1400MHz) were 247, 790, and 1336 minutes. The overall cost of these simulations is governed by the fact that each time step requires the solution of up to $1 + ( 2 \times (\text{[number of components]} - 1) + 1) \times \text{[number of iterations]}$ ($= 36$ for this specific alloy) linear systems. The difference in simulation times for planar, cellular, and dendritic cases is explained by the difference in the total number of grid points required to sample front geometry: in the case of planar growth the number of grid points remains approximately constant through a simulation, while during unstable growth regimes this number is continuously increasing and more so in the dendritic case.}

\begin{figure}[!h]
\begin{center}
\includegraphics[width=\textwidth]{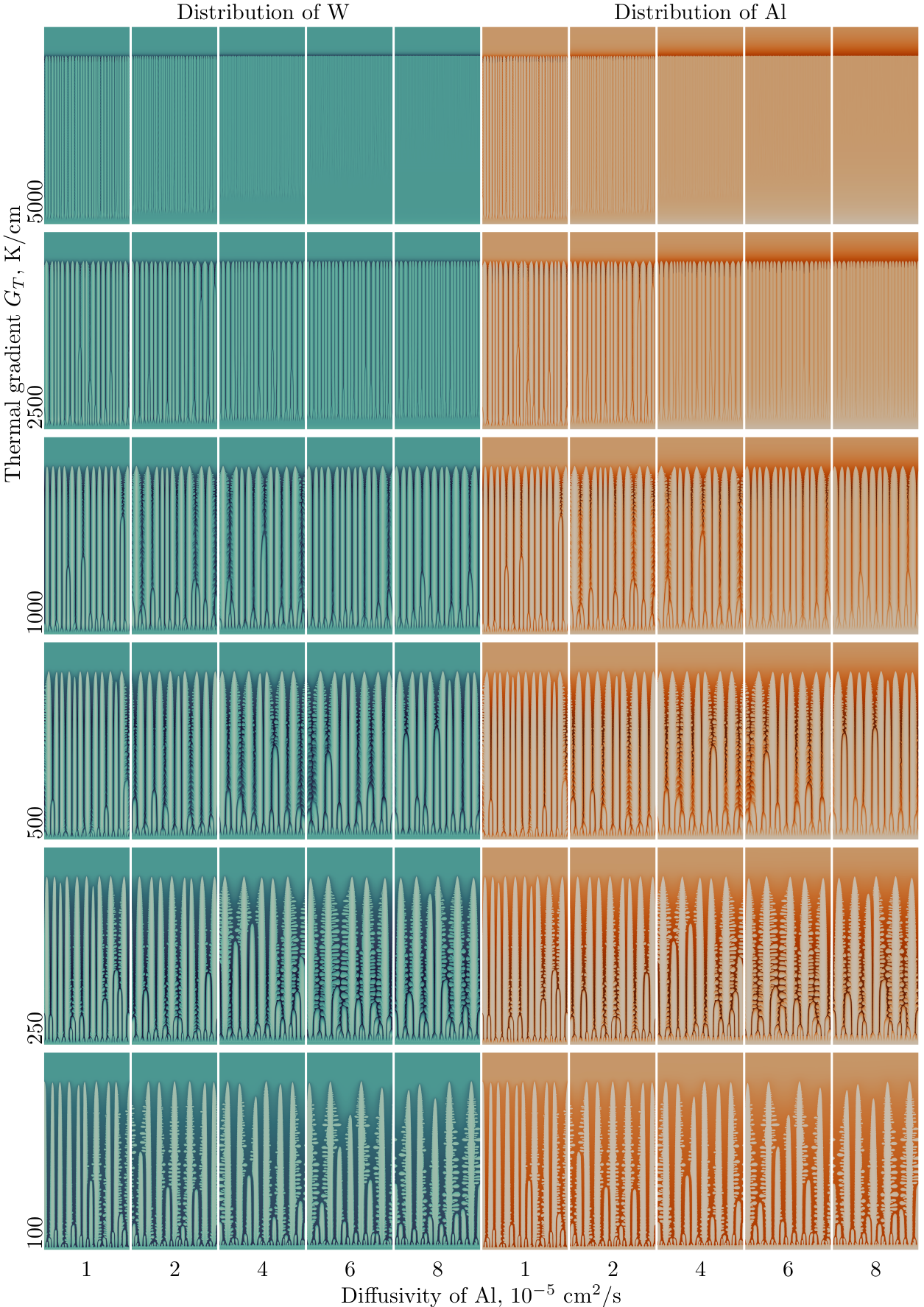}
\end{center}
\vspace{-.3cm}\vspace{-.3cm}\caption{\it Solidification microstructures obtained for values of Al diffusivity from $\D{Al} = 10^{-5}$ cm$^2$/s to $8\cdot10^{-5}$ cm$^2$/s and $G_T$ from 100 K/cm to 5000 K/cm in case of $L = 0.05$ cm and $V = 0.01$ cm/s.}
\label{fig:directional:visual:diff}
\end{figure}

We also perform simulation runs for varying values of the solutes' diffusivity in order to further test the robustness of the computational method as well as to investigate the influence of such alloy parameters on the solidification process. Specifically, we fix the cooling rate parameter to $V = 0.01$ cm/s and the simulation domain scale to $L = 0.05$ cm, and we simulate the solidification process for the values of Al diffusivity $\D{Al}$ in the range from $10^{-5}$ cm$^2$/s to  $8\cdot10^{-5}$ cm$^2$/s (while keeping the W solute diffusivity at $10^{-5}$ cm$^2$/s) and the values of the thermal gradient $G_T$ in the range from $100$ K/cm to $5000$ K/cm. The microstructures so obtained are shown in Figure~\ref{fig:directional:visual:diff}.

\begin{figure}[!h]
\begin{center}
\includegraphics[width=.6\textwidth]{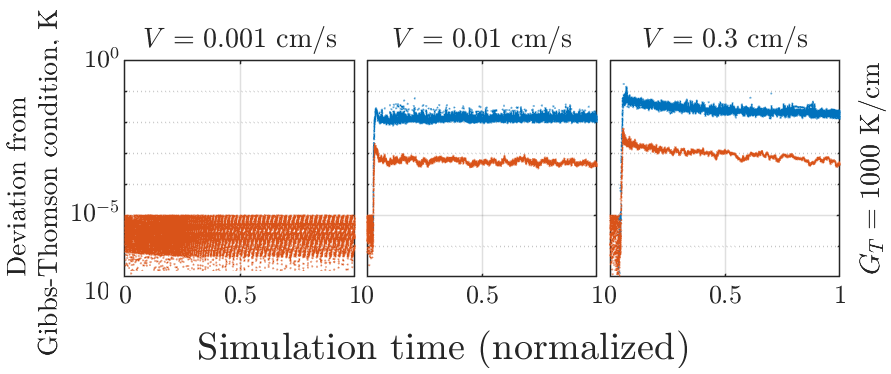}
\end{center}
\vspace{-.3cm}\vspace{-.3cm}\caption{\it Error in satisfying Gibbs-Thomson condition \eqref{eq:model:gibbs} for each time step of three representative simulation runs from Figures \ref{fig:directional:visual:full} (.} \label{fig:directional:bcerror}
\end{figure}

Before discussing the obtained results from the physical perspective, we turn our attention to a few numerical aspects. \hide{rev3:10}\reviewerThree{In order to investigate the effectiveness of the proposed algorithm for solving the non-linear system of PDEs in these challenging conditions, we plot in Figure~\ref{fig:directional:bcerror} the maximum and the average deviations from the Gibbs-Thomson condition along the solidification front for three representative simulation runs: a planar ($V = 0.001$ cm/s), a cellular ($V = 0.01$) cm/s, and a dendritic one ($V=0.3$ cm/s) at $G_T = 1000$ K/cm. Analogous information for all other performed simulation runs are given in Figure~\ref{fig:directional:bcerror:extra}.} As one can see from these results, the more complex is the front geometry, the harder it is for the computational algorithm to reduce the deviation from the Gibbs-Thomson condition to zero in a given number of iteration (specifically, 7 is used in these cases). Still, we see that the maximum deviation is kept at the level around $10^{-2}-10^{-1}$~K and the average deviation at the level of $10^{-4}-10^{-3}$~K across all cases, demonstrating the robustness of the computational method.

\begin{figure}[!h]
\begin{center}
\includegraphics[width=0.3\textwidth]{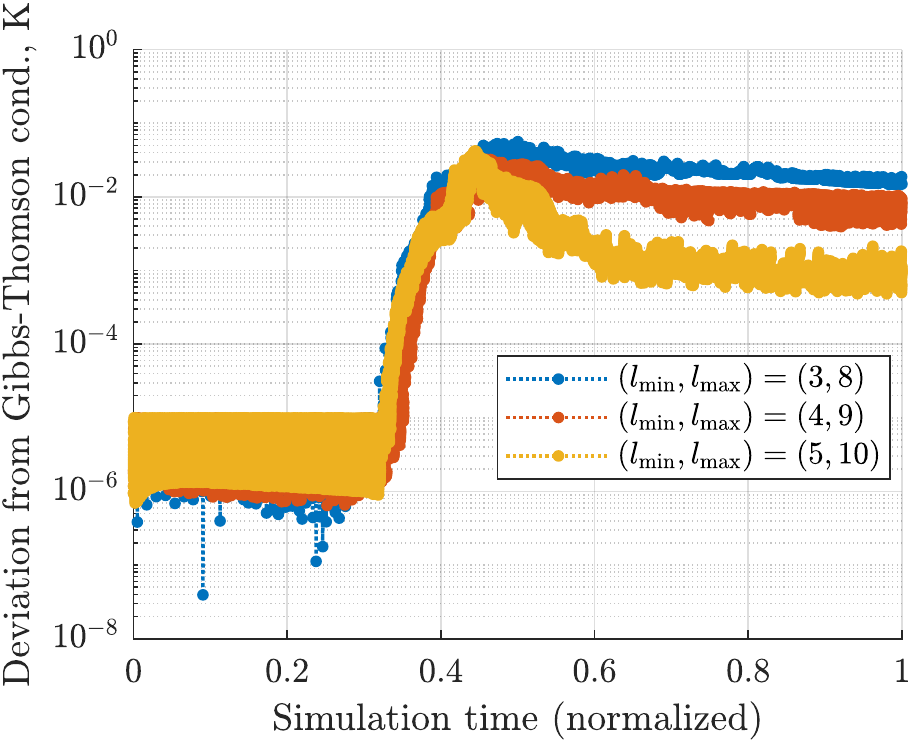}
\includegraphics[width=0.3\textwidth]{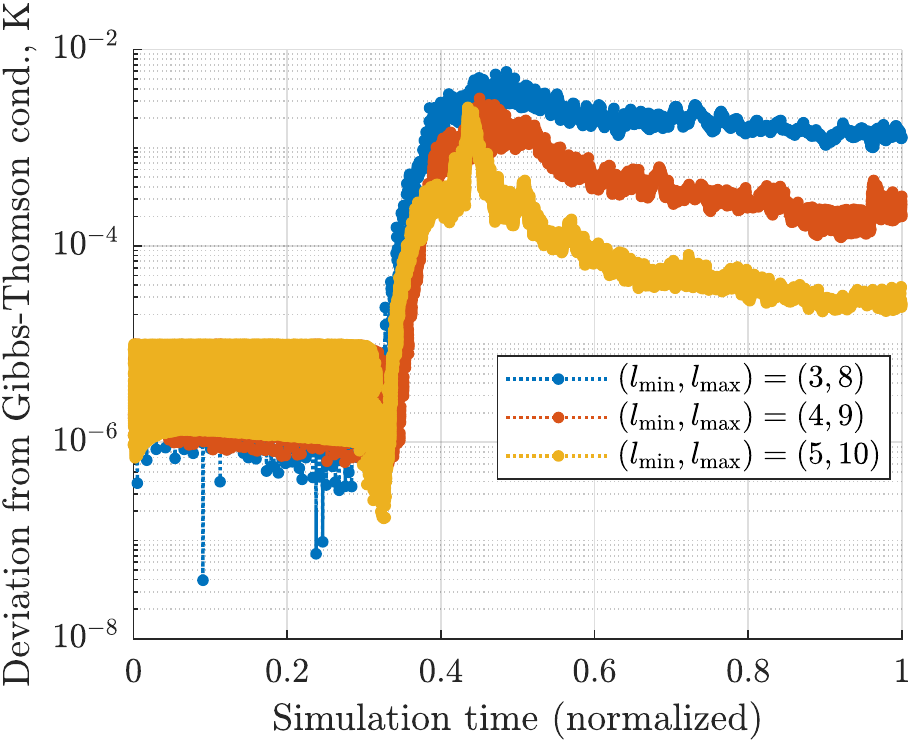}
\end{center}
\vspace{-.3cm}\vspace{-.3cm}\caption{\it Maximum (left) and average (right) error in satisfying Gibbs-Thomson condition \eqref{eq:model:gibbs} for each time step in simulation runs performed on different grid resolutions.}  \label{fig:directional:bcerror:refine}
\end{figure}

In order to investigate whether the observed level of deviations from the Gibbs-Thomson condition is the result of intrinsic limitations of the algorithm or is related to how well the problem geometry is resolved on a computational grid, we perform several simulation runs with different grid resolutions for a specific case of $G_T = 500$ K/cm, $V = 0.03$ cm/s, $L = 0.0075$ cm and a domain aspect ration of $L\times16L$ (i.e, a quarter of the computational domain shown in Figure~\ref{fig:directional:visual:full}). Figure~\ref{fig:directional:bcerror:refine} shows the maximum deviation from the Gibbs-Thomson condition for each time step in these test runs. As one can see, the resulting deviation decreases for increasing grid resolutions suggesting that the observed level of deviations from the Gibbs-Thomson condition is strongly influenced by the level of resolution of the problem geometry on the computational grid. \hide{rev3:11}\reviewerThree{It also follows from these results that the deviation from the Gibbs-Thomson condition improves the slowest in early stages of the growth, that is, during the planar-to-dendritic transition. However, this transition itself typically occupies a negligible amount of experimental samples compared to the majority of solidified material which corresponds to the steady-state like dendritic growth at latter stages, which, in its turn, is a result of the dendrite coarsening/splitting process after the planar-to-dendritic transition.} 

\clearpage
\subsubsection{Analysis of numerical experiments}

Figure~\ref{fig:directional:visual:full} clearly demonstrates the expected dependence of the solidification regime on the thermal gradient and on the rate of cooling, that is, the increase in the ratio of $V/G_T$ leads to transition from planar to cellular to dendritic growth regimes. From the results obtained using different values of the solute diffusivity (Figure~\ref{fig:directional:visual:diff}), one can note that the increase in the diffusivity of the third alloy component (Al) has a similar effect to decreasing the value of $V/G_T$, that is, the solidification front ``flattens''. It is especially clearly evident in the cases near the planar/cellular transition ($G_T = 2500$ K/cm and $G_T = 5000$ K/cm). Qualitatively, this can be explained by the fact that a higher diffusion of the third component from the interdendritic regions leads to a lower local concentration of the third component in those regions and, \reviewerThree{as a result}\hide{rev3:7}, an increased local freezing temperature of the alloy, which, in its turn, corresponds to a ``flatter'' front (given that the isotherms are almost horizontal lines).

In order to obtain more qualitative information about the obtained results, we compute several characteristics.

\begin{figure}[!h]
\begin{center}
\includegraphics[width=\textwidth]{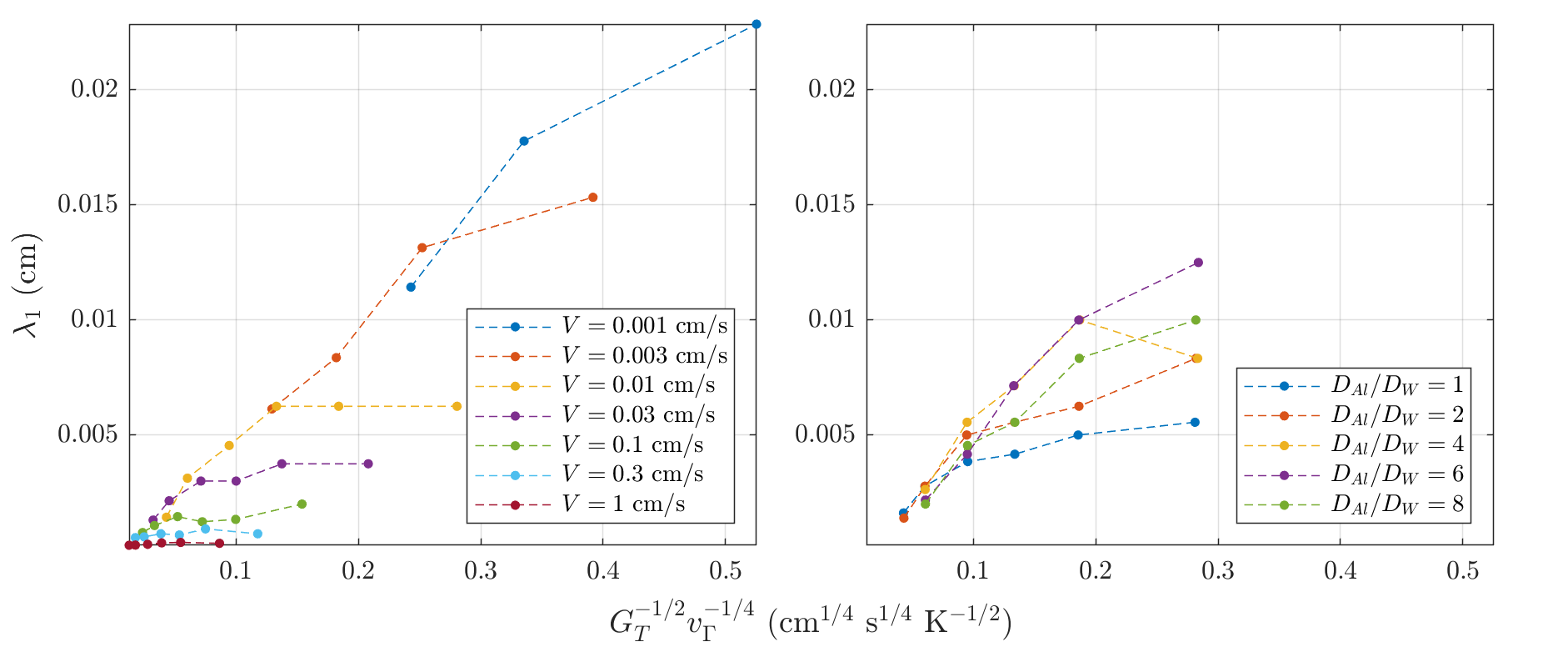}
\end{center}
\vspace{-.3cm}\vspace{-.3cm}\caption{\it Dependence of the primary arm spacing on the tip velocity $v_\Gamma$ and thermal gradient $G_T$ for each of the cases shown in Figure~\ref{fig:directional:visual:full} and Figure~\ref{fig:directional:visual:diff}. } \label{fig:directional:spacing}
\end{figure}

First, we compute the primary arm spacings for each of the cases and analyze its dependence on the quantity $G_T^{-1/2} v_\Gamma^{-1/4}$, where $v_\Gamma$ is the dendrite tip velocity. Simplified models for binary alloys predict a linear dependence between these two quantities (see, e.g., \cite{rappaz2009solidification}). From Figure~\ref{fig:directional:spacing} one can see that not all obtained data follows the single linear dependence. Only cases of low cooling rate parameter $V$ and high imposed thermal gradient $G_T$ fall on the single linear dependence, while in cases of high $V$ and low $G$ the primary arm spacing is almost independent of $G_T^{-1/2} v_\Gamma^{-1/4}$. This, however, should not be interpreted as a breakdown of the scaling relationship on a fundamental level, but as a cumulative effect of geometric dimensions, initial conditions, and processing history. Data presented in Figure~\ref{fig:directional:spacing} (right) also indicates that the diffusivity of the third component has little effect on the resulting tip velocity $v_\Gamma$, however, at the same time, under considered conditions it substantially influences the primary arm spacing, especially, at low values of $G_T$.

\begin{figure}[!h]
\begin{subfigure}{.49\textwidth}
\includegraphics[width=.99\textwidth]{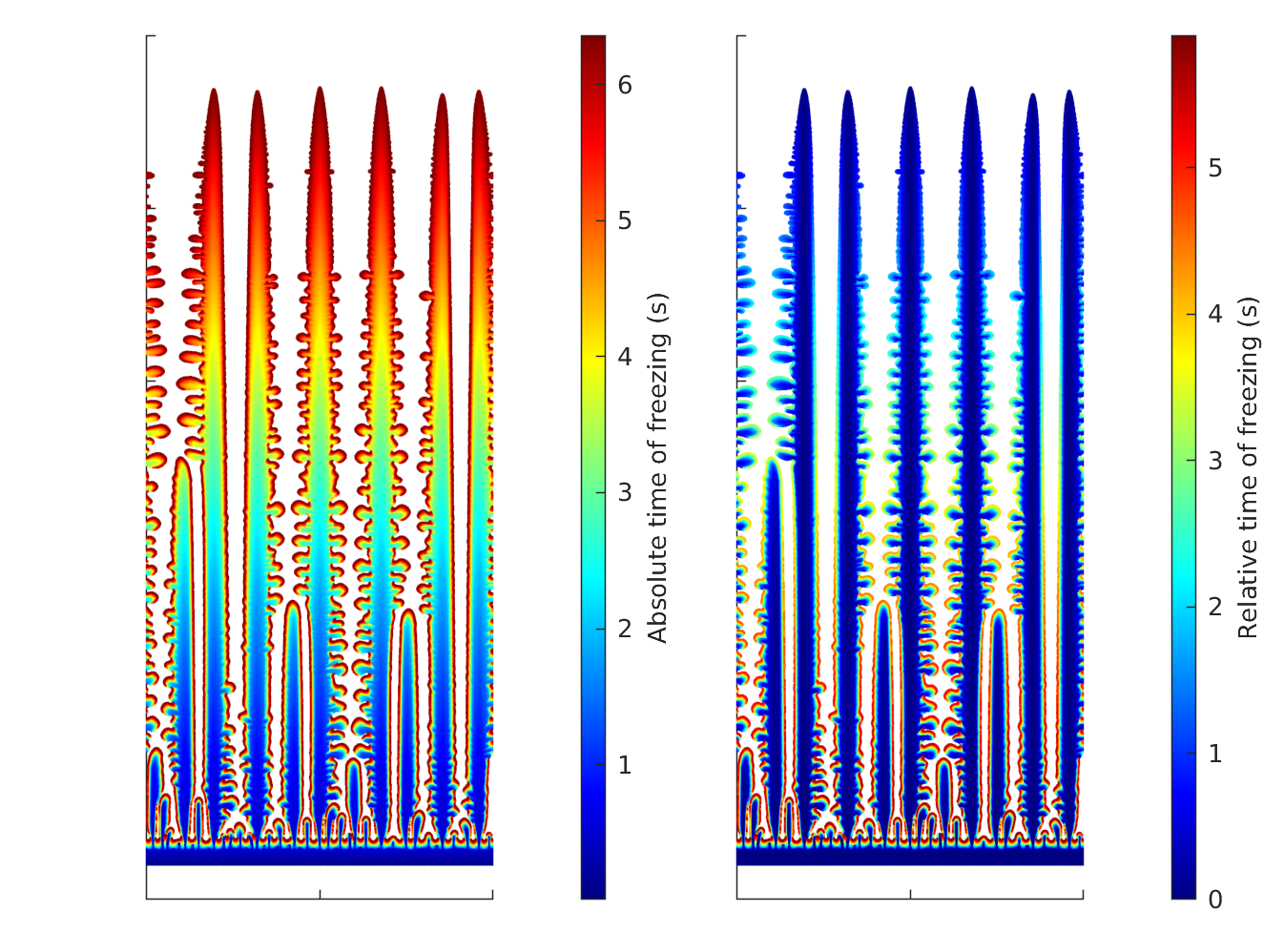}
\end{subfigure}
\begin{subfigure}{.49\textwidth}
\begin{center}
\includegraphics[width=0.5\textwidth]{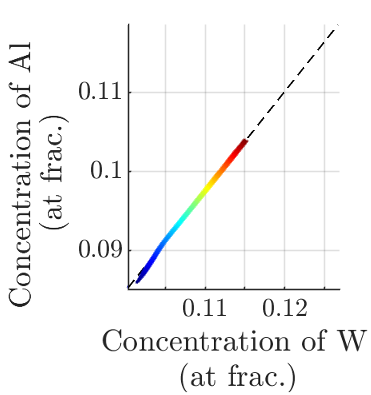}
\end{center}
\end{subfigure}
\vspace{-.3cm}\vspace{-.3cm}\caption{\it \reviewerThree{Left: comparison between the absolute freezing time and the relative freezing time defined as the time it took for a given point in space to turn into the solid phase after the moment of time when the solid phase first reached the $y$-coordinate of the point. Right: concentration path (colored according to the relative freezing time) corresponding to the case $G_T = 1000$ K/cm $V = 0.03$ cm/s from Figure~\ref{fig:directional:visual:full}}.} \label{fig:directional:relativetime}
\end{figure}

Next, we turn our attention to the paths that the concentration of the solid phase follows along the solidus surface as the solidification process proceeds in time. In order to analyze this, we plot the concentration of the solid phase for each point between $y=1.25L$ \reviewerTwo{and $y=1.5L$}\hide{rev2:2} on the $\C{W}$ vs $\C{Al}$ diagram. In order to visualize which area any given point belongs to we color them by the \textit{relative time of freezing}, that is, the time it took for a given point in space to turn into the solid phase after the moment of time when the solid phase first reached the $y$-coordinate of the point (for example, the dendrites' tips have a relative time of freezing of zero). \hide{rev3:9}\reviewerThree{Figure~\ref{fig:directional:relativetime} demonstrates the difference between the absolute time of freezing and the relative time of freezing defined above and an example of concentration path corresponding to the case $G_T = 1000$ K/cm $V = 0.03$ cm/s from Figure~\ref{fig:directional:visual:full}. Concentration path for all other simulation runs from this paper are presented in Figures \ref{fig:directional:solidus:full:extra}-\ref{fig:directional:solidus:diff:extra}}.

%
%

\begin{figure}[!h]
\begin{center}
\includegraphics[width=0.75\textwidth]{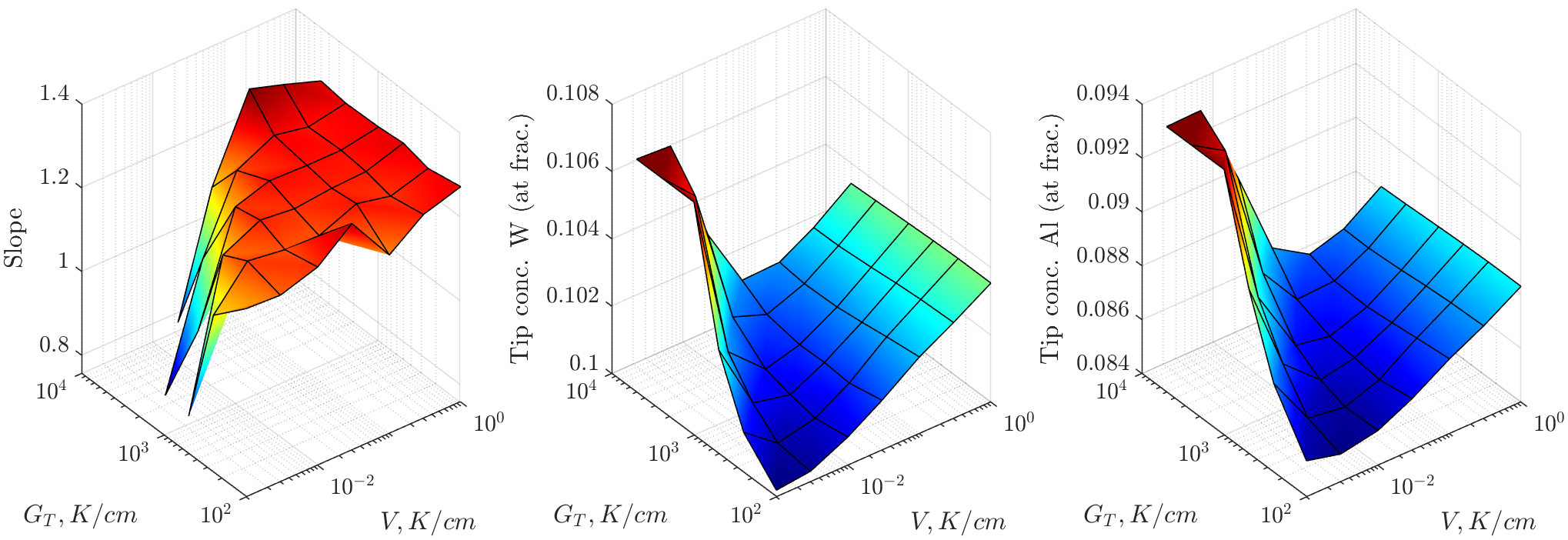}
\end{center}
\vspace{-.3cm}\vspace{-.3cm}\caption{\it \reviewerThree{Extracted characteristics of concentration paths: slope value, concentrations of Al and W at the dendritic tips.}} \label{fig:directional:solidus:full}
\end{figure}

\reviewerThree{To compare solidification paths of different cases we extract certain characteristics of each path.} Specifically, we calculate the slope of each concentration path (using only the second half of available data as some nonlinear behavior is observed for small values of the relative freezing time) and concentrations for the zero relative freezing time (i.e., concentration at the dendritic tips). The slope value characterizes the relative variation of W and Al in the solid phase, that is, the higher its value the faster the Al concentration varies in the solid compared to the W concentration. The tip concentrations characterize the representative absolute concentrations of W and Al. The extracted characteristics are shown in Figure~\ref{fig:directional:solidus:full}. It is seen that the slope of concentration paths varies mildly for all cases up until the processing conditions for solidification approach the planar regime, where the slope value begins to sharply decrease. Correspondingly, the simulations show a non-monotonic behavior of the dendritic tip composition. Going from the planar growth into cellular and dendritic regimes the concentrations of both W and Al initially decrease but start a slight growth as the processing conditions move further into the dendritic growth zone.

\begin{figure}[!h]
\begin{center}
\includegraphics[width=\textwidth]{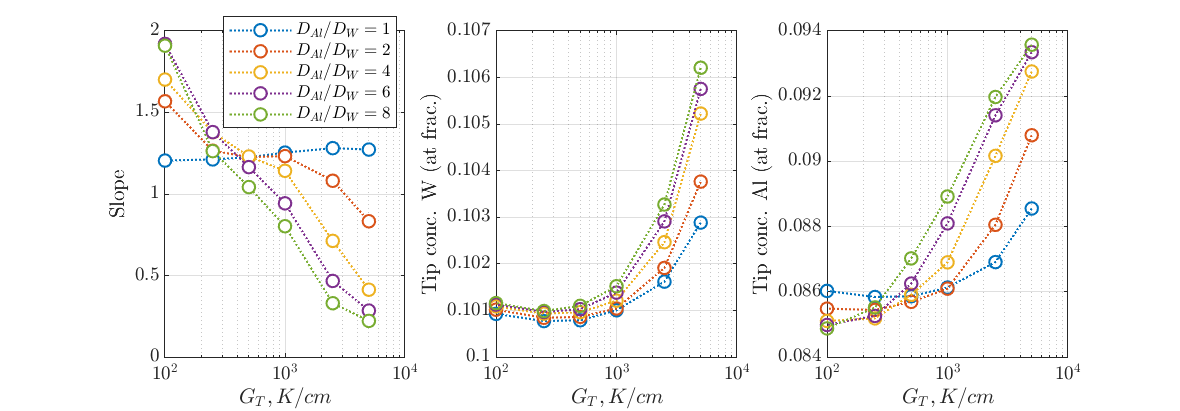}
\end{center}
\vspace{-.3cm}\vspace{-.3cm}\caption{\it \reviewerThree{Extracted characteristics of concentration paths: slope value, concentrations of Al and W at the dendritic tips.}} \label{fig:directional:solidus:diff}
\end{figure}

Figure~\ref{fig:directional:solidus:diff} shows the analogous data obtained for the cases displayed in Figure~\ref{fig:directional:visual:diff}. These results indicate that the sensitivity of the slope value to the imposed thermal gradient $G_T$ varies strongly with the diffusivity of the third component. Specifically, for equal diffusivities of W and Al the slope value is almost insensitive to $G_T$, while for non-equal diffusivity values the higher their ratio the higher sensitivity of the slope value to $G_T$. It is interesting to note that compared to the case of equal diffusivities, the slope value increases (higher variation of Al) for low value of $G_T$ and decreases (higher variation of W) for high value sof $G_T$. As for the dendritic tip compositions, the increase in the ratio of diffusivities $\D{Al}/\D{W}$ leads to a steeper dependence of Al and W concentrations on the thermal gradient $G_T$. Moreover, while the concentration of W increases monotonically with the diffusivity of Al across all values of $G_T$, the concentration of Al increases for low values of $G_T$ and decreases for high value of $G_T$. This steeper dependence indicates an important role for ternary (and higher order) solutes with differential liquid diffusion characteristics in establishing the final structure in high thermal gradient processes, such as those encountered in additive manufacturing. 

\begin{figure}[!h]
\begin{center}
\includegraphics[width=0.6\textwidth]{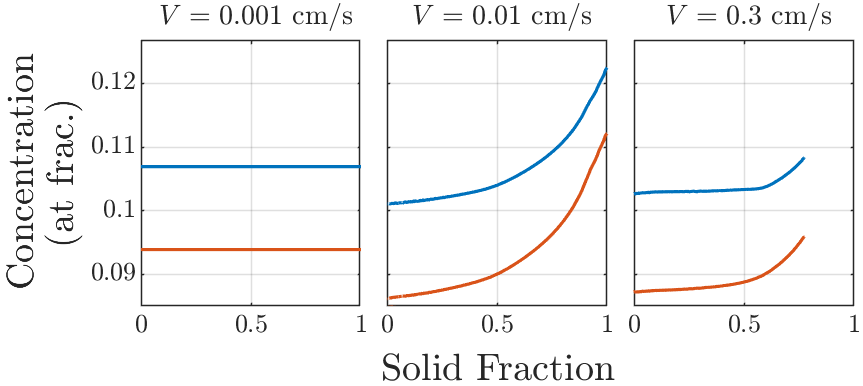}
\end{center}
\vspace{-.3cm}\vspace{-.3cm}\caption{\it Dependence of solutes' concentration on the fraction of material turned into solid phase for cases in Figure~\ref{fig:directional:visual:full}.} \label{fig:directional:segregation:full}
\end{figure}

Finally, in order to analyze the volumetric segregation of alloy components in the solid phase we analyze the dependence of solutes' concentrations on the fraction of the material turned into solid phase. \hide{rev3:13}\reviewerThree{Figure~\ref{fig:directional:segregation:full} shows the dependence of W and Al concentrations on the solid fraction for three representative cases (planar, cellular, and dendritic) shown in Figure~\ref{fig:directional:visual:full}. The analogous figures for all other performed simulation runs are presented in Figures~\ref{fig:directional:segregation:full:extra} and \ref{fig:directional:segregation:diff}}. In cases where solidification occurs in highly dendritic regimes (low $G_T$, high $V$), concentrations of both W and Al stay at approximately the same level up until around 50\% fraction solidified, suggesting of a nearly uniform composition within the dendritic cores. This is followed by a steep rise in concentration. In cases where solidification occurs in the cellular regime, or close to it, the rise in the concentration is more gradual without sharp transition, eventually flattening when processing conditions approach the planar regime. \hide{rev3:14}\reviewerThree{Variation in the diffusivity of the third component does not appear to change this behavior significantly, displaying only two prominent effects that were observed in the above results as well: (1) higher Al diffusion leads to the overall shift in the solidification regime, (2) a change in the ratio of diffusivities shifts the point on the liquidus surface at which solidification occurs. From the results obtained, one can conclude that near the cellular/planar transition (i.e., $G_T = 2500$ K/cm and $5000$ K/cm), the first effect dominates while, on the other end of spectrum, the second effect prevails.} Note that because not all simulation runs are completed until the complete solidification, the obtained result should not be used to compare the maximum reached concentrations. 

\clearpage
\section{Conclusion}

We have presented a sharp-interface computational approach for the simulation of the solidification of multialloy. The main challenge, that is, solving a non-linear system of PDEs, is addressed by a minimization formulation and an approximate \reviewerThree{Newton iteration}\hide{rev3:6} method. Adaptive quad-tree grids are used for domain discretization, the evolution of the solidification front is captured by the level-set method, and a combination of finite-difference and finite-volume methods is used for imposing boundary and interface conditions at the solidification front. The convergence properties of \reviewerThree{Newton method}\hide{rev3:8} is analyzed analytically in one spatial dimension and extensive numerical tests confirm the robustness of the method across all solidification regimes (planar, cellular, dendritic). The accuracy of the overall computational approach was investigated in the case of axially-symmetric solidification and an order of convergence close to second was observed. \daniil{We note that the proposed Newton-type approach for solving the nonlinear system of equations is independent of the particular spatial discretization and can be implemented in other numerical frameworks, for example, interface fitted/unfitted finite element or discontinuous Galerkin discretizations.}

The \reviewerTwo{computational}\hide{rev2:1} method was used to analyze the segregation behavior of an Co-W-Al alloy under a wide range of processing conditions and for different alloy parameters. Differential diffusion rates of solute in the liquid significantly influence the solidification path, influencing the transition from planar to cellular growth at high thermal gradients and the path along the ternary liquidus surface at moderate gradients and interface velocities.

Future work will include improving the Newton's iteration algorithm for faster convergence, developing an implicit time-stepping for the front evolution to remove surface tension induced time-step restriction, and incorporation of additional physical effects such as remelting phenomena, fluid flow, and spontaneous nucleation. 

\section*{Acknowledgement}
The research was partially funded by NSF under the DMREF program DMR-1534264, ONR MURI N00014-11-1-0027, a Department of Defense Vannevar Bush Faculty Fellowship, Grant ONR N00014-18-1-3031. This work used the Extreme Science and Engineering Discovery Environment (XSEDE), which is supported by National Science Foundation grant number ACI-1548562. The authors acknowledge the Texas Advanced Computing Center (TACC) at The University of Texas at Austin for providing HPC resources that have contributed to the research results reported within this paper. Additionally, use was made of computational facilities purchased with funds from the National Science Foundation (CNS-1725797) and administered by the Center for Scientific Computing (CSC). The CSC is supported by the California NanoSystems Institute and the Materials Research Science and Engineering Center (MRSEC; NSF DMR 1720256) at UC Santa Barbara. The authors acknowledge the assistance of S. Murray with thermodynamic simulations. 

\section*{Data availability}
The raw/processed data required to reproduce these findings cannot be shared at this time due to technical or time limitations.

\section*{References}
\bibliographystyle{abbrv}
\addcontentsline{toc}{section}{\refname}
\bibliography{./references_new}

\newpage
\appendix

\section{Functional derivative with respect to $\delta \Cg$}\label{app:derivation}
Consider a generic functional $\mathcal{F}$ defined as an integral of some function $\zeta\of{\vecr}$ over interface $\Gamma$:
\begin{myalign} \label{eq:genericfunctional}
  \mathcal{F} =
  \int_\Gamma
    \zeta\of{\vecr}
    \diff{\Gamma},
\end{myalign}
where function $\zeta\of{\vecr}$ is a combination of solutions $\set{\C{\J}}{\J=1}{N}$, $\set{\T{\phase}}{\phase = \sph,\lph}{}$, and $\vn$ to BVPs \eqref{eq:spliting:c1}-\eqref{eq:spliting:t}. In the case considered in this work:
\begin{myalign*}
  \zeta\of{\vecr} &= \delta\of{\vecr-\vecr_0} \mathcal{E}\of{\vecr}
  \\
  &= \delta\of{\vecr-\vecr_0} (\Tl\of{\vecr} - h_G\of{\vecr} - \Tliq \of{\C{1}\of{\vecr}, \ldots, C_N\of{\vecr}} - \epsilon_v \vn\of{\vecr}).
\end{myalign*}
The deferential of $\mathcal{F}$ with respect to $\Cg$ can be obtained using a Lagrangian multiplier approach. ``Constraints'' \eqref{eq:spliting:c1}-\eqref{eq:spliting:t} can be incorporated into a Lagrangian in different ways. One approach is to treat every equations from \eqref{eq:spliting:c1}-\eqref{eq:spliting:t} as a separate constraint. Another approach is to first obtain variational formulations of BVPs \eqref{eq:spliting:c1}-\eqref{eq:spliting:t} and then use them in the Lagrangian definition. For sufficiently regular functions, the two approaches are equivalent and, roughly speaking, merely define the order of mathematical operations and number of Lagrangian multipliers (the latter approach introduces less multipliers since for each BVP several equations, namely, a PDE and BCs, are combined into a single constraint, a variational form of BVP). For its conceptual simplicity, we employ the former approach. That is, we formally define a Lagrangian as:
\begin{mymultline}
  \mathcal{\lph} =
  \mathcal{F}
  +
  \sum_{\phase=\sph,\lph}
  \big\{
  \omega_{\T{\phase}}
  \cdot
  \left(  
    \text{PDE for $\T{\phase}$}
  \right)
  +
  \beta_{\T{\phase}}
  \cdot
  \left(  
    \text{BC for $\T{\phase}$ on $\partial \Omega$}
  \right)
  \big\}
  \\
  +
  \gamma_{T}
  \cdot
  \left(
    \text{Cond for $\jump{T}$ on $\Gamma$}
  \right)
  +
  \gamma_{\sph}
  \cdot
  \left(
    \text{Cond for $\jump{\heatcond{} \ddn{T}}$ on $\Gamma$}
  \right)
  \\
  +
  \sum_{\J=1}^{N}
  \big\{
  \omega_{\C{\J}}
  \cdot
  \left(  
    \text{PDE for $\C{\J}$}
  \right)
  +
  \gamma_{\C{\J}}
  \cdot
  \left(
    \text{BC for $\C{\J}$ on $\Gamma$}
  \right)
  +
  \beta_{\C{\J}}
  \cdot
  \left(  
    \text{BC for $\C{\J}$ on $\partial \Omega$}
  \right)
  \big\}
  \\
  +
  \gamma_{v}
  \cdot
  \left(
    \text{Equation for $\vn$}
  \right)
\end{mymultline}
where $\set{\omega_{\T{\phase}}, \beta_{\T{\phase}}}{\phase=\sph,\lph}{}$, $\gamma_T$, $\gamma_S$, $\set{\omega_{\C{\J}}, \gamma_{\C{\J}}, \beta_{\C{\J}}}{\J=1}{N}$, and $\gamma_v$ are Lagrangian multipliers. Note that these multipliers are functions of spatial variables with dimensionalities corresponding to constraints they enforce, that is, $\omega_{\Tl}$ and $\set{\omega_{\C{\J}}}{\J=1}{N}$ are defined in $\dom{\lph}$; $\omega_{\Ts}$ is defined in $\dom{\sph}$; $\gamma_T$, $\gamma_S$, $\set{\gamma_{\C{\J}}}{\J=1}{N}$, and $\gamma_v$ are defined on $\Gamma$; $\beta_{\Tl}$ and $\set{\beta_{\C{\J}}}{\J=1}{N}$ are defined on $\partial\Omega \cap \overline{\dom{\lph}}$; finally, $\beta_{\Ts}$ is defined on $\partial\Omega \cap \overline{\dom{\sph}}$. 
Then the differential of $\mathcal{F}$ is equal to a variation of $\mathcal{\lph}$ with respect to $\Cg$:
\begin{myalign*}
	d \mathcal{F} &= \delta_{\Cg} \mathcal{\lph} = \int_\Gamma \fd {\mathcal{\lph}}{\Cg}\of{\vecr} \delta\Cg\of{\vecr} \diff{\Gamma},
\end{myalign*}
provided the following conditions are satisfied:
\begin{myalign}
  \label{eq:lagrange:cond}
  \begin{aligned}
  	\delta_{\omega_{\T{\phase}}} \mathcal{\lph} &= 0 \quad \forall\ \delta \omega_{\T{\phase}}, \quad \phase = \sph,\lph, &\quad
  	\delta_{\T{\phase}} \mathcal{\lph} &= 0 \quad \forall\ \delta \T{\phase}, \quad \phase = \sph,\lph, 
  	\\
  	\delta_{\omega_{\C{\J}}} \mathcal{\lph} &= 0 \quad \forall\ \delta \omega_{\C{\J}}, \quad \J \in \range{1}{N}, &\quad
  	\delta_{\C{\J}} \mathcal{\lph} &= 0 \quad \forall\ \delta \C{\J}, \quad \J \in \range{1}{N}, 
  	\\
  	\delta_{\gamma_{v}} \mathcal{\lph} &= 0 \quad \forall\ \delta \gamma_{v}. &\quad
  	\delta_{\vn} \mathcal{\lph} &= 0 \quad \forall\ \delta \vn, 
    \\
  	\delta_{\gamma_{\C{\J}}} \mathcal{\lph} &= 0 \quad \forall\ \delta \gamma_{\C{\J}}, \quad \J \in \range{1}{N},
    \\
  	\delta_{\gamma_{T}} \mathcal{\lph} &= 0 \quad \forall\ \delta \gamma_{T},
    \\
  	\delta_{\gamma_{\sph}} \mathcal{\lph} &= 0 \quad \forall\ \delta \gamma_{\sph},
    \\
  	\delta_{\beta_{\T{\phase}}} \mathcal{\lph} &= 0 \quad \forall\ \delta \beta_{\T{\phase}}, \quad \phase = \sph,\lph,
    \\
  	\delta_{\beta_{\C{\J}}} \mathcal{\lph} &= 0 \quad \forall\ \delta \beta_{\C{\J}}, \quad \J \in \range{1}{N},
  \end{aligned}
\end{myalign}
Conditions from the left column above ensure that ``constraints'' for $\{\C{\J}\}_{\J=1}^N$, $\{\T{\phase} \}_{\phase = \sph,\lph}$ and $\vn$ are satisfied, while conditions from the right column give equations for Lagrangian multipliers.

The explicit expression for the Lagrangian can be written as:
\begin{mymultline*}
  \mathcal{\lph} =
  \int_\Gamma
    \zeta\of{\vecr}
    \diff{\Gamma}
  +
  \sum_{\phase=\sph,\lph}
  \bigg\{
  \int_{\dom{\phase}}
    \obcolor{\auxColor}{
    \left(  
      \heatcapv{\phase} \T{\phase} - \heatcond{\phase} \lap \T{\phase} - f_{\T{\phase}}
    \right)
    }{\text{PDE for } \T{\phase}}
    \omega_{\T{\phase}}
    \diff{\Omega}
  +
  \int_{\partial \Omega \cap \overline{\dom{\lph}}}
    \obcolor{\auxColor}{
    \left(  
      \heatcond{\phase} \ddn[\phase]{\T{\phase}} - g_{\T{\phase}}
    \right)
    }{\text{BC for } \T{\phase}}
    \beta_{\T{\phase}}
    \diff{\Gamma}
  \bigg\}
  \\
  +
  \int_{\Gamma}
    \ubcolor{\auxColor}{
    \left(
      \Ts-\Tl - h_T
    \right)
    }{\text{Cond for } \jump{T}}
    \gamma_{T}
    \diff{\Gamma}
  +
  \int_{\Gamma}
    \ubcolor{\auxColor}{
    \left(
      \heatcond{\sph} \ddn[s]{\Ts} + \heatcond{\lph} \ddn[l]{\Tl} - L\vn - h_S
    \right)
    }{\text{Cond for } \jump{\heatcond{} \ddn{T}}}
    \gamma_{\sph}
    \diff{\Gamma}
  \\
  +
  \sum_{\J=1}^{N}
  \bigg\{
  \int_{\dom{\lph}}
    \ubcolor{\auxColor}{
    \left(  
      a \C{\J} - \D{\J} \lap \C{\J} - f_{\C{\J}}
    \right)
    }{\text{PDE for } \C{\J}}
    \omega_{\C{\J}}
    \diff{\Omega}
  +
  \int_{\partial \Omega \cap \overline{\dom{\lph}}}
    \ubcolor{\auxColor}{
    \left(  
      \D{\J} \ddn[l]{\C{\J}} - g_{\C{\J}}
    \right)
    }{\text{BC for $\C{\J}$ on $\partial\Omega$}}
    \beta_{\C{\J}}
    \diff{\Gamma}
  \bigg\}
  \\
  +
  \int_{\Gamma}
    \ubcolor{\auxColor}{
    \left(
      \C{1} - \Cg
    \right)
    }{\text{BC for $\C{1}$ on $\Gamma$}}
    \gamma_{\C{1}}
    \diff{\Gamma}
  +
  \sum_{\J=2}^{N}
  \int_{\Gamma}
    \ubcolor{\auxColor}{
    \left(
      \D{\J} \ddn[l]{\C{\J}} - (1-\partcoeff_\J) \vn \C{\J} - h_{\C{\J}}
    \right)
    }{\text{BC for $\C{\J}$ on $\Gamma$}}
    \gamma_{\C{\J}}
    \diff{\Gamma}
  \\
  +
  \int_{\Gamma}
    \ubcolor{\auxColor}{
    \left(
      \D{1} \ddn[l]{\C{1}} - (1-\partcoeff_1) \vn \C{1} - h_{\C{1}}
    \right)
    }{\text{Equation for } \vn}
    \gamma_{v}
    \diff{\Gamma}
\end{mymultline*}

Taking variations of the Lagrangian with respect to multipliers $\set{\omega_{\T{\phase}}, \beta_{\T{\phase}}}{\phase=\sph,\lph}{}$, $\gamma_T$, $\gamma_S$, $\set{\omega_{\C{\J}}, \gamma_{\C{\J}}, \beta_{\C{\J}}}{\J=1}{N}$, and $\gamma_v$, it is trivial to confirm that conditions from the left column of \eqref{eq:lagrange:cond} lead to equations \eqref{eq:spliting:c1}-\eqref{eq:spliting:t}.

Variations with respect to $\{\C{\J}\}_{\J=1}^N$, $\{\T{\phase} \}_{\phase = \sph,\lph}$ and $\vn$ are equal to:
\begin{myalign*}
  \delta_{\C{1}} \mathcal{\lph} =&
  \int_{\Gamma}
    \zeta^\prime_{\C{1}}
    \delta \C{1}
    \diff{\Gamma}
  +
  \int_{\dom{\lph}}
    \left(  
      a \delta \C{1} - \D{1} \lap \delta \C{1}
    \right)
    \omega_{\C{1}}
    \diff{\Omega}
  +
  \int_{\partial \Omega \cap \overline{\dom{\lph}}}
    \beta_{\C{1}}
    \D{1} \ddn[l]{\delta \C{1}}
    \diff{\Gamma} 
  \\&
  +
  \int_{\Gamma}
    \left( \gamma_{\C{1}} - (1-\partcoeff_1) \vn \gamma_v \right)
    \delta \C{1}
    \diff{\Gamma}
  +
  \int_{\Gamma}
    \gamma_{v}
    \D{1} \ddn[l]{\delta \C{1}}
    \diff{\Gamma},
  \\
  \delta_{\C{\J}} \mathcal{\lph} =&
  \int_{\Gamma}
    \zeta^\prime_{\C{\J}}
    \delta \C{\J}
    \diff{\Gamma}
  +
  \int_{\dom{\lph}}
    \left(  
      a \delta \C{\J} - \D{\J} \lap \delta \C{\J}
    \right)
    \omega_{\C{\J}}
    \diff{\Omega}
  +
  \int_{\partial \Omega \cap \overline{\dom{\lph}}}
    \beta_{\C{\J}}
    \D{\J} \ddn[l]{\delta \C{\J}}
    \diff{\Gamma} 
  \\&
  -
  \int_{\Gamma}
    \gamma_{\C{\J}}
    (1-\partcoeff_\J) \vn \delta \C{\J}
    \diff{\Gamma}
  +
  \int_{\Gamma}
    \gamma_{\C{\J}}
    \D{\J} \ddn[l]{\delta \C{\J}}
    \diff{\Gamma}
  ,\quad i = 2, \ldots, N,
  \\
  \delta_{\Ts} \mathcal{\lph} =&
  \int_{\Gamma}
    \zeta^\prime_{\Ts}
    \delta \Ts
    \diff{\Gamma}
  +
  \int_{\dom{\sph}}
    \left(  
      \heatcapv{\sph} \delta \Ts - \heatcond{\sph} \lap \delta \Ts
    \right)
    \omega_{\Ts}
    \diff{\Omega}
  +
  \int_{\partial \Omega \cap \overline{\dom{\sph}}}
    \beta_{\Ts}
    \heatcond{\sph} \ddn[s]{\delta \Ts}
    \diff{\Gamma} 
  \\&
  +
  \int_{\Gamma}
    \gamma_T
    \delta \Ts
    \diff{\Gamma}
  +
  \int_{\Gamma}
    \gamma_S
    \heatcond{\sph} \ddn[s]{\Ts}
    \diff{\Gamma},
  \\
  \delta_{\Tl} \mathcal{\lph} =&
  \int_{\Gamma}
    \zeta^\prime_{\Ts}
    \delta \Ts
    \diff{\Gamma}
  +
  \int_{\dom{\lph}}
    \left(  
      \heatcapv{\lph} \delta \Tl - \heatcond{\lph} \lap \delta \Tl
    \right)
    \omega_{\Tl}
    \diff{\Omega}
  +
  \int_{\partial \Omega \cap \overline{\dom{\lph}}}
    \beta_{\Tl}
    \heatcond{\lph} \ddn[l]{\delta \Tl}
    \diff{\Gamma} 
  \\&
  -
  \int_{\Gamma}
    \gamma_T
    \delta \Tl
    \diff{\Gamma}
  +
  \int_{\Gamma}
    \gamma_S
    \heatcond{\lph} \ddn[l]{\Tl}
    \diff{\Gamma},
  \\
  \delta_{\vn} \mathcal{\lph} =&
  \int_{\Gamma}
    \left(
      \zeta^\prime_{\vn}
      -
      (1-\partcoeff_1) \C{1} \gamma_v
      -
      \sum_{\J=2}^{N} (1-\partcoeff_\J) \C{\J} \gamma_{\C{\J}}
      -
      L \gamma_S
    \right)
    \delta \vn
    \diff{\Gamma},
\end{myalign*}
where $\zeta^\prime_{\T{\phase}}$, $\zeta^\prime_{\C{\J}}$ and $\zeta^\prime_{\vn}$ denote classical partial derivatives of function $\zeta$ with respect to ${\T{\phase}}$, ${\C{\J}}$ and ${\vn}$, correspondingly.
Using the Green's second identity the first four expressions can be transformed into:
\begin{myalign*}
  \delta_{\C{1}} \mathcal{\lph} =&
  \int_{\dom{\lph}}
    \left(  
      a \omega_{\C{1}}  - \D{1} \lap \omega_{\C{1}}
    \right)
    \delta \C{1}
    \diff{\Omega}
  \\&
  +
  \int_{\partial \Omega \cap \overline{\dom{\lph}}}
    \left( \beta_{\C{1}} - \omega_{\C{1}} \right)
    \D{1} \ddn[l]{\delta \C{1}}
    \diff{\Gamma} 
  +
  \int_{\partial \Omega \cap \overline{\dom{\lph}}}
    \D{1} \ddn[l]{\omega_{\C{1}}}
    \delta \C{1}
    \diff{\Gamma}
  \\&
  +
  \int_{\Gamma}
    \left( \gamma_{\C{1}} - (1-\partcoeff_1) \vn \gamma_v + \D{1} \ddn[l]{\omega_{\C{1}}} + \zeta^\prime_{\C{1}} \right)
    \delta \C{1}
    \diff{\Gamma}
  +
  \int_{\Gamma}
    \left( \gamma_{v} - \omega_{\C{1}} \right)
    \D{1} \ddn[l]{\delta \C{1}}
    \diff{\Gamma},
  \\
  \delta_{\C{\J}} \mathcal{\lph} =&
  \int_{\dom{\lph}}
    \left(  
      a \omega_{\C{\J}} - \D{\J} \lap \omega_{\C{\J}}
    \right)
    \delta \C{\J}
    \diff{\Omega}
  \\&
  +
  \int_{\partial \Omega \cap \overline{\dom{\lph}}}
    \left( \beta_{\C{\J}} - \omega_{\C{\J}} \right)
    \D{\J} \ddn[l]{\delta \C{\J}}
    \diff{\Gamma} 
  +
  \int_{\partial \Omega \cap \overline{\dom{\lph}}}
    \D{\J} \ddn[l]{\omega_{\C{\J}}}
    \delta \C{\J}
    \diff{\Gamma} 
  \\&
  +
  \int_{\Gamma}
    \left( \D{\J} \ddn[l]{\omega_{\C{\J}}} - (1-\partcoeff_\J) \vn \gamma_{\C{\J}} + \zeta^\prime_{\C{\J}} \right)
    \delta \C{\J}
    \diff{\Gamma}
  +
  \int_{\Gamma}
    \left( \gamma_{\C{\J}} - \omega_{\C{\J}} \right)
    \D{\J} \ddn[l]{\delta \C{\J}}
    \diff{\Gamma}
  ,\quad i = 2, \ldots, N,
  \\
  \delta_{\Ts} \mathcal{\lph} =&
  \int_{\dom{\sph}}
    \left(  
      \heatcapv{\sph} \omega_{\Ts} - \heatcond{\sph} \lap \omega_{\Ts}
    \right)
    \delta \Ts
    \diff{\Omega}
  \\&
  +
  \int_{\partial \Omega \cap \overline{\dom{\sph}}}
    \left( \beta_{\Ts} - \omega_{\Ts} \right)
    \heatcond{\sph} \ddn[s]{\delta \Ts}
    \diff{\Gamma} 
  +
  \int_{\partial \Omega \cap \overline{\dom{\sph}}}
    \heatcond{\sph} \ddn[s]{\omega_{\Ts}}
    \delta \Ts
    \diff{\Gamma} 
  \\&
  +
  \int_{\Gamma}
    \left( \gamma_T + \heatcond{\sph} \ddn[s]{\omega_{\Ts}} + \zeta^\prime_{\Ts} \right)
    \delta \Ts
    \diff{\Gamma}
  +
  \int_{\Gamma}
    \left( \gamma_S - \omega_{\Ts} \right)
    \heatcond{\sph} \ddn[s]{\Ts}
    \diff{\Gamma},
  \\
  \delta_{\Tl} \mathcal{\lph} =&
  \int_{\dom{\lph}}
    \left(  
      \heatcapv{\lph} \omega_{\Tl} - \heatcond{\lph} \lap \omega_{\Tl}
    \right)
    \delta \Tl
    \diff{\Omega}
  \\&
  +
  \int_{\partial \Omega \cap \overline{\dom{\lph}}}
    \left( \beta_{\Tl} - \omega_{\Tl} \right)
    \heatcond{\lph} \ddn[l]{\delta \Tl}
    \diff{\Gamma} 
  +
  \int_{\partial \Omega \cap \overline{\dom{\lph}}}
    \heatcond{\lph} \ddn[l]{\omega_{\Tl}}
    \delta \Tl
    \diff{\Gamma} 
  \\&
  +
  \int_{\Gamma}
    \left( - \gamma_T + \heatcond{\lph} \ddn[l]{\omega_{\Tl}} + \zeta^\prime_{\Tl} \right)
    \delta \Tl
    \diff{\Gamma}
  +
  \int_{\Gamma}
    \left( \gamma_S - \omega_{\Tl} \right)
    \heatcond{\lph} \ddn[l]{\Tl}
    \diff{\Gamma},
\end{myalign*}
It is easy to see now that conditions from the right column of \eqref{eq:lagrange:cond} lead to the following equations for Lagrangian multipliers:
\begin{myalign}
  \label{eq:lagrange:raw_multipliers}
  \begin{aligned}
  \delta_{\C{1}} \mathcal{\lph} = 0\ \forall\ \delta \C{1}
  \quad
  &\Rightarrow
  \quad
  \left\{
  \begin{aligned}
    \left( a - \D{1} \lap \right) \omega_{\C{1}} &= 0 &&\text{in } \dom{\lph}
    \\
    \omega_{\C{1}} &= \gamma_{v} &&\text{on } \Gamma
    \\
    \D{1} \ddn[l]{\omega_{\C{1}}} &= 0 &&\text{on } \partial\Omega \cap \overline{\dom{\lph}}
    \\
    \gamma_{\C{1}} &= (1-\partcoeff_1) \vn \gamma_v - \zeta^\prime_{\C{1}} - \D{1} \ddn[l]{\omega_{\C{1}}} &&\text{on } \Gamma
    \\
    \beta_{\C{1}} &= \omega_{\C{1}} &&\text{on } \partial\Omega \cap \overline{\dom{\lph}}
  \end{aligned}
  \right.
  \\
  \delta_{\C{\J}} \mathcal{\lph} = 0\ \forall\ \delta \C{\J}
  \quad
  &\Rightarrow
  \quad
  \left\{
  \begin{aligned}
    \left( a  - \D{\J} \lap \right) \omega_{\C{\J}} &= 0 &&\text{in } \dom{\lph}
    \\
    \D{\J} \ddn[l]{\omega_{\C{\J}}} - (1-\partcoeff_\J) \vn \gamma_{\C{\J}} &= -\zeta^\prime_{\C{\J}} &&\text{on } \Gamma
    \\
    \D{\J} \ddn[l]{\omega_{\C{\J}}} &= 0 &&\text{on } \partial\Omega \cap \overline{\dom{\lph}}
    \\
    \gamma_{\C{\J}} &= \omega_{\C{\J}} &&\text{on } \Gamma
    \\
    \beta_{\C{\J}} &= \omega_{\C{\J}} &&\text{on } \partial\Omega \cap \overline{\dom{\lph}}
  \end{aligned}
  \right.
  \\
  \delta_{\Ts} \mathcal{\lph} = 0\ \forall\ \delta \Ts
  \quad
  &\Rightarrow
  \quad
  \left\{
  \begin{aligned}
    \left( \heatcapv{\sph} - \heatcond{\sph} \lap \right) \omega_{\Ts} &= 0 &&\text{in } \dom{\sph}
    \\
    \omega_{\Ts} &= \gamma_S &&\text{on } \Gamma
    \\
    \heatcond{\sph} \ddn[s]{\omega_{\Ts}} &= - \gamma_T - \zeta^\prime_{\Ts} &&\text{on } \Gamma
    \\
    \heatcond{\sph} \ddn[s]{\omega_{\Ts}} &= 0 &&\text{on } \partial\Omega \cap \overline{\dom{\lph}}
    \\
    \beta_{\Ts} &= \omega_{\Ts} &&\text{on } \partial\Omega \cap \overline{\dom{\sph}}
  \end{aligned}
  \right.
  \\
  \delta_{\Tl} \mathcal{\lph} = 0\ \forall\ \delta \Tl
  \quad
  &\Rightarrow
  \quad
  \left\{
  \begin{aligned}
    \left( \heatcapv{\lph} - \heatcond{\lph} \lap \right) \omega_{\Tl} &= 0 &&\text{in } \dom{\lph}
    \\
    \omega_{\Tl} &= \gamma_S &&\text{on } \Gamma
    \\
    \heatcond{\lph} \ddn[l]{\omega_{\Tl}} &= \gamma_T - \zeta^\prime_{\Tl} &&\text{on } \Gamma
    \\
    \heatcond{\lph} \ddn[l]{\omega_{\Tl}} &= 0 &&\text{on } \partial\Omega \cap \overline{\dom{\lph}}
    \\
    \beta_{\Tl} &= \omega_{\Tl} &&\text{on } \partial\Omega \cap \overline{\dom{\lph}}
  \end{aligned}
  \right.
  \\
  \delta_{\vn} \mathcal{\lph} = 0\ \forall\ \delta \vn
  \quad
  &\Rightarrow
  \quad
  \gamma_v = \frac{1}{(1-\partcoeff_1) \C{1}}
    \left(
      \zeta^\prime_{\vn}
      -
      \sum_{\J=2}^{N} (1-\partcoeff_\J) \C{\J} \gamma_{\C{\J}}
      -
      L \gamma_S
    \right)
  \quad \text{on } \Gamma
  \end{aligned}
\end{myalign}
which after several eliminations and rearrangements can be expressed as:
\begin{myalign*}
  &
  \left\{
  \begin{aligned}
    \left( \heatcapv{\phase} - \heatcond{\phase} \lap \right) \omega_{\T{\phase}} &= 0 &&\text{in } \dom{\phase}, \quad \phase = \sph,\lph
    \\
    \left[ \omega_T \right] &= 0 &&\text{on } \Gamma
    \\
    \left[ \heatcond{} \ddn{\omega_T} \right] &= -\left( \zeta^\prime_{\Tl} + \zeta^\prime_{\Ts}\right) &&\text{on } \Gamma
    \\
    \heatcond{\phase} \ddn[\phase]{\omega_{\T{\phase}}} &= 0 &&\text{on } \partial\Omega \cap \overline{\dom{\phase}}, \quad \phase = \sph,\lph
  \end{aligned}
  \right.
  \\
  \\&
  \left\{
  \begin{aligned}
    \left( a - \D{\J} \lap \right) \omega_{\C{\J}} &= 0 &&\text{in } \dom{\lph}
    \\
    \D{\J} \ddn[l]{\omega_{\C{\J}}} - (1-\partcoeff_\J) \vn \omega_{\C{\J}} &= - \zeta^\prime_{\C{\J}}  &&\text{on } \Gamma
    \\
    \D{\J} \ddn[l]{\omega_{\C{\J}}} &= 0 &&\text{on } \partial\Omega \cap \overline{\dom{\lph}}
  \end{aligned}
  \right.
  \\
  \\&
  \left.
  \begin{aligned}
    \gamma_v = \frac{1}{(1-\partcoeff_1) \Cg} \left( \zeta^\prime_{\vn} - \sum_{\J=2}^{N} (1-\partcoeff_\J) \omega_{\C{\J}} \C{\J} - L \omega_{\Tl} \right)  &&\text{on } \Gamma
  \end{aligned}
  \right.
  \\
  \\&
  \left\{
    \begin{aligned}
    \left( a - \D{1} \lap \right) \omega_{\C{1}} &= 0 &&\text{in } \dom{\lph}
    \\
    \omega_{\C{1}} &= \gamma_{v} &&\text{on } \Gamma
    \\
    \D{1} \ddn[l]{\omega_{\C{1}}} &= 0 &&\text{on } \partial\Omega \cap \overline{\dom{\lph}}
    \end{aligned}
  \right.
\end{myalign*}

Finally, by taking variation of the Lagrangian with respect to $\Cg$ we obtain the full differential of the cost functional:
\begin{myalign*}
  d \mathcal{F} =
  \delta_{\Cg} \mathcal{\lph} =
  -
  \int_\Gamma 
    \gamma_{\C{1}}
    \delta \Cg
    \diff{\Gamma},
\end{myalign*}
or, after taking into account equations \eqref{eq:lagrange:raw_multipliers}:
\begin{myalign*}
  d \mathcal{F} =
  \delta_{\Cg} \mathcal{\lph} =
  \int_\Gamma 
    \left(
    \zeta^\prime_{\C{1}}
    + \D{1} \ddn[l]{\omega_{\C{1}}}
    - (1-\partcoeff_1) \vn \omega_{\C{1}}
    \right)
    \delta \Cg
    \diff{\Gamma},
\end{myalign*}
thus, the functional derivative of $\mathcal{F}$ with respect to $\Cg$ is 
\begin{myalign*}
  \fd{\mathcal{F}}{\Cg} = 
    \zeta^\prime_{\C{1}}
    + \D{1} \ddn[l]{\omega_{\C{1}}}
    - (1-\partcoeff_1) \vn \omega_{\C{1}}.
\end{myalign*}
Note that for the functional considered in \ref{sec:derivation:nonlinear:newton}:
\begin{myalign*}
  \zeta^\prime_{\Tl} &= \delta\of{\vecr-\vecr_0},
  \\
  \zeta^\prime_{\Ts} &= 0,
  \\
  \zeta^\prime_{\C{\J}} &= - \pd{\Delta T_C}{\C{\J}} \delta\of{\vecr-\vecr_0},
  \\
  \zeta^\prime_{\vn} &= - \epsilon_v \delta\of{\vecr-\vecr_0}.
\end{myalign*}

\section{Directional derivative with respect to $\delta \Cg$}\label{app:derivation2}

The expressions derived in \ref{app:derivation} predict the change in a functional in response to any perturbation of $\Cg$, however, they require solution of an adjoint system of PDEs for every point on the boundary, which is hardly achievable in practice. Instead, sometimes it is necessary to only know the derivative of a functional along a given perturbation in $\Cg$, e.g. as in the approximate Newton described in \ref{sec:derivation:nonlinear:newton}. 

Let us again consider a generic functional $\mathcal{F}$ defined in \eqref{eq:genericfunctional}. Let us consider system of equations \eqref{eq:spliting:c1}-\eqref{eq:spliting:t} for $\Cg$ and $\Cg + \varepsilon \delta \Cg$, where $\varepsilon \ll 1$. It is easy to show that solutions of \eqref{eq:spliting:c1}-\eqref{eq:spliting:t} in these two cases are related to each other as:
\begin{myalignat*}{2}
\at{\C{\J}}{{\Cg} + \varepsilon \delta \Cg} &= \at{\C{\J}}{{\Cg}} + \varepsilon \lam_{\C{\J}} + \Oof{\varepsilon^2}, \quad &&\J \in \range{1}{N} \\
\at{\T{\phase}}{{\Cg} + \varepsilon \delta \Cg} &= \at{\T{\phase}}{{\Cg}} + \varepsilon \lam_{\T{\phase}} + \Oof{\varepsilon^2}, \quad &&\phase = \sph,\lph\\
\at{\vn}{{\Cg} + \varepsilon \delta \Cg} &= \at{\vn}{{\Cg}} + \varepsilon \lam_{\vn} + \Oof{\varepsilon^2},
\end{myalignat*}
where $\set{\lam_{\C{\J}}}{\J=1}{N}$, $\set{\lam_{\T{\phase}}}{\phase = \sph,\lph}{}$, and $\lam_{\vn}$ satisfy the following adjoint system of equations:
\begin{myalign*}
\left\{
\begin{alignedat}{2}
  \left(a - \D{1} \nabla^2 \right) \lam_{\C{1}} &= 0, \quad &&\textrm{in } \dom{\lph}, \\
  \lam_{\C{1}} &= \delta \Cg, \quad &&\textrm{on } \Gamma, \\
  \D{1} \ddn[l]{\lam_{\C{1}}} &= 0, \quad &&\textrm{on } \partial\Omega\cap\dom{\lph}.
\end{alignedat}
\right.
\end{myalign*}
\begin{myalign*}
  \lam_{\vn} = \frac{1}{(1-\partcoeff_1) \C{1}} \left( \D{1} \ddn[l]{\lam_{\C{1}}} - \vn (1-\partcoeff_1) \lam_{\C{1}} \right) 
\end{myalign*}
\begin{myalign*}
\left\{
\begin{alignedat}{2}
  \left(a - \D{\J} \nabla^2 \right) \lam_{\C{\J}} &= 0, \quad &&\textrm{in } \dom{\lph}, \\
  \D{\J} \ddn[l]{\lam_{\C{\J}}} - (1-\partcoeff_\J) \vn \lam_{\C{\J}} &= (1-\partcoeff_\J) \lam_{\vn} \C{\J}, \quad &&\textrm{on } \Gamma, \\
  \D{\J} \ddn[l]{\lam_{\C{\J}}} &= 0, \quad &&\textrm{on } \partial\Omega\cap\dom{\lph}.
\end{alignedat}
\right.
\end{myalign*}
\begin{myalign*}
\left\{
\begin{aligned}
	\left( \heatcapv{\phase} - \heatcond{\phase} \lap \right) \lam_{\T{\phase}} &= 0 &&\text{in } \dom{\phase}, \quad \phase = \sph,\lph
	\\
	\jump{\lam_{T}} &= 0 &&\text{on } \Gamma
	\\
	\jump{\heatcond{} \ddn{\lam_{T}}} &= \latheat \lam_{\vn} &&\text{on } \Gamma
	\\
	\heatcond{\phase} \ddn[\phase]{\lam_{\T{\phase}}} &= 0 &&\text{on } \partial\Omega \cap \overline{\dom{\phase}}, \quad \phase = \sph,\lph
\end{aligned}
\right.
\end{myalign*}
Using this result the derivative of $\mathcal{F}$ in the direction $\delta \Cg$ can be computed as:
\begin{myalign*}
\int_\Gamma \fd{\mathcal{F}}{\Cg} \delta\Cg \diff{\Gamma}
=
\lim\limits_{\varepsilon \rightarrow 0}
\frac{1}{\varepsilon}
\left(
\at{\mathcal{F}}{{\Cg} + \varepsilon \delta \Cg} - \at{\mathcal{F}}{{\Cg}} 
\right)
= 
\int_{\Gamma} 
\left( 
\sum_{\phase=\sph,\lph} \zeta^\prime_{\T{\phase}} \lam_{\T{\phase}} + 
\sum_{\J=1}^{N} \zeta^\prime_{\C{\J}} \lam_{\C{\J}} + 
\zeta^\prime_{\vn} \lam_{\vn}
\right) \diff{\Gamma}.
\end{myalign*}

\section{Details of linear stability analysis of iterative schemes for solving nonlinear system of PDEs}
\label{app:stability}

In the simple case of a planar geometry considered in section \ref{sec:derivation:nonlinear:stability} the fixed-point and approximate Newton iterations can be explicitly written as:
\begin{myalign}\label{eq:iterationsexplicit}
   \Cg^{(q+1)} \of{x} &= \Cg^{(q)} \of{x} - \frac{\Tl^{(q)}\of{x,0} - \sum_{\J=1}^N \ml{\J} \C{\J}^{(q)}\of{x,0} - h_G\of{x}}{\ml{1}}, \\
   \Cg^{(q+1)} \of{x} &= \Cg^{(q)} \of{x} - \frac{\Tl^{(q)}\of{x,0} - \sum_{\J=1}^N \ml{\J} \C{\J}^{(q)}\of{x,0} - h_G\of{x}}{\lam_{\Tl}\of{x,0} - \sum_{\J=1}^N \ml{\J} \lam_{\C{\J}}\of{x,0}}.
\end{myalign}  
It is straightforward to show that the solution of \eqref{eq:spliting:c1}-\eqref{eq:spliting:t} corresponding to a perturbed boundary concentration of the form $$\Cg^{(q)} = \tilde{\Cg} + r^q \delta_C \exp \of{-i \omega_x x} $$ can be found in the form:
\begin{myalignat*}{2}
  \C{\J}^{(q)}\of{x,y} &= \tilde{\C{\J}}\of{y} + A_\J r^\nu \delta_0 \exp \of{ -i \omega_x x } \exp \of{ -\Omega_\J y } + \Oof{\delta_C^2}, \quad &\J &\in \range{1}{N}, \\
  \T{\phase}^{(q)}\of{x,y} &= \tilde{\T{\phase}}\of{y} + B_{\phase} r^\nu \delta_0 \exp \of{ -i \omega_x x } \exp \of{ -\Omega_{\phase} y } + \Oof{\delta_C^2}, \quad &\phase &= \sph,\lph \\
  \vn^{(q)}\of{x} &= \tilde{\vn} + E r^\nu \delta_0 \exp \of{ -i \omega_x x } + \Oof{\delta_C^2},
\end{myalignat*}
where $\Omega_\J = \Omega_\J \of{\omega_x}$, $\J\in\range{1}{N}$, and $\Omega_{\phase} = \Omega_{\phase} \of{\omega_x}$, $\phase = \lph,\sph$, satisfy \eqref{eq:frequencies} and  
\begin{myalign*} 
  E &= \frac{\D{1} \Omega_1 - \tilde{\vn} (1-\partcoeff_1)}{(1-\partcoeff_1)\tilde{\C{1}}\of{0}},
  \\
  A_\J &= 
  \left(
  \frac{\tilde{\C{\J}}\of{0}}{\tilde{\C{1}}\of{0}}
  \right) 
  \left(
  \frac{1-\partcoeff_\J}{1-\partcoeff_1}
  \right)
  \left(
  \frac{\D{1} \Omega_1 - \tilde{\vn} (1-\partcoeff_1)}{\D{\J} \Omega_\J - \tilde{\vn} (1-\partcoeff_\J)}
  \right), \quad \J = 1, \ldots, N,
\\
  B_{\sph} = B_{\lph} &= 
  \frac{\latheat}{(1-\partcoeff_1)\tilde{\C{1}}\of{0}} 
  \frac{\D{1} \Omega_1 - \tilde{\vn} (1-\partcoeff_1)}{\heatcond{\sph} \dom{\sph} + \heatcond{\lph} \dom{\lph}}.
\end{myalign*}
The adjoint system of equation \eqref{eq:adjoint:c1}-\eqref{eq:adjoint:t} in this case has the solution:
\begin{myalign*}
  \lam_{\C{1}}^{(q)} \of{x,y} &= \exp \of{ -\Omega_1\of{0} y } + \Oof{\delta_C}, \\
  \lam_{v}^{(q)} \of{x} &= \frac{\D{1} \Omega_1\of{0} - (1-\partcoeff_1) \tilde{\vn}}{(1-\partcoeff_1)\tilde{\C{1}}\of{0}} + \Oof{\delta_C}, \\
  \lam_{\C{\J}}^{(q)} \of{x,y} &= 
  \left( \frac{1-\partcoeff_\J}{1-\partcoeff_1} \right)
  \left( \frac{\tilde{\C{\J}}\of{0}}{\tilde{\C{1}}\of{0}} \right)
  \left(
  \frac{\D{1} \Omega_1\of{0} - (1-\partcoeff_1) \tilde{\vn}}{\DD{\lph}{\J} \Omega_\J\of{0} - (1-\partcoeff_\J) \tilde{\vn}}
  \right)
  \exp \of{ -\Omega_\J \of{0} y } + \Oof{\delta_C},\quad \J = 2,\ldots,N \\
  \lam_{\Tl} &=
  \frac{\latheat}{(1-\partcoeff_1)\tilde{\C{1}}\of{0}}
  \frac{\D{1} \Omega_1\of{0} - \tilde{\vn} (1-\partcoeff_1)}{\heatcond{\sph} \dom{\sph}\of{0} + \heatcond{\lph} \dom{\lph}\of{0} }
  \exp \of{ -\dom{\lph}\of{0} y } + \Oof{\delta_C}, \\
  \lam_{\Ts} &=
  \frac{\latheat}{(1-\partcoeff_1)\tilde{\C{1}}\of{0}} 
  \frac{\D{1} \Omega_1\of{0} - \tilde{\vn} (1-\partcoeff_1)}{\heatcond{\sph} \dom{\sph}\of{0} + \heatcond{\lph} \dom{\lph}\of{0} }
  \exp \of{ \dom{\sph}\of{0} y } + \Oof{\delta_C}.
\end{myalign*}
Note that it is not necessary to obtain linear correction while solving the adjoint system of equations because 
\begin{myalign*}
  \Tl^{(q)}\of{x,0} - \sum_{\J=1}^N \ml{\J} \C{\J}^{(q)}\of{x,0} - h_G\of{x} = \Oof{\delta_C}
\end{myalign*}
Substitution of the above expressions into \eqref{eq:iterationsexplicit} produces amplification factors \eqref{eq:amplification:fp} and \eqref{eq:amplification:an}

\section{Removing extremely underresolved regions}\label{app:removing}
In order to ensure the robustness of numerical simulations, we use the following two-pass strategy to regularize underresolved geometries.

During the first pass, narrow gaps of liquid material of the size less than two grid spacings are ``bridged''. Specifically, we compute an auxiliary level-set function $\phi_{\textrm{aux}}$ as the signed distance to the $\phi_{n}= - \Delta x$ isocontour of the original level-set function $\phi_{n}$ and shift it back by $\Delta x$. In the case when the geometry is sufficiently resolved, $\phi_{\textrm{aux}}$ and $\phi_{n}$ have coinciding signs on all grid nodes and very close in values. In the case when narrow regions of liquid material of width less than $2 \Delta x$ are present  $\phi_{\textrm{aux}}$ and $\phi_{n}$ will have different signs on grid nodes in such regions (more precisely, $\phi_{\textrm{aux}} > 0$ and $\phi_{n} < 0$). Substituting values of $\phi_{n}$ with values of $\phi_{\textrm{aux}}$ effectively eliminates such under-resolved regions. Note that $\phi_{n}$ remains unchanged whenever the front's geometry is sufficiently resolved.

\begin{figure}[!h]
  \centering  
  \begin{subfigure}{.3\textwidth}
    \centering
    \includegraphics[width=0.7\textwidth]{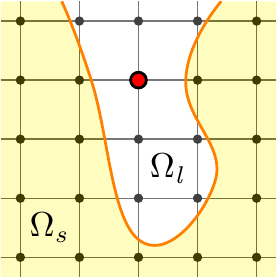}
    \caption{}
  \end{subfigure}
  \begin{subfigure}{.3\textwidth}
    \centering
    \includegraphics[width=0.7\textwidth]{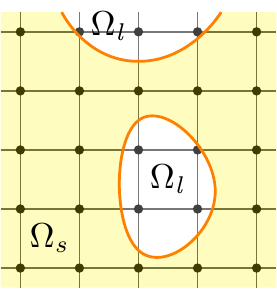}
    \caption{}
  \end{subfigure}
  \begin{subfigure}{.3\textwidth}
    \includegraphics[width=0.7\textwidth]{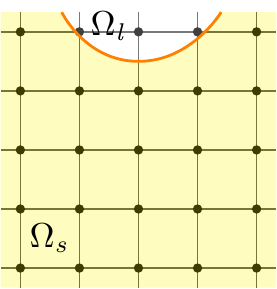}
    \centering
    \caption{}
  \end{subfigure}
  \caption{Illustration of the procedure used for removing extremely under-resolved regions of liquid: (a) identification of grid nodes at which locally distinct parts of $\dom{\sph}$ are separated just by a single node; (b) ``bridging'' narrow liquid gaps at such grid nodes; (c) identification and ``solidification'' of liquid pools left behind by the previous step.}
  \label{fig:underresolved}
\end{figure}

During the second pass, isolated pools of liquid created as a result of such ``bridging'' procedure (if any) are identified and ``solidified'' as well. The identification of isolated pools on distributed computational grids is done using the parallel ``island counting'' algorithm described in \cite{mistani2018island}.

We have observed that such a procedure ensures an excellent robustness of the computational scheme across all crystal growth regimes. It is especially useful for simulating the cellular regime and its transition to the planar growth.

\section{Similarity solution for the solidifying infinite cylinder due to a heat sink} \label{app:analytical}

Let us consider the axisymmetric solidification of an infinite cylinder from a line heat sink of strength $Q$ located at the cylinder's center into an infinite liquid alloy of composition $\set{\C{\J}^\infty}{\J=1}{N}$ and temperature $T^\infty$. In this case spatial distributions of temperature $\T{\phase} = \T{\phase} \of{t,r}$, $\phase = \sph,\lph$, and concentrations $\Cl{\J} = \Cl{\J} \of{t,r}$, $\J \in \range{1}{N}$, are only functions of time $t$ and distance from the cylinder's center $r$, which without loss of generality can be assumed  at $r=0$. In addition to assumptions made in section \ref{sec:model}, we further assume that the constitutional undercooling has a linear dependence on the composition ($\Tliq =  T_m + \sum_{\J=1}^{N} \ml{\J} \Cl{\J}$), that the kinetic and curvature undercoolings are negligible ($\epsilon_c = \epsilon_v = 0$) and that the partition coefficients $\set{\partcoeff_{i}}{\J=1}{N}$ are constant. Denote the cylinder's radius as $R\of{t}$. Mathematically such a problem can be formulated as:
\begin{myalign}\label{eq:similarity:system}
\begin{aligned}
\text{\textbf{Governing equations}:} \\
\text{\textit{Heat transport:}} &&
\den{\sph} \heatcap{\sph} \ddt{\Ts} - \heatcond{\sph} \lap \Ts &= 0 \quad \text{for } 0<r<R\of{t},	\\
&&
\den{\lph} \heatcap{\lph} \ddt{\Tl} - \heatcond{\lph} \lap \Tl &= 0 \quad \text{for } R\of{t}<r<\infty,	\\
\text{\textit{Species transport:}} &&
	\ddt{\Cl{\J}} - \DD{\lph}{\J} \lap \Cl{\J} &= 0 \quad \text{for } R\of{t}<r<\infty, \\
&&&\quad \J \in \range{1}{N}
\\
\text{\textbf{Interface conditions}:}
\\
\text{\textit{Temperature continuity:}} &&
\jump{ T } &= 0,
\\
\text{\textit{Stefan condition:}} &&
\jump{\heatcond{} \del{r}T } &= \vn L ,
\\
\text{\textit{Gibbs-Thomson:}} &&
\Tl &= T_m + \sum_{\J=1}^{N} \ml{\J} \C{\J} ,
\\
\text{\textit{Solute-rejection:}} &&
\DD{\lph}{\J} \del{r} \C{\J} + (1-\partcoeff_\J) \vn \C{\J} &= 0, \quad \J \in \range{1}{N},
\\
\text{\textbf{Boundary conditions}:} \\
\text{\textit{Line source:}} &&
\lim\limits_{r \rightarrow 0} \left( 2\pi r \heatcond{\sph} \del{r} \Ts \right) &= Q, \\
\text{\textit{Temperature:}} &&
\lim\limits_{r \rightarrow \infty} \Tl &= T^\infty, \\
\text{\textit{Composition:}} &&
\lim\limits_{r \rightarrow \infty} \Cl{\J} &= \C{\J}^\infty, \quad \J \in \range{1}{N}, \\
\text{\textbf{Initial conditions}:} \\
\text{\textit{Front location:}} &&
\at{R}{t=0} &= 0, \\
\text{\textit{Temperature:}} &&
\at{\Tl}{t=0} &= T^\infty, \\
\text{\textit{Composition:}} &&
\at{\Cl{\J}}{t=0} &= \C{\J}^\infty, \quad \J \in \range{1}{N}, \\
\end{aligned}
\end{myalign}
where the same notation as in section \ref{sec:model} is used. Note that in the axisymmetric case the Laplace operator has the form:
\begin{myalign*}
  \lap &= \frac{1}{r} \partial_r \left( r \partial_r \right).
\end{myalign*}

It can be shown that similarly to other Stefan-type problems the considered problem admits a similarity solution of the form:
\begin{myalign}\label{eq:similarity:form}
\begin{aligned}
R\of{t} &= 2 \sqrt{\theta t}, \\
\vn\of{t} &= \sqrt{\frac{\theta}{t}}, \\
\Ts\of{t,r} &= A^s + B^s E_1  \left( \frac{r^2}{4 \heatdiff{\sph} t} \right), \\
\Tl\of{t,r} &= A^l + B^l E_1  \left( \frac{r^2}{4 \heatdiff{\lph} t} \right), \\
\Cl{\J}\of{t,r} &= A_\J + B_\J E_1  \left( \frac{r^2}{4 \DD{\lph}{\J} t} \right), \quad \J \in \range{1}{N},
\end{aligned}
\end{myalign}
where $\heatdiff{\sph} = \frac{\heatcond{\sph}}{\den{\sph} \heatcap{\sph}}$, $\heatdiff{\lph} = \frac{\heatcond{\lph}}{\den{\lph} \heatcap{\lph}}$ are thermal diffusivities and $E_1$ denotes the exponential integral:
\begin{myalign*}
  E_1(z) = \int_x^\infty \frac{e^{-s}}{\sph} ds,\quad z > 0.
\end{myalign*}
Indeed, using direct substitution one can show that \eqref{eq:similarity:form} is the solution to \eqref{eq:similarity:system} provided the values of constants $A^s$, $B^s$, $A^l$, $B^l$, $A_\J$, $B_\J$, $\J \in \range{1}{N}$, are given by:
\begin{myalignat*}{2}
  A^s &= T^\ast \of{\theta} + \frac{Q}{4\pi \heatcond{\sph}} E_1 \left( \frac{\theta}{\alpha_s} \right), & \quad
  B^s &= -\frac{Q}{4\pi \heatcond{\sph}}, \\
  A^l &= T^\infty, &\quad
  B^l &= \frac{T^\ast\of{\theta} - T^\infty}{E_1 \left( \frac{\theta}{\heatdiff{\lph}} \right)}, \\
  A_\J &= \C{\J}^\infty, \quad \J \in \range{1}{N}, &\quad
  B_\J &= \frac{C^\ast_\J\of{\theta} - \C{\J}^\infty}{E_1 \left( \frac{\theta}{\DD{\lph}{\J}} \right)}, \quad \J \in \range{1}{N},
\end{myalignat*}
and quantity $\theta$, called the growth constant, satisfies the nonlinear algebraic equation:
\begin{myalign*}
  T^\ast \of{\theta} = T_m + \sum_{\J=1}^{N} \ml{\J} \C{\J}^\ast \of{\theta}
\end{myalign*}
where
\begin{myalign*}
  C^\ast_\J\of{\theta} &= \frac{\C{\J}^\infty}{1-(1-\partcoeff_\J) \frac{\theta}{\DD{\lph}{\J}} \exp \left( \frac{\theta}{\DD{\lph}{\J}} \right) E_1 \left( \frac{\theta}{\DD{\lph}{\J}} \right)}, \quad \J \in \range{1}{N} \\
  T^\ast\of{\theta} &= T^\infty + \frac{\theta}{\heatdiff{\lph}} \exp \left( \frac{\theta}{\heatdiff{\lph}} \right) E_1 \left( \frac{\theta}{\heatdiff{\lph}} \right) \left(  \frac{\lph}{\den{\lph} \heatcap{\lph}} - \frac{Q}{4\pi \heatcond{\sph}} \frac{\den{\sph} \heatcap{\sph}}{\den{\lph} \heatcap{\lph}}  \frac{1}{\frac{\theta}{\alpha_s} \exp \left( \frac{\theta}{\alpha_s} \right)}\right).
\end{myalign*}
Note that:
\begin{myalignat*}{2}
  \lim\limits_{\theta \rightarrow 0} \C{\J}^\ast &= \C{\J}^\infty, \quad &\lim\limits_{\theta \rightarrow \infty} \C{\J}^\ast &= \frac{\C{\J}^\infty}{\partcoeff_\J}, \quad \J \in \range{1}{N}, \\
  \lim\limits_{\theta \rightarrow 0} T^\ast &= -\infty, \quad &\lim\limits_{\theta \rightarrow \infty} T^\ast &= T^\infty + \frac{\lph}{\den{\lph} \heatcap{\lph}}.
\end{myalignat*}
Thus, for $Q>0$ and $0 < \partcoeff_\J < 1$ such a solution exists as long as:
\begin{myalign*}
  T^\infty > T_m + \sum_{i = 1}^{N} \ml{\J} \frac{\C{\J}^\infty}{\partcoeff_\J} - \frac{\lph}{\den{\lph} \heatcap{\lph}}.
\end{myalign*}

Existing of such an analytical solution allows creating of a non-trivial benchmark test for a multidimensional solidification code. Specifically, in this work we consider an annular region with internal and external radii $R_\textrm{in}$ and $R_\textrm{out}$. We start with initial conditions given by the analytical solution at some initial time $t_0$, such that $R_\textrm{in} < R\of{t_0} < R_\textrm{out}$, and impose time-dependent boundary conditions (Dirichlet or Neumann) on the inner and outer boundaries of the region based on the analytical solution.

Interesting features of this similarity solution are that the values of temperature and concentration at the solidification front are constant throughout the entire solidification process, that is:
\begin{myalign*}
  \Tl = \Ts \of{t,R\of{t}} = T^\ast \quad \textrm{and} \quad \C{\J}^l\of{t,R\of{t}} = C^\ast_\J, \, \J \in \range{1}{N}, \, \forall\,t>0,
\end{myalign*}
and that the ratio of compositional and thermal gradients at the solidification front is constant as well
\begin{myalign*}
  M = \at{\frac{\sum_{\J=1}^{N} \ml{\J} \del{r} \C{\J}^l}{\del{r} \Tl}}{r=R\of{t}} = \sum_{\J=1}^{N} \ml{\J} \frac{\C{\J}^\infty-\C{\J}^\ast}{T^\infty-T^\star} \frac{\exp \left( \frac{\theta}{\heatdiff{\lph}} \right) E_1 \left( \frac{\theta}{\heatdiff{\lph}} \right)}{\exp \left( \frac{\theta}{\DD{\lph}{\J}} \right) E_1 \left( \frac{\theta}{\DD{\lph}{\J}} \right)}.
\end{myalign*}
Recall that the solidification front is expected to be stable for $M < 1$ and unstable for $M > 1$ (compositional undercooling ahead of the front). For purposes of verification of multidimensional solidification codes it is desired to consider stable processes, otherwise due to unavoidable numerical errors the numerical solution  would quickly diverge from the symmetric configuration predicted by the analytical solution and a comparison would not be possible. For this reason it is more convenient to select an analytical solution based on the value of the gradients' ratio $M=M_0$ instead of imposing the strength of heat sink $Q$ and the temperature value at infinity $T^\infty$. In addition, to more easily and independently select the characteristic front velocity we impose a specific value of the front velocity $\vn\of{t_0} = v_0$ at the beginning of simulation $t_0$ when the seed radius is equal to a given $R\of{t_0} = R_0$. Thus, alternatively to \eqref{eq:similarity:system}, boundary conditions can be formulated as:
\begin{myalign*}
  M &= M_0, \\
  \vn\of{t_0} &= v_0, \textrm{ where $t_0$ is such that } R\of{t_0} = R_0 \\
  \lim\limits_{r\rightarrow \infty} \C{\J}\of{t,r} &= \C{\J}^\infty, \quad \J \in \range{1}{N}.
\end{myalign*}
In this case the integration constants in the analytical solution are given by:
\begin{myalignat*}{2}
  \theta &= \hf v_0 R_0 \\
  A_\J &= \C{\J}^\infty,  \quad \J \in \range{1}{N}, &\quad 
  B_\J &= \frac{C^\ast_\J\of{\theta} - \C{\J}^\infty}{E_1 \left( \frac{\theta}{\DD{\lph}{\J}} \right)},  \quad \J \in \range{1}{N},\\
  A^l &= T^\ast\of{\theta} - B^l E_1\left(\frac{\theta}{\heatdiff{\lph}}\right), &\quad
  B^l &= \frac{1}{M_0} \sum_{\J=1}^{N} \ml{\J} B_\J \frac{\exp \left(\frac{\theta}{\heatdiff{\lph}} \right)}{\exp \left(\frac{\theta}{\DD{\lph}{\J}} \right)}, \\
  A^s &= T^\ast\of{\theta} - B^s E_1\left(\frac{\theta}{\alpha_s}\right), & \quad
  B^s &= B^l \frac{\den{\lph} \heatcap{\lph}}{\den{\sph} \heatcap{\sph}} \frac{ \frac{\theta}{\heatdiff{\sph}} \exp \left( \frac{\theta}{\heatdiff{\sph}} \right)}{ \frac{\theta}{\heatdiff{\lph}} \exp \left( \frac{\theta}{\heatdiff{\lph}} \right)} - \frac{\lph}{\den{\sph} \heatcap{\sph}} \frac{\theta}{\heatdiff{\sph}} \exp \left( \frac{\theta}{\heatdiff{\sph}} \right), 
\end{myalignat*}
where
\begin{myalign*}
  \C{\J}^\ast\of{\theta} &= \frac{\C{\J}^\infty}{1-(1-\partcoeff_\J) \frac{\theta}{\DD{\lph}{\J}} \exp \left( \frac{\theta}{\DD{\lph}{\J}} \right) E_1 \left( \frac{\theta}{\DD{\lph}{\J}} \right)}, \quad \J \in \range{1}{N}, \\
  T^\ast\of{\theta} &= T_m + \sum_{\J=1}^{N} \ml{\J} \C{\J}^\ast\of{\theta}.
\end{myalign*}
Note that the initial moment of time for setting up numerical simulations is given by:
\begin{myalign*}
  t_0 = \hf \frac{R_0}{v_0}.
\end{myalign*}

It is also easy to extend the above similarity solution to the case of a nonlinear liquidus surface $\Tliq = \Tliq\of{\CC{\lph}{1}^\ast, \ldots, \CC{\lph}{N}^\ast}$ and nonconstant partition coefficients $\set{\partcoeff_\J = \partcoeff_\J \of{\CC{\lph}{1}^\ast, \ldots, \CC{\lph}{N}^\ast}}{\J=1}{N}$. In such a case the interfacial concentration are found by solving the nonlinear algebraic system of equations (we found that the fixed-point iteration suffices):
\begin{myalign*}
  \C{\J}^\ast\of{\theta} &= \frac{\C{\J}^\infty}{1-(1-\partcoeff_\J \of{\CC{\lph}{1}^\ast, \ldots, \CC{\lph}{N}^\ast}) \frac{\theta}{\DD{\lph}{\J}} \exp \left( \frac{\theta}{\DD{\lph}{\J}} \right) E_1 \left( \frac{\theta}{\DD{\lph}{\J}} \right)}, \quad \J \in \range{1}{N},
\end{myalign*}
and the interfacial temperature is simply given by:
\begin{myalign*}
  T^\ast\of{\theta} &= \Tliq\of{\CC{\lph}{1}^\ast, \ldots, \CC{\lph}{N}^\ast}.
\end{myalign*}

\newpage

\section{Additional visualizations for presented simulation runs} \label{app:extra}
\begin{figure}[!h]
\begin{center}
\includegraphics[width=\textwidth]{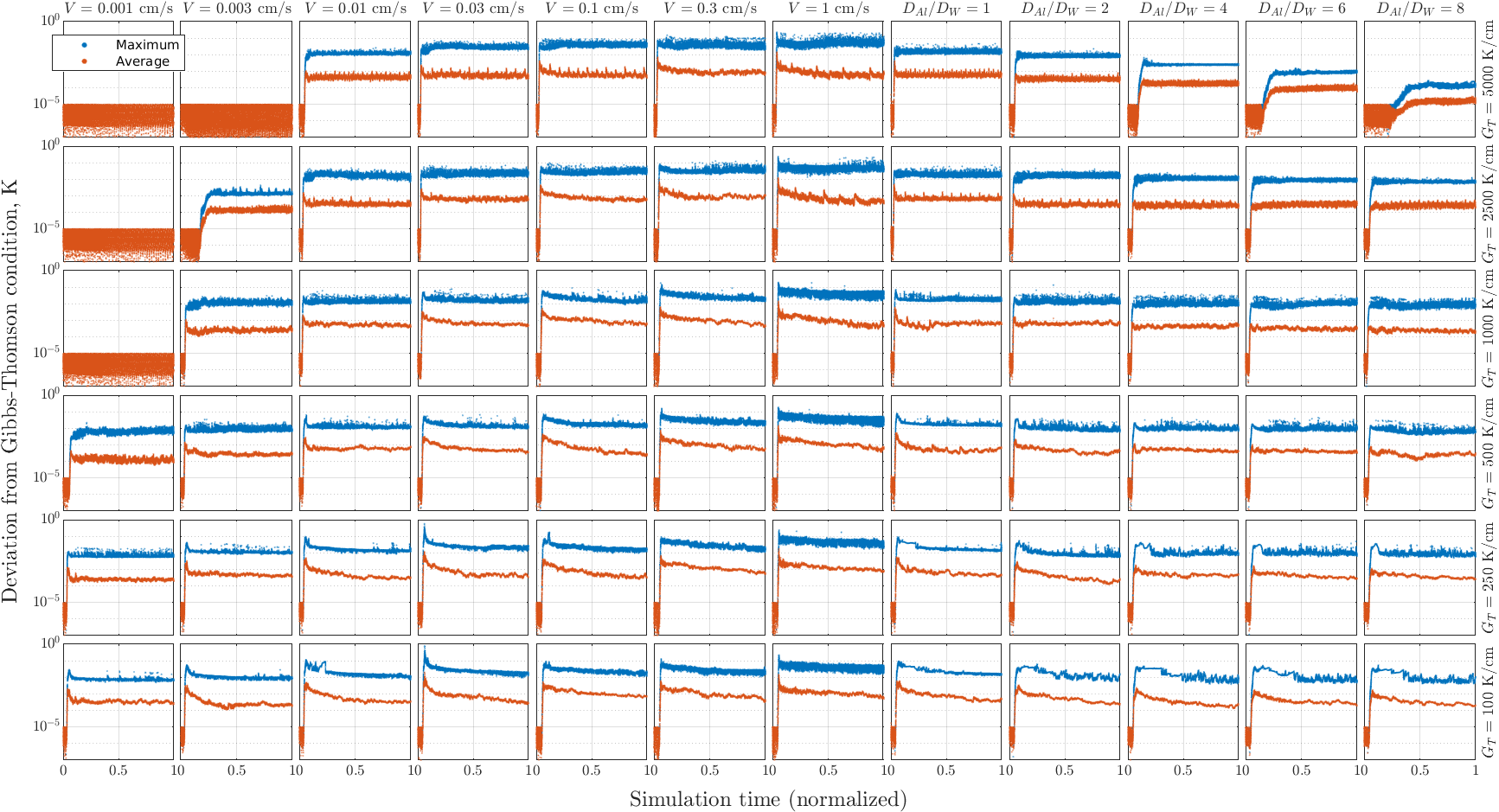}
\end{center}
\vspace{-.3cm}\vspace{-.3cm}\caption{\it Error in satisfying Gibbs-Thomson condition \eqref{eq:model:gibbs} for each time step of simulation runs shown in Figures \ref{fig:directional:visual:full} and \ref{fig:directional:visual:diff}.} \label{fig:directional:bcerror:extra}
\end{figure}

\begin{figure}[!h]
\begin{center}
\includegraphics[width=\textwidth]{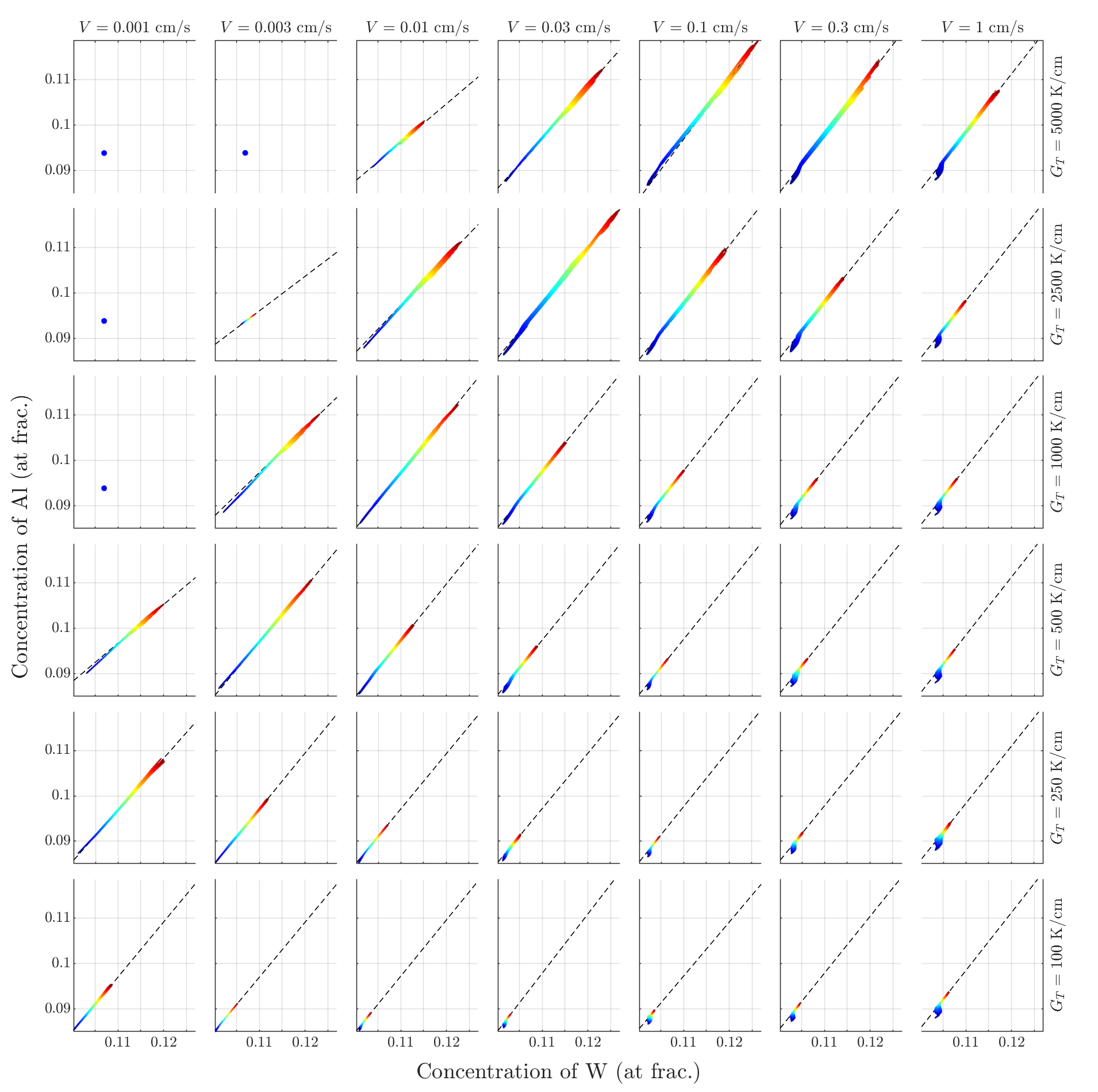}
\end{center}
\vspace{-.3cm}\vspace{-.3cm}\caption{\it Concentration paths (colored according to the relative freezing time) corresponding to cases in Figure~\ref{fig:directional:visual:full}.} 
\label{fig:directional:solidus:full:extra}
\end{figure}

\begin{figure}[!h]
\begin{center}
\includegraphics[width=\textwidth]{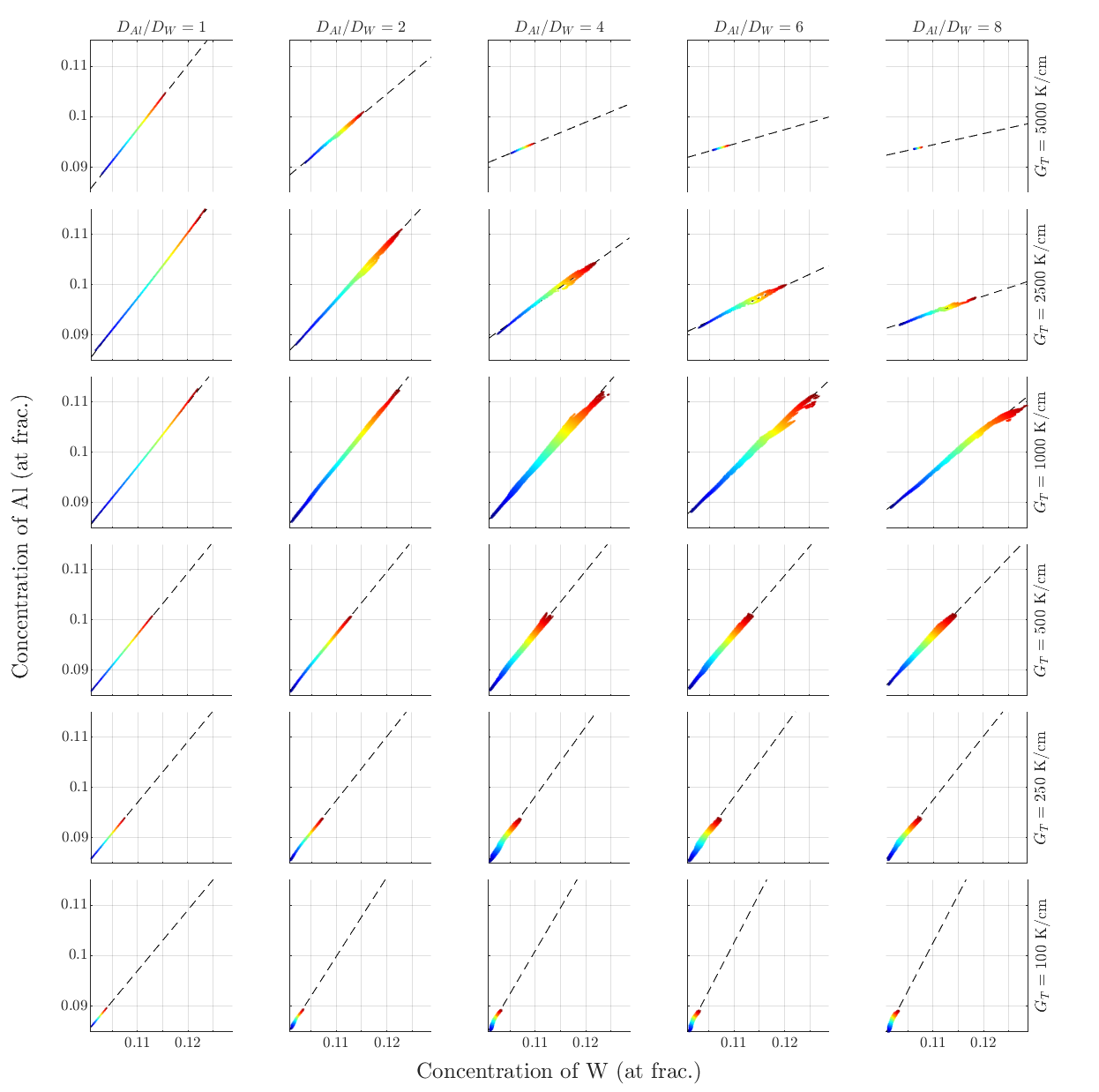}
\end{center}
\vspace{-.3cm}\vspace{-.3cm}\caption{\it Concentration paths (colored according to the relative freezing time) corresponding to cases in Figure~\ref{fig:directional:visual:diff}.}
\label{fig:directional:solidus:diff:extra}
\end{figure}

\begin{figure}[!h]
\begin{center}
\includegraphics[width=\textwidth]{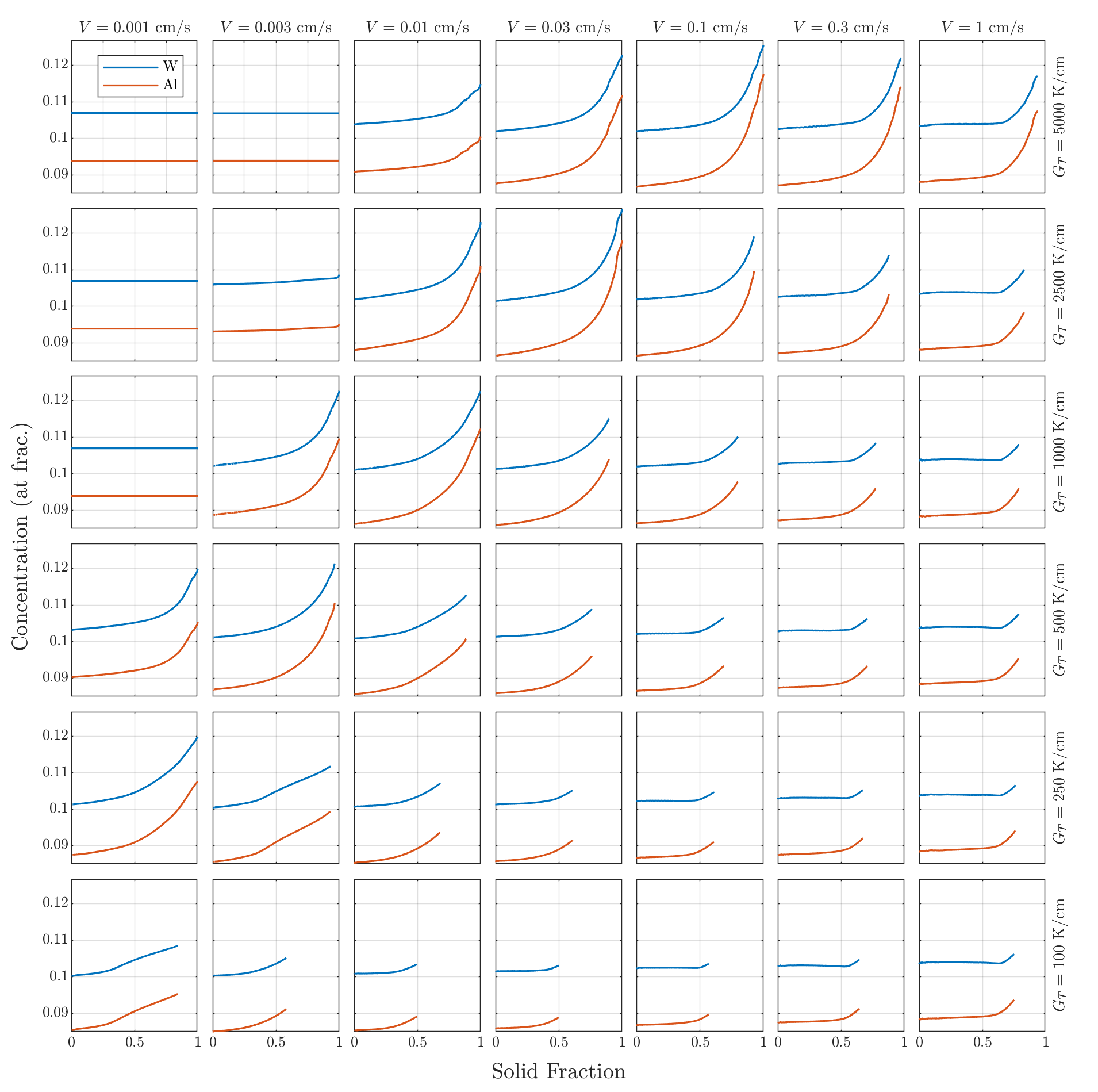}
\end{center}
\vspace{-.3cm}\vspace{-.3cm}\caption{\it Dependence of solutes' concentration on the fraction of material turned into solid phase for cases in Figure~\ref{fig:directional:visual:full}.} 
\label{fig:directional:segregation:full:extra}
\end{figure}
\begin{figure}[!h]
\begin{center}
\includegraphics[width=0.45\textwidth]{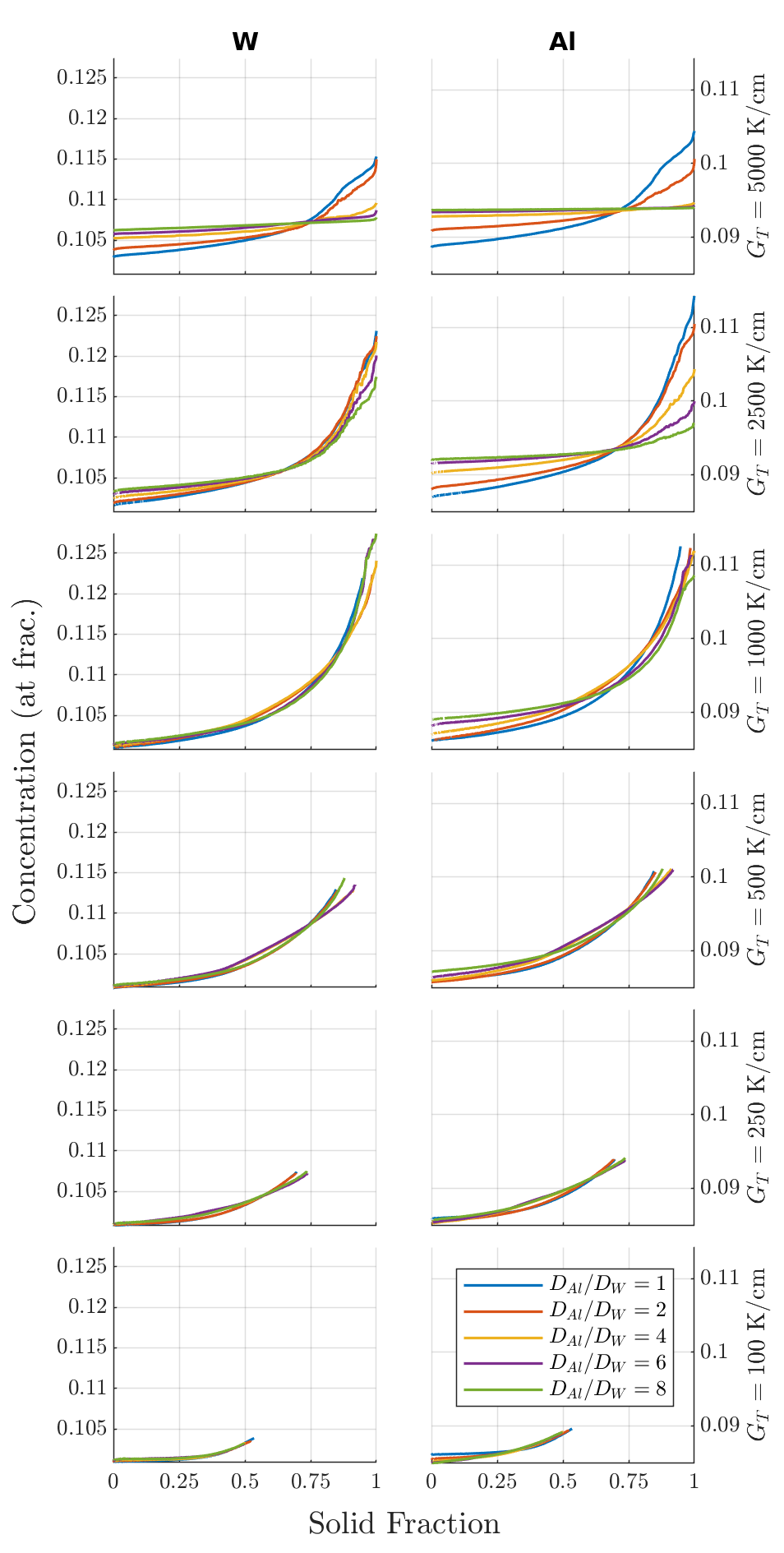}
\includegraphics[width=0.45\textwidth]{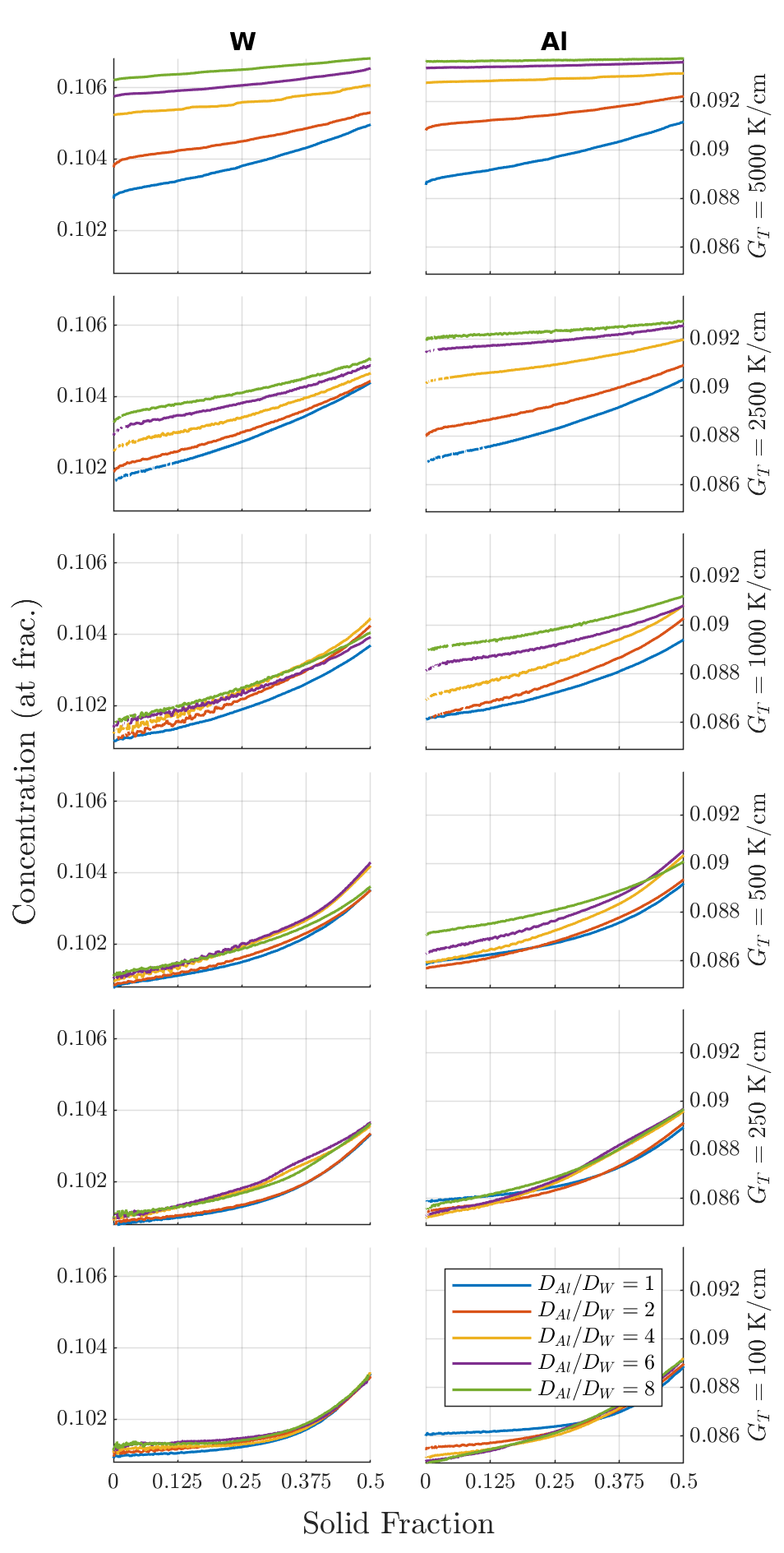}
\end{center}
\vspace{-.3cm}\vspace{-.3cm}\caption{\it Dependence of solutes' concentration on the fraction of material turned into solid phase for cases in Figure~\ref{fig:directional:visual:diff} (left: full curves; right: a zoom-in for solid fractions from 0 to 0.5).} \label{fig:directional:segregation:diff}
\end{figure}

\end{document}